\documentclass[sigconf,nonacm]{acmart}

\usepackage{textcomp}
\usepackage[ruled,linesnumbered,vlined]{algorithm2e}
\usepackage{multirow}
\usepackage{siunitx}
\usepackage{color}
\usepackage{soul}
\usepackage{makecell}
\usepackage{enumitem}
\usepackage{fontawesome}
\usepackage{hyperref}

\DeclareMathOperator*{\argmin}{arg\,min}

\newcommand{\emptycirc}{{\scriptsize\faCircleO}}
\newcommand{\halfcirc}{{\scriptsize\faAdjust}}
\newcommand{\fullcirc}{{\scriptsize\faCircle}}

\newcommand\name{\textsf{PhantomSeal}}

\begin{document}

\title[\name]{\textit{\name}: Proactive Deepfakes Defense with Identity/Context Protection and Forensic Tracing}


\author{Liangqin Ren}
\affiliation{%
  \institution{The University of Kansas}
  \city{Lawrence}
  \state{KS}
  \country{USA}}
\email{liangqinren@ku.edu}

\author{Zeyan Liu}
\affiliation{%
  \institution{University of Louisville}
  \city{Louisville}
  \state{KY}
  \country{USA}}
\email{z0liu020@louisville.edu}

\author{Ye Wang}
\affiliation{%
  \institution{The University of Kansas}
  \city{Lawrence}
  \state{KS}
  \country{USA}}
\email{yeah\_wong@ku.edu}

\author{Yuxin Chen}
\affiliation{%
  \institution{The University of Chicago}
  \city{Chicago}
  \state{IL}
  \country{USA}}
\email{chenyuxin@uchicago.edu}

\author{Fengjun Li}
\affiliation{%
  \institution{The University of Kansas}
  \city{Lawrence}
  \state{KS}
  \country{USA}}
\email{fli@ku.edu}

\author{Bo Luo}
\affiliation{%
  \institution{The University of Kansas}
  \city{Lawrence}
  \state{KS}
  \country{USA}}
\email{bluo@ku.edu}


\begin{abstract}
{Deepfakes, especially face-swapping attacks, pose significant challenges to authenticity, security, and ethics across science, engineering, and society. While most existing detection/tracing approaches operate \textit{post hoc}, proactive defenses that aim to intervene before deepfake generation remain limited in terms of real-world effectiveness.} In this paper, we present \name, the first proactive defense to simultaneously protect both the identity and the context of users' images from being used in face-swapping attacks, while supporting forensic tracing. We present a novel cloaking technique that embeds a selected identity as a stealthy identifier. This mechanism steers the deepfake generation process toward producing content that resembles the chosen cloak identity, thereby preventing successful face-swapping while enabling effective feature-based forensic analysis. The effectiveness and robustness of \name\ is demonstrated in extensive experiments across different face-swapping architectures and models.  For example, it reduces the attack success rate of {\em SimSwap}, an advanced deepfake model, to 0.30\%, and correctly identifies 97.97\% of manipulated content. The source code is available at \url{https://github.com/LiangqinRen/PhantomSeal}.
\end{abstract}



\maketitle

\section*{Publication Notice}

This document is the extended author version of the paper published in the
\textit{Proceedings of the 2026 ACM SIGSAC Conference on Computer and Communications Security (CCS '26)}.
The Version of Record is available at
\url{https://doi.org/10.1145/3830454.3832581}.
This version includes additional technical details and appendices that were omitted from the conference version due to space limitations.

\section{Introduction}\label{sec:intro}

The idiom, ``Seeing is believing'' reflects people’s natural trust in visual content, such as photos and videos. However, advances in Deep Learning (DL) enabled the creation of deepfakes--highly realistic, fabricated multimedia that closely mimics genuine content \cite{nguyen2022deep, li2024adversarial, mirza2014conditional}. Although deepfakes have shown potential in education \cite{lee2019deepfake}, film production \cite{9115874}, and art \cite{itzkoff2016rogue}, they are most often exploited for harmful purposes, such as generating fake intimate images \cite{cole2017ai} and spreading fake news \cite{temir2020deepfake}, raising widespread concerns.
For example, a 2024 investigation by Channel 4 News analyzed the five most trafficked deepfake websites and found that non-consensual intimate deepfakes had been generated using over 4,000 public figures, including Channel 4 presenter Cathy Newman \cite{cathy_newman}. Such platforms, specializing in sexual deepfakes, collectively attracted over 100 million views within a span of three months.

Due to the widespread misuse of face-swapping technologies, the corresponding defense has attracted research attention. Existing face-swapping defenses can be broadly categorized into four categories, as summarized in Table~\ref{tab:summary_of_defense}. 
Detection methods~\cite{8638330, zhao2021multi, huang2021initiative} aim to identify artifacts introduced by generative models to determine whether an image is synthetically generated. Such approaches offer no guaranteed protection, as they fundamentally rely on detectable artifacts that can be deliberately suppressed during model training or inference. This inherent limitation places detection methods and face-swapping models in a continual cat-and-mouse game. 
To improve accountability, watermarking techniques~\cite{lin2022source, wu2023sepmark, yu2021artificial} embed invisible markers into deepfake generation for post hoc traceability, while restoration-based approaches~\cite{shen2024hiding, ai2023deepfake, yu2024dfrec} further integrate watermarking with facial content to enhance interpretability.
Nevertheless, these \textit{ex post facto} countermeasures only take effect after deepfake content has already been generated and disseminated. Consequently, proactive defenses that prevent harm \emph{before} dissemination are highly desirable.

For proactive defense, disruption-based approaches leverage adversarial examples~\cite{zhang2019adversarial} to interfere with the deepfake generation pipeline. Existing approaches either target early processing stages, such as face detection~\cite{zhu2024hiding, segalis2020ogan, zhang2023disrupting} and landmark extraction~\cite{li2024landmarkbreaker}, to prevent identity feature extraction, or manipulate latent representations and encoding processes~\cite{he2022defeating, dong2021visually, shim2023leat, guan2022defending, tang2024feature} to distort face-swapping outputs. In practice, when the deepfake has visually evident distortions or strong noise, an attacker is unlikely to persist with that sample. Instead, they may switch to alternative, unprotected images and relaunch the attack.
Therefore, we argue that effective identity protection should induce identity deviations rather than merely introducing obvious artifacts.

Furthermore, face-swapping generally involves two roles: the {\em source identity}, which provides the face appearance, and the {\em target context}, which contributes attributes like pose, expression, and background. Existing proactive defenses protect an image either from being used as the source (identity) or the target (context) in face swapping. None of them provides comprehensive protection from {\bf {\em both}} aspects. We argue that both aspects are valuable targets in deepfake attacks, and users cannot reliably predict which one attackers will exploit. Therefore, a method capable of protecting both simultaneously is highly desirable.

To address these challenges and provide a practical defense against face-swapping deepfakes, we present \name, a comprehensive proactive defense framework that simultaneously protects both identity and context while enabling forensic tracing.
\name\ selects appropriate \emph{cloak} images to guide image protection and control face-swapping outcomes. It carefully controls visual similarity, preserving a sufficient degree of resemblance to the victim to mislead attackers and deter repeated attempts, while enabling reliable differentiation between genuine and manipulated outputs through rigorous similarity metrics.
In addition, the cloak identity serves as an effective, interpretable anchor for forensic tracing. Meanwhile, \name~retains the same capability as traditional disruption-based defenses: by selecting cloak images that are maximally dissimilar to the victim, it can force the face-swapped output to visibly deviate from the victim’s identity. Experimental results show that \name~effectively provides identity protection, context protection, and forensic tracing across a wide range of face-swapping models under the white-box setting. \name~also outperforms state-of-the-art solutions in identity protection in the black-box setting. Furthermore, \name~is robust to common real-world transformations, adaptive perturbation removal, and standard pre- and post-processing operations.

In summary, our contributions are:

\begin{table}[t]
\small
  \centering
  \captionsetup{skip=3 pt}
  \caption{Existing deepfake defenses. \textit{$G_1$}: Identity protection, \textit{$G_2$}: Context protection, \textit{$G_3$}: Tracing, \textit{$G_4$}: Enhancing detection. \fullcirc: satisfy, \halfcirc: partially satisfy, \emptycirc: not satisfy.}
  \label{tab:summary_of_defense}
  
  \setlength\tabcolsep{3.1 pt}
  \begin{tabular}{l|l|c c c c}
    \hline
    \multicolumn{2}{c|}{Approaches}  & \textit{$G_1$} & \textit{$G_2$} & \textit{$G_3$} & \textit{$G_4$} \\
    \hline
    \multirow[c]{2}*{\makecell{Detection}} & Direct: \cite[etc.]{li2020identification,nataraj2019detecting,carlini2020evading,huang2020fakepolisher,liu2021spatial,frank2020leveraging}  & \emptycirc & \emptycirc & \emptycirc & \halfcirc \\
    & Fingerprint: \cite[etc.]{beuve2023waterlo,liu2023bifpro,yang2021faceguard,lan2024facial,asnani2022proactive,wang2024lampmark}  & \emptycirc & \emptycirc & \emptycirc & \fullcirc \\
    \hline
    \multirow[c]{2}*{Tracing} & Training-based: \cite{yu2021artificial, sun2023faketracer}  & \emptycirc & \emptycirc & \fullcirc & \halfcirc \\
    & Testing-based: \cite[etc.]{guarnera2020fighting, lin2022source, wu2023sepmark, hu2023draw} & \emptycirc & \emptycirc & \fullcirc & \halfcirc \\
    \hline
    \multirow[c]{2}*{\makecell{Restoration}} & Hiding-based: \cite{shen2024hiding}  & \emptycirc & \halfcirc & \fullcirc & \emptycirc \\
    & Inference-based: \cite[etc.]{ai2023deepfake, yu2024dfrec, chen2021magdr,ai2023deepreversion}  & \emptycirc & \emptycirc & \fullcirc & \emptycirc \\
    \hline
    \multirow[c]{3}{*}{\makecell{Disruption}} & Landmark / Face: \cite[etc.]{cherepanova2021lowkey, li2024landmarkbreaker, zhu2024hiding, zhang2023disrupting}  & \fullcirc & \emptycirc & \emptycirc & \emptycirc \\
    & Latent code: \cite[etc.]{shim2023leat, guan2022defending, tang2024feature, dong2021visually,he2022defeating, wang2025nullswap}  & \emptycirc & \fullcirc & \emptycirc & \fullcirc \\
    & Model-based: IDGuard \cite{dai2024idguard}  & \fullcirc & \emptycirc & \emptycirc & \emptycirc \\
    \hline
    \multicolumn{2}{c|}{\name\ (ours)} & \fullcirc & \fullcirc & \fullcirc & \fullcirc \\
    \hline
  \end{tabular}
  \footnotesize
  \vspace{-3mm}
\end{table}

\begin{itemize}[itemsep=0.5pt,topsep=0pt,leftmargin=*]
  \item We propose the first proactive protection framework that protects {\em both} the identity and the context of the users' images against face-swapping deepfakes.
  \item We introduce a novel ``cloaking'' mechanism that 
purposefully misguides the face-swapping process to offer traceability for forensic analysis. \name\ is the first defense that achieves \textbf{both} deepfake disruption and forensic tracing simultaneously. 
  \item We evaluate \name\ on various face-swapping architectures and models to demonstrate its effectiveness. We also conduct a user study to evaluate the human perception of \name's usability and protection effectiveness. 
\end{itemize}

In the rest of the paper, we introduce the background and threat model in Sections~\ref{sec:background} and~\ref{sec:problem_statement}, followed by the technical details and experiment results in Sections~\ref{sec:simswap} to \ref{sec:robustness}. We discuss the limitations of \name~in Section~\ref{sec:limitations}, and conclude the paper in Section~\ref{sec:conclusion}.
\section{Background and Related Work}\label{sec:background}

\subsection{Deep Learning Models for Face-swapping}
\label{subsec:face_swap_architecture}

\noindent\textbf{Autoencoder-based Deepfakes. }Autoencoder-based face swapping~\cite{kingma2019introduction, wang2022anti} uses an encoder to convert input images into latent representations, which the decoder uses to reconstruct the output images. The attacker trains a shared encoder ($\mathcal{E}$) and two identity-specific decoders, $\mathcal{D}_A$ and $\mathcal{D}_B$, for identities \textit{A} and \textit{B}. An image of \textit{A} is encoded by $\mathcal{E}$ and decoded by $\mathcal{D}_B$, which transfers identity \textit{B} onto the original context of \textit{A}. Autoencoder-based deepfakes can only swap faces seen during training; hence, they have been superseded by GAN- and diffusion-based attacks. Nevertheless, to demonstrate the broader applicability of \name, we still present our defense against autoencoder-based deepfakes in Appendix~\ref{app:autoencoder}.

\noindent\textbf{GAN- and Diffusion-based Deepfakes.} Modern face-swapping models decouple identity representation from identity-specific decoders. As a result, state-of-the-art deepfake systems increasingly adopt GAN- and diffusion-based architectures, which enable identity manipulation in a more flexible latent space. For example, SimSwap \cite{chen2020simswap} introduces an identity injection module between the encoder and decoder, allowing identity features to be extracted from arbitrary source images and injected into target images. This design supports face swapping between arbitrary identities without requiring identity-specific training.

Diffusion-based deepfakes generally yield higher image quality than GAN-based ones. However, defending against GAN-based attacks remains both technically and practically important for several reasons:
(1) GAN-based face-swapping tools remain highly accessible. Some of the most-starred deepfake projects on GitHub, e.g., Deep-Live-Cam~\cite{deep_live_cam}, are still GAN-based.
(2) GAN-based attacks require significantly fewer resources, e.g., SimSwap~\cite{chen2020simswap} can be trained on a single 3GB GPU, while DiffFace~\cite{kim2025diffface} requires eight 80GB A100 GPUs. 
(3) Although commercial diffusion platforms (e.g., BrandStudio~\cite{brandstudio} and Herahaven~\cite{herahaven}) lower the technical barrier for novice attackers, their strict policies can limit large-scale misuse.

\subsection{Adversarial Examples}
\label{subsec:adversarial_examples}

Adversarial examples are input samples deliberately and subtly modified at inference time to mislead a DL model into making incorrect predictions \cite{goodfellow2014explaining}. Popular approaches include FGSM \cite{goodfellow2014explaining}, BIM \cite{kurakin2016adversarial}, and PGD \cite{madry2017towards}. For example, given a DL model $f_{\theta}$, an input $\mathbf{x}$, and its ground truth label $y$, FGSM constructs an adversarial example by approximately maximizing the loss function $\mathcal{L}\left(f_{\theta}, \mathbf{x}, y\right)$ under an $L_{\infty}$ constraint \cite{goodfellow2014explaining}:
$\mathbf{x}_{adv} = \mathbf{x} + \alpha \cdot \mathrm{sign}\left(\nabla_{\mathbf{x}}\mathcal{L}\left(f_{\theta}, \mathbf{x}, y\right)\right)$.
PGD extends FGSM by iteratively applying gradient-based updates and projecting the perturbed input back into a valid $L_\infty$-norm ball $\mathcal{B}_{\infty}(\mathbf{x}, \epsilon)$ to ensure the perturbation remains bounded, i.e., $\|\mathbf{x}_{adv} - \mathbf{x}\|_{p} \leq \epsilon$. The iterative update is formalized as:
\begin{align}
\hspace{-1mm}
\mathbf{x}_{adv}^{T+1} = \Pi_{\mathcal{B}_{p}\left(\mathbf{x}, \epsilon\right)}\left(\mathbf{x}_{adv}^{T} + \alpha \cdot \mathrm{sign}\left(\nabla_{\mathbf{x}_{adv}^{T}}\mathcal{L}\left(f_{\theta}, \mathbf{x}_{adv}^{T}, y\right)\right)\right)
\label{eqa:pgd_general}
\end{align}
\noindent where $T$ is the current step, $\alpha$ denotes the step size of each PGD update, and $\Pi$ denotes the projection operator.
Although our approach is not an adversarial attack, it adopts an iterative projected-gradient update similar in spirit to PGD as a general tool for generating bounded, imperceptible perturbations (see Section~\ref{subsec:methodology}).

\subsection{Defenses against Face-swapping Deepfakes}
\label{subsec:deepfake_defense}

Existing defenses against deepfakes follow four directions: detection, disruption, fingerprint-based tracing, and restoration. Table \ref{tab:summary_of_defense} summarizes the SOTA approaches and their capabilities with regard to four main protection goals: (G1) to protect the identities of the source images, (G2) to protect the contexts of the target images, (G3) to support traceability of source images (see detailed definition in Section~\ref{subsec:evaluation_metrics}), and (G4) to improve deepfake detection by embedding additional information/artifacts into protected images.

Existing defenses either directly examine deepfakes from the spatial domain \cite{li2020identification,nataraj2019detecting,carlini2020evading,chai2020makes,hu2021exposing,huang2022fakelocator} or frequency domain \cite{huang2020fakepolisher,liu2021spatial,frank2020leveraging,barni2020cnn,tan2024frequency}, or embed fingerprints for detection~\cite{beuve2023waterlo,liu2023bifpro, yang2021faceguard,lan2024facial,asnani2022proactive,wang2024lampmark}, e.g., Deepsonar \cite{wang2020deepsonar} amplifies and identifies deepfake traits; fragile watermarks \cite{beuve2023waterlo,liu2023bifpro} and identity fingerprints ~\cite{yang2021faceguard,lan2024facial,asnani2022proactive,wang2024lampmark} indicate deepfake interference. Such detectors identify deepfake \textit{ex post facto}. They cannot \textit{prevent} deepfake generation.

The disruption approaches inject perturbations to disturb different stages of deepfake generation: face detection \cite{zhu2024hiding, li2021obstructing, li2019hiding, segalis2020ogan, zhang2023disrupting}, landmark localization \cite{li2024landmarkbreaker}, latent code extraction \cite{shim2023leat, guan2022defending, dong2021visually, tang2024feature, he2022defeating}, and attention \cite{jeong2024faceshield, liu2024disrupting}. They cause deepfake outputs to appear stained/distorted, or add warning signs \cite{zhai2023defending}. IDGuard \cite{dai2024idguard} embeds an identity extractor to the swapping model to disrupt face swapping when a protected identity is present. These approaches protect either the source identity or the target context, but not both. Hence, the protected images are still vulnerable to deepfakes.

Another thrust is to trace the origins of the deepfakes by embedding fingerprints into training \cite{yu2021artificial, sun2023faketracer} or testing data \cite{guarnera2020fighting, sun2023faketracer, lin2022source, wu2023sepmark, li2024dual, hu2023draw, wu2024watermarks, thakkar2024deepfakes}. For example, SepMark \cite{wu2023sepmark}, combines fragile and robust watermarks to facilitate both detection and tracing. Last, the restoration approaches recover information for tracing, e.g., inference-based approaches~\cite{ai2023deepfake, yu2024dfrec, chen2021magdr, zhang2021restore, niudiffusion, ai2023deepreversion} reconstruct original facial features that survived through deepfake, while hiding-based approaches \cite{shen2024hiding} embed the original face in the context and restore it after face swapping. However, these approaches only take effect after face swaps but cannot prevent it from being generated.

As summarized in Table~\ref{tab:summary_of_defense}, \name\ is the first defense in literature that simultaneously achieves four protection goals: protecting both the identity and the context by disrupting deepfake generation, providing forensic traceability through ``cloaking'', and enhancing deepfake detection using AI-generated cloaks. 
\section{The Threat Model and Solution Overview}\label{sec:problem_statement}

\subsection{The Threat Model} \label{sec:threat_model}

\begin{figure}[t]
    \includegraphics[width=0.40\textwidth]{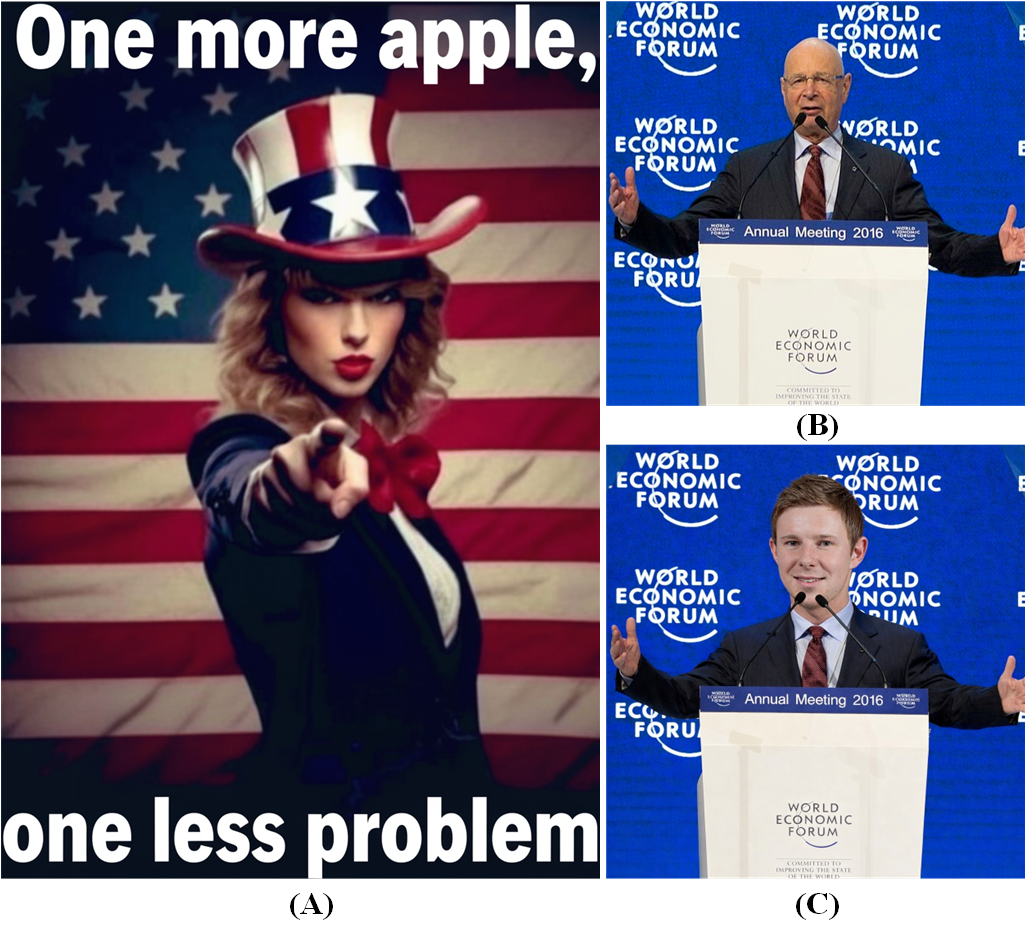}
    \captionsetup{skip=-2 pt}
    \caption{Identity-stealing (A) and context-stealing (B, C)}
    \Description{}
    \vspace{-3mm}
    \label{fig:deepfake_defense_story}
\end{figure}

We consider a generic face-swapping attack $\mathfrak{T}(\mathbf{s}, \mathbf{t})$ that transfers the identity from a source image $\mathbf{s}$ to a target image $\mathbf{t}$ while preserving the target’s contextual attributes (e.g., expression, pose, and gaze). Figure~\ref{fig:deepfake_defense_story} illustrates representative cases of identity and context theft. In Image~(A), an attacker steals the identity of \textit{Taylor Swift} to fabricate the statement ``one more apple, one less problem,'' while similar manipulations can be repurposed for political propaganda~\cite{walker2024merging}. 
Image~(B) shows a genuine photograph of \textit{Klaus Schwab}, founder of the World Economic Forum, during a forum presentation~\cite{wikimediaFileDrKlaus}. In contrast, Image~(C) illustrates a context-stealing attack in which the attacker, Charlie (from the FFHQ dataset~\cite{karras2019style}), swaps his face into the scene while preserving \textit{Klaus Schwab}’s contextual cues, creating the false impression that Charlie is presenting at the forum—an effect that can be exploited for potential scams.

In this paper, we follow the practice in the deepfake literature (e.g., \cite{li2019faceshifter,chen2020simswap,cui2023face}) to only operate on (i.e., swap) the face area of the source and target images. We consider the following roles in the scenario and use them throughout the paper:

\noindent$\bullet~$\textbf{Alice} {is the victim, who shares images on the Internet or through other channels, e.g., Alice posts a photo of herself attending a highly prestigious award ceremony.}

\noindent$\bullet~$\textbf{Bob} {attempts to steal the \textit{identity} from Alice's images (\textbf{identity-stealing attacks}), i.e., Bob extracts Alice's face from her images and uses it to replace his own face in a different context, misleading others into believing that Alice engaged in actions she did not commit. For example, Bob uses Alice's identity to publish political messages to manipulate public opinion.}

\noindent$\bullet~$\textbf{Charlie} {attempts to steal the \textit{context} from Alice’s images (\textbf{context-stealing attacks}) by inserting his own face into Alice’s photos to create the illusion that he shares the same context, e.g., Charlie swaps his face into Alice’s award-ceremony photo to fabricate the appearance of having attended the event.}

To protect both the identity and context in an image, Alice applies \name\ prior to public dissemination. For real-world deployment, we suggest integrating \name\ into photo-sharing platforms to enable automatic protection of all published images. 
We assume that neither Bob nor Charlie has access to Alice’s unprotected images, which are outside the protection scope of \name. In addition, under our recommended deployment workflow of (e.g., platform-side protection with AI-generated cloak images discussed in Section~\nameref{ethic:cloak_image_selection}), attackers do not have access to the cloak images used for protection.

\name\ is designed to defend against diffusion-,  GAN-, and autoencoder-based face-swapping attacks that provide the attacker with precise control over whose identity is inserted into which context. In contrast, multimodal generative approaches, e.g., \cite{wang2024instantid}, synthesize new images from identity references and text prompts, following a fundamentally different threat model that does not rely on explicit target images. Even when target images are provided, they often fail to preserve the context consistently, producing changes in background semantics, clothing, and environmental lighting, which is referred to as ``background contamination'' \cite{li2026inject}. Since they implement a substantially different generation pipeline that jointly models identity references, text prompts, pose control, layout constraints, and generative priors, a different protection mechanism would be needed. Therefore, prompt-based generative models are considered outside the scope of \name.

\subsection{Defense Model and Evaluation Metrics}
\label{subsec:evaluation_metrics}

\noindent\textbf{Protection Goals.} In this paper, we present a proactive defense instead of postmortem detection. To prevent attackers \textit{Bob} and \textit{Charlie} from successfully using her photos in a deepfake attack, the defender \textit{Alice} aims to achieve the following goals:

\noindent$\bullet~$\textbf{Identity protection.} Alice aims to prevent Bob from successfully swapping her identity into any other image or context. That is, when Bob performs the identity-stealing deepfake attack, the resulting image does not resemble Alice. 

\noindent$\bullet~$\textbf{Forensic tracing.} Additionally, by employing a {\em cloak image} in the protection mechanism (discussed in Section~\ref{subsec:solution_overview}), the face-swapped output generated from the protected image will resemble the identity of the cloak. Therefore, it becomes possible to trace the source of the swapped image using forensic evidence.

\noindent$\bullet~$\textbf{Context protection.} {Alice aims to disrupt the contextual integrity of face-swapping outputs, thereby preventing Charlie from embedding his identity into her context. In GAN-based models, such context disruption introduces conspicuous artifacts, whereas in diffusion-based models, it interferes with facial landmark detection, rendering face-swapping generation ineffective.}

\noindent$\bullet~$\textbf{Minimal perturbation.} {Alice ensures that the protected images remain visually indistinguishable from the original, allowing her to share the images as usual.}

\noindent\textbf{Evaluation Metrics.} To assess the extent to which Alice achieves her objectives, we follow the metrics proposed in \cite{li2023unganable} to evaluate two key aspects: effectiveness and utility. 

\noindent$\bullet~$\textbf{Effectiveness} measures the performance of deepfake defense and forensic tracing. {In quantitative evaluation, a deepfake attack is considered successful if a face recognition model identifies the swapped face as the attacker’s intended identity. Specifically, an identity-stealing attack succeeds when the generated face is recognized as Alice, instead of Bob, whereas a context-stealing attack succeeds when it is recognized as Charlie, instead of Alice. We define the matching rate (attack success rate) as the fraction of successful swaps among all attack attempts. For forensic analysis, an identity-stealing attack is successfully traced if the generated face is recognized as the pre-embedded cloak identity, not Alice or Bob. We measure \textbf{tracing effectiveness} as the matching rate between swapped faces and their corresponding cloak identities.}

\noindent$\bullet~${\textbf{Utility} quantifies the visual similarity between an image and its protected version. As no single metric universally measures perceptual similarity, we evaluate utility with four complementary measures. We adopt mean squared error (MSE) and peak signal-to-noise ratio (PSNR) to assess pixel-level distortion, and use the structural similarity index (SSIM)~\cite{wang2004image} and learned perceptual image patch similarity (LPIPS)~\cite{zhang2018unreasonable} to capture perceptual and structural differences. Together, these metrics better capture human visual perception and are more robust to lighting and contrast variations. Accordingly, the \textbf{utility of a protected image} is quantified by its similarity to the original image, where higher similarity (e.g., lower MSE or higher SSIM) indicates better utility.}

\begin{figure*}[t]
    \centerline{\includegraphics[width=0.95\textwidth]{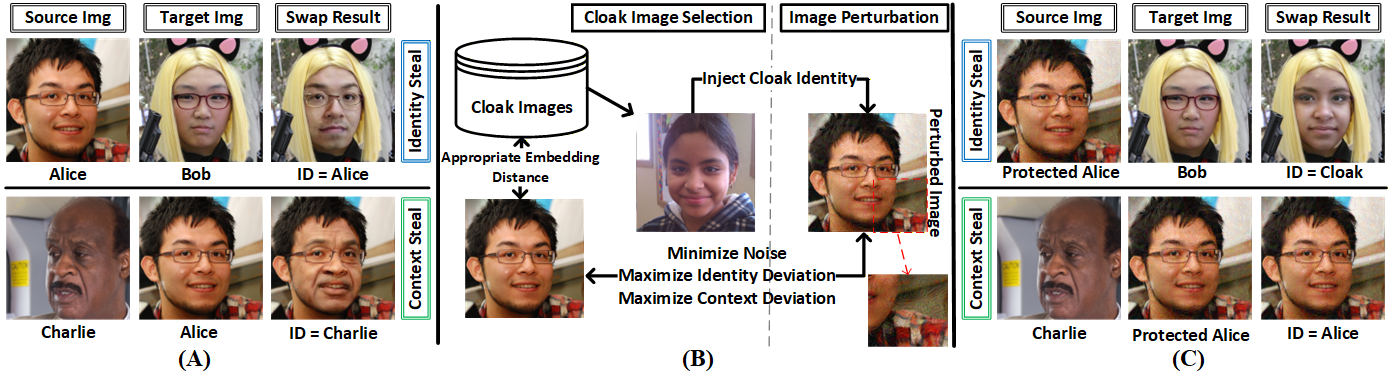}}
    \captionsetup{skip=0pt}
    \caption{An overview of the \name\ workflow. (A) Without protection, Bob and Charlie successfully perform face swapping on Alice’s image. (B) A suitable cloak image is selected and applied to protect Alice. (C) With protection, face swapping fails: Bob’s output resembles the cloak identity, while Charlie’s result fails to integrate his face into Alice’s image.}
    \Description{}
    \label{fig:framework}
    \vspace{-3mm}
\end{figure*}

\subsection{Overview of the Proposed Solution}
\label{subsec:solution_overview}

\name~injects invisible perturbations into the original images to protect them when used as either the source or the target in face-swapping pipelines. In particular, when the protected image is used as the source identity, \name~incorporates \textbf{\textit{cloak images}} as anchors to guide perturbation generation, shifting the protected identity toward the cloak identity. As a result, if an identity-stealing attack occurs, the face-swapped output generated from the protected image resembles the cloak identity.

Figure \ref{fig:framework} overviews the workflow of the proposed \name. Without protection, Bob and Charlie successfully perform face swap attacks to steal Alice's identity and context, i.e., attack outputs in Figure \ref{fig:framework} (A) are recognized as Alice and Charlie, respectively. \name\ proposes a unified defense mechanism against both attacks, which comprises two key steps in Figure \ref{fig:framework}(B): 

\noindent\textbf{1.} Cloak image selection: A cloak image is selected based on the strategies discussed in Section \ref{key:strategy}.

\noindent\textbf{2.} Image perturbation: A perturbed image is generated such that: ({\em i}) it is visually very similar to the original, ({\em ii}) it generates an identity vector similar to the cloak, and ({\em iii}) it produces a latent code that is different from the original. The details are presented in Section \ref{subsec:methodology}.

As shown in Figure \ref{fig:framework} (C), Alice's protected image is immune to both identity and context stealing: When attacker Bob swaps the identity from the perturbed image to a different context, the output image resembles the identity of the cloak. Meanwhile, if Charlie reuses the latent code from the perturbed image, he cannot swap his face into Alice’s context.

\section{The \name\ Defense}\label{sec:simswap}

We now formalize the cloak-selection and image perturbation problems (introduced in Section~\ref{subsec:solution_overview}). In a nutshell, given an image $\mathbf{x}\in[0,1]^{H\times W\times C}$ and a candidate cloak set $\mathcal{X}_c$, we seek a protected image $\tilde{\mathbf{x}}=\mathbf{x}+\Delta\mathbf{x}$, bounded by perception-aware constraints, that
(i) steers identity features toward a chosen cloak and away from the identity features of Alice, and (ii) induces sufficient context feature deviation to disrupt face swapping.

\subsection{Methodology}\label{subsec:methodology}

We adopt SimSwap~\cite{chen2020simswap} and DiffFace~\cite{kim2025diffface} as representative GAN-based and diffusion-based face-swapping models, respectively, as they are among the most widely used face-swapping methods.

\noindent\textbf{SimSwap Defense.}
SimSwap extracts an identity vector from the source image and injects it into the latent representation of a target image through an identity injection module.
This process generates a deepfake by embedding the source identity into the target image’s context, which can be expressed as:
\begin{equation}
\begin{aligned}
\mathbf{z}_{s} = \mathcal{E}_{id}\left(\mathbf{x}_{s}\right); & \qquad
\mathbf{z}_{t} = \mathcal{E}_{ctx}\left(\mathbf{x}_t\right); & \qquad
\hat{\mathbf{x}} = & \mathcal{D}\left(\mathbf{z}_{s} \oplus \mathbf{z}_{t}\right)
\label{eqa:simswap}
\end{aligned}
\end{equation}
where $\mathbf{x}_{s}, \mathbf{x}_{t} \in \mathbb{R}^{H \times W \times C}$ denote the source and target images; $\mathbf{z}_{s}$ and $\mathbf{z}_{t}$ denote the corresponding encoded feature vectors; $\hat{\mathbf{x}}$ denotes the generated deepfake image; and $\oplus$ denotes the fusion operation between two vectors.
$\mathcal{E}_{id}$, $\mathcal{E}_{ctx}$, and $\mathcal{D}$ denote the identity extractor, context extractor, and decoder, respectively.

Our defense targets two critical components in SimSwap: the \textit{identity vector} and the \textit{latent features}.
By introducing carefully crafted, imperceptible perturbations that alter these representations, we disrupt the deepfake generation process. Moreover, for source images, we introduce a \emph{cloak identity} $\tilde{I}_c$ to replace the original identity $I_s$ in the generated output $\hat{\mathbf{x}}$.

Specifically, \name\ addresses two tasks: \textbf{Context Protection} ($\mathcal{T}_{context}$) and \textbf{Identity Protection} ($\mathcal{T}_{identity}$).
Context protection applies to the scenario defined in Section~\ref{sec:threat_model}, where Alice’s image $\mathbf{x}$ is used as the target by Charlie. To protect $\mathbf{x}$, we perturb it into a protected version $\tilde{\mathbf{x}}$ such that its latent representation deviates sufficiently from the original.
Concretely, we aim to maximize the latent deviation $\Delta \mathbf{z}_t = \tilde{\mathbf{z}}_{t} - \mathbf{z}_{t}$, where $\tilde{\mathbf{z}}_{t} = \mathcal{E}_{ctx}(\tilde{\mathbf{x}})$.
Meanwhile, the perturbation $\Delta \mathbf{x} = \tilde{\mathbf{x}} - \mathbf{x}$ should be minimized to preserve utility.
Accordingly, we aim to minimize the following combined loss:
\begin{align}
\mathcal{L}_{context} = \lambda_{util} \|\Delta \mathbf{x}\| - \lambda_{ctx} \|\Delta \mathbf{z}_t\| 
\label{eqa:pgd_simswap_target}
\end{align}
We adopt a similar strategy for identity protection when Alice’s image is used as the source by Bob. Our objective is two-fold: 
(1) we shift the protected image to imitate a pre-selected cloak identity $\tilde{I}_c$ by minimizing its distance to the cloak $\mathbf{x}_c$ in the identity space, i.e., we minimize $\Delta \mathbf{z}_c=\tilde{\mathbf{z}}_{s}-\mathbf{z}_{c}$ where $\tilde{\mathbf{z}}_{s}{=}\mathcal{E}_{id}\left(\tilde{\mathbf{x}}\right)$ and $\mathbf{z}_{c}{=}\mathcal{E}_{id}\left(\mathbf{x}_{c}\right)$. 
(2) To prevent the face-swapping result from resembling the source identity, we move the protected identity to deviate sufficiently from the original identity, i.e., $\Delta \mathbf{z}_s=\tilde{\mathbf{z}}_{s}-\mathbf{z}_{s}$.
Meanwhile, the perturbation $\Delta\mathbf{x}$ is regularized to preserve utility.
In SimSwap, $\mathbf{z}_{c}$ is represented as a tensor of shape $\mathbb{R}^{3 \times 512}$, while $\mathbf{z}_{t}$ is a tensor of shape $\mathbb{R}^{3 \times 512 \times 28 \times 28}$. The identity protection loss function is defined as:
\begin{equation}
\mathcal{L}_{identity}= \lambda_{util} \|\Delta \mathbf{x}\| + \lambda_{cloak} \|\Delta \mathbf{z}_c\| - \lambda_{id} \|\Delta \mathbf{z}_s\| 
\label{eqa:pgd_simswap_source}
\end{equation}
For Alice's image, \name~simultaneously protects both context and identity by integrating the corresponding objectives in Equation~(\ref{eqa:pgd_simswap_target}) and Equation~(\ref{eqa:pgd_simswap_source}) into a unified optimization process:
\begin{equation}
\begin{aligned}
\mathcal{L}
&= \lambda_{util} \|\Delta \mathbf{x}\| - \lambda_{ctx} \|\Delta \mathbf{z}_t\| + \lambda_{cloak} \|\Delta \mathbf{z}_c\| - \lambda_{id} \|\Delta \mathbf{z}_s\|
\label{eqa:pgd_simswap_all}
\end{aligned}
\end{equation}

\noindent\textbf{DiffFace Defense.}
DiffFace also uses identity/context extractors $\mathcal{E}_{id}$ and $\mathcal{E}_{ctx}$, and uses the extracted identity and contextual features to guide the denoising process during diffusion-based generation.
Accordingly, similar to SimSwap, our defense against DiffFace is applied to both the \textit{identity vector} and the \textit{latent features}.

For identity protection, our defense against DiffFace follows the same procedure as for SimSwap and adopts the identical loss function in Eq.~\ref{eqa:pgd_simswap_source}.
In contrast, context protection differs due to how diffusion-based models represent facial structure. Unlike GAN-based methods, which encode the entire facial context into a single latent embedding, diffusion-based approaches decompose a face into multiple structural components, such as facial landmarks (e.g., eyes, nose, and lips) and non-facial regions (e.g., hair, clothing, and background), via semantic segmentation.
Accordingly, when deviating DiffFace’s latent features, i.e., $\Delta \mathbf{z}_{t} = \tilde{\mathbf{z}}_{t} - \mathbf{z}_{t}$, the objective is not merely to increase numerical differences in feature space, but to induce deviations that mislead landmark detectors into interpreting regions as different facial landmarks.
As a result, without correct landmark guidance, when Charlie attempts to swap his face onto Alice’s protected image, the generated output remains visually identical to the protected image $\tilde{\mathbf{x}}$.
The overall context protection loss follows SimSwap’s formulation in Eq.~\ref{eqa:pgd_simswap_target}, and the final objective adopts the same structure as Eq.~\ref{eqa:pgd_simswap_all}.

\noindent\textbf{Methodological Contributions.}
\name~differs from previous deepfake-disruption mechanisms in two main aspects:
\textbf{(1)} \name~jointly protects identity and context while enabling tracing (Eq.~\ref{eqa:pgd_simswap_all}), whereas previous works, e.g., \cite{cherepanova2021lowkey, wang2025nullswap}, only prevents identity stealing.
\textbf{(2)} From a methodological perspective, \name~introduces the \textit{cloak} to guide the identity embedding toward a pre-selected cloak identity, instead of aimlessly deviating it away from the original. That is, \name~leverages the cloak identity to simultaneously \textit{push} the source identity away from the original and \textit{pull} it toward the cloak. This \textit{push-and-pull co-optimization} mechanism achieves both tracing and stronger identity protection. The parameter sensitivity analysis further demonstrates that the cloak-guided disruption is substantially more effective than simple deviation for identity protection (see Section~\ref{sec:ablation_studies} for details).

\noindent\textbf{Human-Perception-Aware Noise Bounds.} To constrain the perturbations, we adopt channel-specific RGB limits $\delta_{RGB}$.
Motivated by the higher sensitivity of the human vision to green~\cite{hsieh2014combining}, we impose tighter bounds on the green channel and allow larger perturbations on red and blue channels.

\noindent\textbf{Constrained Optimization with Priority.} Balancing the conflicting objectives of minimizing visual noise and maximizing protection is challenging, as the optimization is highly sensitive to hyperparameters, which can cause one objective to dominate and suppress the gradients of the other.
To address this issue, we adopt a strategy of \textit{constrained and prioritized optimization}. 
We first prioritize protection effectiveness by assigning large weights to $\lambda_{id}$ and $\lambda_{ctx}$ to maximize identity and context deviation. This phase continues until the deviations $|\Delta \mathbf{z}_s|$ and $|\Delta \mathbf{z}_t|$ reach their predefined upper bounds, $\delta_{id}$ and $\delta_{ctx}$, respectively.
Once these constraints are met, the optimization shifts its focus to minimizing the perturbation magnitude to preserve the visual quality of the protected image.

\noindent\textbf{Composite Protection Score.} We introduce perception-aware noise bounds and constrained optimization to limit perturbations and balance multiple protection objectives.
However, protection effectiveness may vary across objectives: a strategy may induce stronger identity deviation but weaker context protection or tracing, while another may exhibit the opposite trade-off.
To evaluate whether—and to what extent—a protection scheme meets the design objectives of \name, we define a composite protection score $\mathcal{S}$ that provides a unified measure of overall protection quality:
\begin{equation}
\label{eqa:total_point}
\mathcal{S} = w_{id} (1-ASR_{id}) + w_{ctx} (1-ASR_{ctx}) + w_{cloak} TSR
\end{equation}
{$w_{id}$, $w_{ctx}$, and $w_{cloak}$ denote the weights for identity protection, context protection, and tracing, respectively.
$ASR_{id}$ and $ASR_{ctx}$ denote the identity and context attack success rates, respectively, and $TSR$ denotes the tracing success rate.
Therefore, the range of $\mathcal{S}$ is between 0 and 3, where a higher $\mathcal{S}$ indicates stronger overall protection performance. In practice, effective protection typically achieves a high $TSR$ (e.g., ${\approx}90\%$) while maintaining low $ASR_{id}$ and $ASR_{ctx}$ (e.g., ${<}5\%$ each), resulting in a $\mathcal{S}$ of approximately 2.8. An improvement of $\Delta \mathcal{S} = 0.01$ is considered meaningful, as it approximately corresponds to a 1\% change in identity/context attack success rates or tracing success rate.
Finally, we integrate the constrained optimization and deviation bounds into the iterative process, and select the cloak whose protected image $\tilde{\mathbf{x}}$ maximizes the composite protection score $\mathcal{S}$. 
This outer-loop cloak selection, together with the inner-loop constrained optimization, naturally forms a bilevel optimization problem, as shown in Equation~\ref{eqa:bilevel}.} 
\begin{align}
    \max_{\mathbf{x}_c \in \mathcal{X}_c}
    \mathcal{S}\!\left(\tilde{\mathbf{x}}^{*}(\mathbf{x}_c)\right), \quad 
    \tilde{\mathbf{x}}^{*}(\mathbf{x}_c) 
    = \argmin_{\tilde{\mathbf{x}} \in \mathcal{B}_{\infty}^{\text{RGB}}(\mathbf{x}, \delta_{RGB})} \mathcal{L}(\tilde{\mathbf{x}},\mathbf{x}_c)
    \label{eqa:bilevel}
\end{align}
where $\mathcal{L}$ is the loss defined in Equation~\ref{eqa:pgd_simswap_all}, and $\mathcal{B}_{\infty}^{\text{RGB}}$ denotes the $L_\infty$ ball centered at $\mathbf{x}$. The full workflow for generating protected images is summarized in Algorithm~\ref{alg:whole_process}. The list of notations used in the optimization and algorithm is provided in Appendix Table~\ref{tab:notations}.

\begin{algorithm}[t]
\small
\caption{Protecting image against deepfake}\label{alg:whole_process}
\SetKwInput{Input}{Input}
\SetKwInput{Output}{Output}
\newcommand{\blueline}[1]{\textcolor{blue}{\textit{// #1}}}
\newcommand{\blackfollow}[1]{\hfill \textcolor{black}{\textit{// #1}}}
\Input{
    {Image $\mathbf{x}$; cloak image set $\mathcal{X}_c = \{ \mathbf{x}_c^{(1)}, \dots, \mathbf{x}_c^{(N)} \}$}
}
\Output{protected image $\tilde{\mathbf{x}}$}

\textbf{Initialization:} $\tilde{\mathbf{x}}^{(0)} \gets \mathbf{x}$,\quad
$\mathcal{S}_{\text{best}} \gets 0$,\quad
$\tilde{\mathbf{x}}_{\text{best}} \gets \mathbf{x}$ \\
{$\mathbf{z}_s \gets \mathcal{E}_{id}(\mathbf{x})$, $\mathbf{z}_t \gets \mathcal{E}_{ctx}(\mathbf{x})$ \blackfollow{image identity/context feature} \\}

\For{$i\gets 1$ \KwTo $N$} {
    $\mathbf{z}_c \gets \mathcal{E}_{id}(\mathbf{x}_c^{(i)})$ \blackfollow{cloak identity vector.}\\

    \For{$t\gets 1$ \KwTo $T$}{
        \blueline{1. extract the identity \& context features} \\
        $\tilde{\mathbf{z}}_s^{(t)} \gets \mathcal{E}_{id}(\tilde{\mathbf{x}}^{(t-1)})$ \blackfollow{current identity}\\
        $\tilde{\mathbf{z}}_t^{(t)} \gets \mathcal{E}_{ctx}(\tilde{\mathbf{x}}^{(t-1)})$ \blackfollow{current context}\\
        
        \blueline{2. calculate the feature differences} \\
        $\Delta \mathbf{x} \gets \tilde{\mathbf{x}}^{(t-1)} - \mathbf{x} $ \blackfollow{perturbation noise} \\
        $\Delta \mathbf{z}_s \gets \tilde{\mathbf{z}}_s^{(t)} - \mathbf{z}_s$ \blackfollow{identity deviation} \\
        $\Delta \mathbf{z}_c \gets \tilde{\mathbf{z}}_s^{(t)} - \mathbf{z}_c$ \blackfollow{cloak identity difference} \\
        $\Delta \mathbf{z}_t \gets \tilde{\mathbf{z}}_t^{(t)} - \mathbf{z}_t$ \blackfollow{latent deviation} \\
        
        \blueline{3. loss (subject to deviation bounds)} \\
        $\mathcal{L}^{(t)} = \lambda_{util} \|\Delta \mathbf{x}\|
        \;-\; \lambda_{ctx}\min\!\big(\|\Delta \mathbf{z}_t\|,\ \delta_{ctx}\big)
        \;+\; \lambda_{cloak} \|\Delta \mathbf{z}_c\| - \lambda_{id} \min\!\big( \|\Delta \mathbf{z}_s\|,\ \delta_{id}\big)$

        \blueline{4. update the image (subject to RGB bounds)} \\
        $\tilde{\mathbf{x}}^{(t)} \gets \tilde{\mathbf{x}}^{(t-1)} - \eta \cdot \text{sign}\big(\nabla_{\tilde{\mathbf{x}}}\mathcal{L}^{(t)}\big)$
        
        {$\tilde{\mathbf{x}}^{(t)} \gets \operatorname{clip}\!\Big(\Pi_{\mathcal{B}_{\infty}^{RGB}\left(\mathbf{x},\delta_{RGB}\right)}\left(\tilde{\mathbf{x}}^{(t)}\right),0,1\Big)$}

    \blueline{5: find the best protected image by protection score} \\
    $\mathcal{S}_{i} \gets $ evaluate($\tilde{\mathbf{x}}^{T}$) \\

    \If{$\mathcal{S}_{i} > \mathcal{S}_{\text{best}}$}{
        $\mathcal{S}_{best} \gets \mathcal{S}_i$,\quad
        $\tilde{\mathbf{x}}_{\text{best}} \gets \tilde{\mathbf{x}}^{(T)}$
    }
  }
}
\Return{$\tilde{\mathbf{x}}_{\text{best}}$}
\end{algorithm}

\subsection{Cloak Image Selection}\label{key:strategy}

\name\ shifts the identity vector of Alice’s image toward a pre-selected cloak identity, causing identity-stealing deepfakes to resemble the cloak. Hence, cloak image selection is critical. If the cloak identity is too close to Alice, it increases the risk for the protected swap being recognized as Alice (higher $ASR_{id}$), weakening protection. Meanwhile, a substantially different cloak drives the swap further away from Alice, reducing the likelihood of being recognized as Alice, but may weaken the resemblance to the cloak itself, reducing tracing effectiveness.

To balance the objectives of maximizing $TSR$ while minimizing $ASR$, we introduce $\mathcal{S}$ (Equation~\ref{eqa:total_point}) and use Algorithm~\ref{alg:whole_process} to select the cloak image that maximizes $\mathcal{S}$. We next discuss the sources of cloak images and the corresponding selection strategies.

\noindent\textbf{Data Source.}
Cloak images can be selected from any face image dataset. With the rapid progress of AI-generated content (AIGC), high-quality \textit{synthetic faces} have become effectively unlimited, making them particularly suitable for real-world deployment of \name. In our experiments, we use real image datasets (e.g., VGGFace2) to evaluate \name\ under a controlled setting where both the protected and cloak images are drawn from the same dataset. We adopt StyleGAN3-generated faces as synthetic cloaks.

In Algorithm~\ref{alg:whole_process}, we iterate over all cloak images to generate protected outputs, evaluate protection metrics, and compute the composite score to select the optimal cloak and its corresponding protected result. However, exhaustively evaluating all cloak candidates is time-consuming. As discussed earlier, the identity distance between Alice and a cloak image plays a critical role in protection effectiveness. Therefore, in real-world deployment, a distance-based pre-selection strategy can be adopted to identify suitable cloaks and avoid exhaustive search. A detailed analysis of identity distance is provided in Section~\ref{subsec:exp_simswap}.

\noindent\textbf{Selection Strategies.} We present three strategies for computing the composite protection score and guiding cloak image selection.

\noindent$\bullet~$\textbf{The Baseline.} Alice considers identity protection, context protection, and tracing equally important. Accordingly, she sets the weights $w_{id}$, $w_{ctx}$, and $w_{cloak}$ to 1, leading \name\ to provide balanced protection across all three objectives.

\noindent$\bullet~$\textbf{A Balanced Heuristic.} 
Empirical evidence suggests that gender strongly affects facial perception. Even at the same identity distance, i.e., the Euclidean distance between face embeddings extracted by a face recognition model (e.g., FaceNet~\cite{schroff2015facenet}), opposite-gender cloaks tend to appear more visually distinct. Accordingly, Alice retains the baseline weights but restricts cloak candidates to the opposite gender to induce larger perceptual differences.

\noindent$\bullet~$\textbf{Prioritizing Protection.} When Alice’s primary objective is to defend against identity and context stealing, she retains the default settings $w_{id}=w_{ctx}=1$ and sets $w_{cloak}=0$. This configuration maximizes protection against both attack types.
\section{Experiment Results}\label{sec:exp}

\subsection{Experimental Setup}\label{subsec:exp_setup}

\noindent\textbf{Experimental Environment.} We implement \name\ using Python~3.10, PyTorch~2.8.0, and CUDA~12.8. All experiments are conducted on a workstation with an AMD Ryzen~9~9950X CPU and NVIDIA RTX~5090 GPU, running Ubuntu~24.04~LTS.

\noindent\textbf{Dataset.} For GAN-based face-swapping, we adopt VGGFace2~\cite{cao2018vggface2}. To ensure identity diversity, we randomly sample 6,000 identities from VGGFace2 and select one image per identity. These 6,000 images are evenly split into validation and test sets.
For the diffusion-based model DiffFace, we follow its official implementation and use FFHQ~\cite{karras2019style} as the evaluation dataset. We randomly select 3,000 images for validation and another 3,000 images for testing.
When selecting cloak images from VGGFace2 or FFHQ (as discussed in Section~\ref{key:strategy}), we ensure that all cloaks are \textit{not} drawn from either the validation or test sets.
We randomly sample 3,000 image pairs from the test set to assess protection effectiveness.

\noindent\textbf{Face Recognition Models.} We adopt the following state-of-the-art face recognition (FR) models:
(1) \textit{FaceNet-512}~\cite{schroff2015facenet}, which has been widely used in prior research (e.g., UnGANable~\cite{li2023unganable}). We use the same decision threshold as DeepFace~\cite{serengil2020lightface, serengil2024lightface}.
(2) \textit{Megvii Face++}~\cite{faceplusplus}, a large-scale commercial computer vision service platform.
(3) In Section~\ref{subsec:more_FR_and_human_evaluation}, we further evaluate \name\ using two additional FR models: the widely adopted open-source \textit{Face Recognition} library~\cite{face_recognition} and Amazon \textit{AWS Rekognition}~\cite{aws_rekognition}.

\noindent\textbf{Efficiency. } \name\ generates a protected image in under 2 seconds using 1{,}000 epochs on NVIDIA RTX~5090. Reducing to 300 epochs lowers the processing time to about 0.5 seconds, with minor performance degradation (e.g., a 5\% drop in $TSR$).

\subsection{Defending against GAN-based Deepfakes}\label{subsec:exp_simswap}\label{subsec:gan_exp}

\noindent\textbf{Parameters/Settings.} The default weights of SimSwap in Eq.~\ref{eqa:pgd_simswap_all} are set to $\lambda_{util}{=}1000$, $\lambda_{ctx}{=}0.1$, and $\lambda_{id}{=}\lambda_{cloak}{=}10000$, and all losses are computed using the $\ell_2$ norm. The deviation upper bounds for identity and context are set to $\delta_{id}{=}0.003$ and $\delta_{ctx}{=}25$, respectively. The human-perception-aware noise bounds $\delta_{RGB}$ are set to 0.075, 0.03, and 0.075 for the R, G, and B channels, respectively. 
To construct the cloak image set, we start with the official sample images provided by StyleGAN3, consisting of 216 synthesized faces generated by StyleGAN3 trained on FFHQ (130 male and 86 female). We then randomly sample the same number of real face images from VGGFace2 and FFHQ, following the same gender distribution, to form a real-image cloak set.
Using these cloak images and hyperparameters, we evaluate all cloak candidates by computing the protection score $\mathcal{S}$ on the validation set. The best-performing cloak achieves a score of 2.924 (FaceNet-512). 
With the identity-distance-based selection strategy, we observe that when the identity distance between the cloak image and Alice’s image $\mathbf{x}$ is approximately 1.20, the resulting protection score reaches 2.923, only 0.001 below the optimum. Crucially, distance-based selection requires generating the protected image only once, reducing computation to 0.463\% while retaining 99.965\% of the optimal protection performance.
Based on this observation, we adopt 1.20 as the identity-distance criterion for selecting cloak images in all subsequent test-set evaluations.

\setlength{\textfloatsep}{20.0pt plus 2.0pt minus 4.0pt}
\begin{table}[t]
  \centering
  \small
  \captionsetup{skip=3 pt}
  \caption{Examples of \name\ protection. \textbf{w/o Pt.}: SimSwap outputs without protection. \textbf{w/ Pt.}: SimSwap outputs with \name\ protection. $\mathcal{T}_{identity}$: identity protection, $\mathcal{T}_{context}$: context protection}
  \label{tab:simswap_protection}
  \setlength\tabcolsep{1.7 pt}
  \begin{tabular}{c  c  c  c  c  c  c}
    \hline
    \multicolumn{1}{c|}{\multirow{2}{*}{\makecell{$\mathbf{x}$ \\ (``Alice'')}}} & \multicolumn{1}{c|}{\multirow{2}{*}{\makecell{``Bob'' or \\ ``Charlie''}}} & \multicolumn{3}{c|}{$ \mathcal{T}_{identity}$} & \multicolumn{2}{c}{$\mathcal{T}_{context}$} \\
    \cline{3-7}
    \multicolumn{1}{c|}{} & \multicolumn{1}{c|}{} & w/o Pt. & Cloak & \multicolumn{1}{c|}{w/ Pt.} & w/o Pt. & w/ Pt. \\
    \toprule
    \raisebox{-0.5\height}{\includegraphics[scale=0.138]{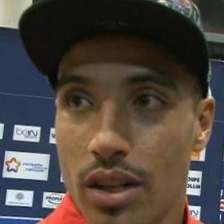}} & 
    \raisebox{-0.5\height}{\includegraphics[scale=0.138]{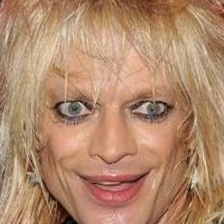}} & 
    \raisebox{-0.5\height}{\includegraphics[scale=0.138]{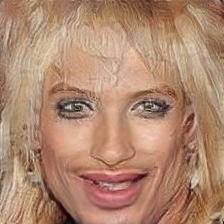}} &
    \raisebox{-0.5\height}{\includegraphics[scale=0.138]{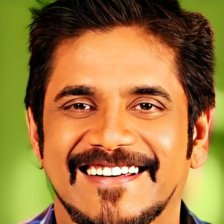}} &
    \raisebox{-0.5\height}{\includegraphics[scale=0.138]{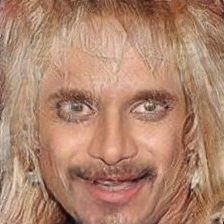}} &
    \raisebox{-0.5\height}{\includegraphics[scale=0.138]{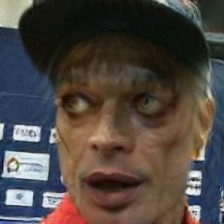}} &
    \raisebox{-0.5\height}{\includegraphics[scale=0.138]{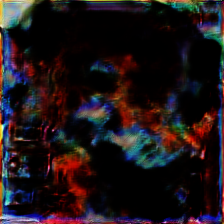}} \\
    \bottomrule
    \end{tabular}
    \vspace{-6 mm}
\end{table}

\noindent\textbf{Qualitative Results.}
We show examples of defenses against SimSwap in Table~\ref{tab:simswap_protection}. In identity-stealing, face-swapping generated from unprotected images (column~3) are classified as Alice by FR models, whereas those generated from protected images (col.~5) are classified as the corresponding cloak identities, confirming that \name\ effectively prevents identity stealing.
Similarly, in context-stealing, Charlie fails to swap his face into the protected images. The resulting outputs exhibit pronounced blurring and dotted artifacts, arising from the induced deviations in the latent representation. 
Overall, \name\ prevents SimSwap from identity- and context-stealing attacks and enables forensic tracing.

\begin{table*}[t]
  \centering
  \small
  \captionsetup{skip=2 pt}
  \caption{Evaluation of \name\ on SimSwap and DiffFace. $\mathcal{S}$: composite protection score, $ASR_o$: original attack success rate (without protection). $ASR_p$: attack success rate under protection. $\uparrow$: higher is better; $\downarrow$: lower is better.
  }\label{tab:protection_performance}
  \setlength\tabcolsep{2.1 pt}
  \begin{tabular}{c | c  c  c  c |  c  c  c @{\extracolsep{4pt}} c c @{\extracolsep{0pt}} c c | c  c  c @{\extracolsep{4pt}} c  c @{\extracolsep{0pt}} c c  }
    \hline    
    \multirow{3}*{Model} & \multicolumn{4}{c}{Utility} \vline &  \multicolumn{7}{c}{Protection performance on FaceNet-512} \vline & \multicolumn{7}{c}{Protection performance on Megvii Face++}  \\
    \cline{2-19}
    ~ & \multicolumn{4}{c}{$\mathbf{x}$ and $\tilde{\mathbf{x}}$ } \vline  &   \multirow{2}{*}{$\tilde{\mathbf{x}}\uparrow$} & \multicolumn{3}{c}{$\mathcal{T}_{identity}$} &\multicolumn{2}{c}{$\mathcal{T}_{context}$} & \multirow{2}{*}{$\mathcal{S}{\uparrow}$} & \multirow{2}{*}{$\tilde{\mathbf{x}}\uparrow$} & \multicolumn{3}{c}{$\mathcal{T}_{identity}$} &\multicolumn{2}{c}{$\mathcal{T}_{context}$} & \multirow{2}{*}{$\mathcal{S}{\uparrow}$} \\
    \cline{2-5} \cline{7-9}\cline{10-11} \cline{14-16}\cline{17-18}
     ~ & MSE $\downarrow$ & PSNR $\uparrow$ & SSIM $\uparrow$ & LPIPS $\downarrow$ & & $ASR_o$ & $ASR_p{\downarrow}$ & $TSR{\uparrow}$ & $~ASR_o~$ & $~ASR_p{\downarrow}~$ & & & $ASR_o$ & $ASR_p{\downarrow}$ & $TSR{\uparrow}$ & $~ASR_o~$ & $~ASR_p{\downarrow}~$ &  \\
    \hline
    SimSwap & 106 & 27.88 & 0.68 & 0.25 & 100.00 & 94.13 & 0.73 & 94.93 & 94.87 & 1.90 & 2.92 & 99.90 & 98.20 & 0.30 & 97.97 & 98.80 & 3.20 & 2.95 \\ 
    DiffFace & 57 & 30.67 & 0.77 & 0.28 & 99.50 & 69.67 & 0.43 & 68.97 & 70.30 & 1.00 & 2.68 & 100.00 & 70.13 & 0.13 & 68.93 & 70.50 & 0.50 & 2.68 \\ 
    \hline
  \end{tabular}  
  \vspace{-3mm}
\end{table*}

\noindent\textbf{Quantitative Results.}
Table~\ref{tab:protection_performance} summarizes the utility and effectiveness of \name.
\textbf{(1)} The protected images ($\tilde{\mathbf{x}}$) preserve high visual quality and structural consistency relative to the originals ($\mathbf{x}$), as reflected by low MSE, high PSNR, and strong perceptual similarity under SSIM and LPIPS, indicating that the perturbations are subtle and imperceptible.
\textbf{(2)} \name\ preserves the original identity, with $\tilde{\mathbf{x}}$ being recognized as $\mathbf{x}$ at a matching rate of 99.9\% by both FaceNet-512 and Face++.
\textbf{(3)} Without protection, identity- and context-stealing attacks succeed at high rates, with swapped outputs recognized as Alice or Charlie, respectively.
\textbf{(4)} When protected images are used in identity-stealing attacks, matching rates drop sharply to 0.73\% (FaceNet-512) and 0.30\% (Megvii Face++), indicating that swapped outputs no longer resemble the protected identity.
\textbf{(5)} Similar effectiveness is observed for context protection, where matching rates decrease to 1.90\% and 3.20\%, preventing the attacker’s identity from being embedded into Alice’s context.

We further evaluate the identity protection performance of \name~against FaceShifter~\cite{li2019faceshifter} and HifiFace~\cite{wang2021hififace}, and use these results to motivate the unified protection strategy described in Section \ref{subsec:black_box}.
We report the results in Table~\ref{tab:protection_comparison_with_lowkey}, which show trends similar to those observed for SimSwap. With only minor perturbations, \name~reduces $ASR_{id}$ to below 2\% and achieves $TSR$ values comparable to the corresponding $ASR_{o}$.

\subsection{Forensic Tracing \& Cloak Image Selection}\label{subsec:cloak_image}

\begin{table*}[t]
  \centering
  \small
  \captionsetup{skip=3 pt}
  \caption{Protection performance of SimSwap and DiffFace under different cloak image selection strategies. Exp \textsc{I}: different cloak image source; Exp \textsc{II}: gender-aware cloak image selection; Exp \textsc{III}: prioritize protection; Exp \textsc{IV}: different cloak image distances.}
  \label{tab:cloak_image_selection}
  \setlength\tabcolsep{2.4 pt}

\begin{tabular}{c|c|cccc|cccc|cccc|cccc}
  \hline
  \multirow{3}{*}{Exp.} &
  \multirow{3}{*}{Setting} &
  \multicolumn{8}{c|}{SimSwap} &
  \multicolumn{8}{c}{DiffFace} \\
  \cline{3-18}
  & &
  \multicolumn{4}{c|}{FaceNet-512} &
  \multicolumn{4}{c|}{Megvii Face++} &
  \multicolumn{4}{c|}{FaceNet-512} &
  \multicolumn{4}{c}{Megvii Face++} \\
  \cline{3-18}
  & &
  $ASR_{id}{\downarrow}$ & $TSR{\uparrow}$ & $ASR_{ctx}{\downarrow}$ & $\mathcal{S}{\uparrow}$ &
  $ASR_{id}{\downarrow}$ & $TSR{\uparrow}$ & $ASR_{ctx}{\downarrow}$ & $\mathcal{S}{\uparrow}$ &
  $ASR_{id}{\downarrow}$ & $TSR{\uparrow}$ & $ASR_{ctx}{\downarrow}$ & $\mathcal{S}{\uparrow}$ &
  $ASR_{id}{\downarrow}$ & $TSR{\uparrow}$ & $ASR_{ctx}{\downarrow}$ & $\mathcal{S}{\uparrow}$ \\
  \hline

  -- & Baseline &
  0.73 & 94.93 & 1.90 & 2.92 &
  0.30 & 97.97 & 3.20 & 2.95 &
  0.43 & 68.97 & 1.00 & 2.68 &
  0.13 & 68.93 & 0.50 & 2.68 \\
  \hline

  I & StyleGAN3 &
  1.03 & 85.13 & 1.90 & 2.82 &
  0.13 & 87.80 & 3.40 & 2.84 &
  0.37 & 64.23 & 1.10 & 2.63 &
  0.00 & 61.23 & 0.33 & 2.61 \\
  \hline

  II & Opposite &
  0.50 & 92.07 & 2.70 & 2.89 &
  0.00 & 96.70 & 3.90 & 2.93 &
  0.30 & 63.77 & 1.13 & 2.62 &
  0.00 & 63.87 & 0.53 & 2.63 \\
  \hline

  III & Protection &
  0.00 & 89.93 & 1.93 & 1.98 &
  0.00 & 96.87 & 3.53 & 1.97 &
  0.00 & 63.53 & 1.10 & 1.99 &
  0.00 & 63.50 & 0.60 & 1.99 \\
  \hline

  \multirow{5}{*}{IV} & distance=1.04 &
  5.97 & 95.63 & 1.37 & 2.88 &
  0.63 & 98.60 & 2.37 & 2.96 &
  8.30 & 69.73 & 1.30 & 2.60 &
  1.10 & 69.90 & 0.53 & 2.68 \\

  & distance=1.25 &
  0.30 & 92.63 & 2.63 & 2.90 &
  0.00 & 96.50 & 3.77 & 2.93 &
  0.27 & 65.67 & 1.21 & 2.64 &
  0.07 & 64.43 & 0.53 & 2.64 \\

  & distance=1.50 &
  0.00 & 86.67 & 3.03 & 2.84 &
  0.00 & 94.80 & 5.67 & 2.89 &
  0.13 & 65.00 & 1.20 & 2.64 &
  0.10 & 64.53 & 0.60 & 2.64 \\

  & distance=1.75 &
  0.00 & 83.80 & 4.13 & 2.80 &
  0.00 & 93.67 & 6.33 & 2.87 &
  0.07 & 64.40 & 1.17 & 2.63 &
  0.00 & 63.53 & 0.57 & 2.63 \\

  & distance=2.00 &
  0.00 & 84.30 & 4.23 & 2.80 &
  0.00 & 93.40 & 7.30 & 2.86 &
  0.00 & 64.33 & 1.10 & 2.63 &
  0.00 & 64.00 & 0.60 & 2.63 \\
  \hline
\end{tabular}
  \vspace{-3mm}
\end{table*}

We evaluate cloak selection strategies discussed in Section~\ref{key:strategy}:

\noindent\textbf{Exp \textsc{I}. Cloak image sources.}
As we have discussed in Section~\ref{key:strategy}, cloak images can be drawn from any face dataset. For real-world deployment, we recommend using AI-generated faces to avoid potential privacy concerns.
As shown in Table~\ref{tab:cloak_image_selection} Exp~\textsc{I}, \name\ achieves strong performance with StyleGAN3-generated cloaks, with similar $ASR_{id}$ and identical $ASR_{ctx}$ as baseline. Tracing effectiveness $TSR$ is slightly lower, which we attribute to domain mismatch, as SimSwap is trained on VGGFace2.

\noindent\textbf{Exp \textsc{II}. Gender-aware cloak image selection.}
The balanced heuristic employs a gender-aware cloak selection strategy, motivated by the observation that an opposite-gender cloak tends to appear more visually distinct to human eyes than a same-gender cloak.
As shown in Table~\ref{tab:cloak_image_selection} Exp~\textsc{II}, with the same identity-distance-based selection, opposite-gender cloaks improve identity protection: $ASR_{id}$ decreases from 0.73\% to 0.50\% (FaceNet-512) and from 0.30\% to 0.00\% (Face++). This gain comes with a small decrease in tracing effectiveness, with $TSR$ dropping from 94.93\% to 92.07\% (FaceNet-512) and from 97.97\% to 96.70\% (Face++).
Context protection also slightly degrades, as $ASR_{ctx}$ increases from 1.90\% to 2.70\% on FaceNet-512 and from 3.20\% to 3.90\% on Face++.

\noindent\textbf{Exp \textsc{III}. Prioritize protection.} 
In the baseline and balanced heuristics, we treat protection and tracing as equally important objectives by setting $w_{id}{=}w_{ctx}{=}w_{cloak}{=}1$. When Alice prioritizes identity and context protection, we set $w_{cloak}=0$ to emphasize protection performance. Under this configuration, validation identifies an optimal identity-distance threshold of 1.35.

On the test set, this configuration achieves $ASR_{id}{=}0\%$ under both FaceNet-512 and Face++, while $ASR_{ctx}$ remains low at 1.93\% and 3.53\%, respectively. By prioritizing protection, identity-stealing attacks are fully eliminated, while context protection remains largely stable, as it is primarily constrained by the perturbation bound. We further discuss this limitation in Section~\ref{sec:ablation_studies}.

\noindent\textbf{Exp \textsc{IV}. Cloak image distances.}
In the previous experiments, we adopted a distance-based strategy to select the cloak image that maximizes the composite protection score $\mathcal{S}$. Here, we further explore how protections vary as the identity distance increases from 1.04 (the identity-matching threshold) to 2.0.
A clear trade-off is observed from the results shown in Table~\ref{tab:cloak_image_selection} Exp~\textsc{IV}: As the distance increases (i.e., the cloak becomes less similar to Alice), $ASR_{id}$ decreases, indicating stronger identity protection; while tracing effectiveness ($TSR$) declines. Meanwhile, $ASR_{ctx}$ increases with larger identity distances. We attribute this trend to the limited perturbation budget: enforcing greater identity deviation consumes more noise budget, leaving insufficient capacity for preserving contextual features and thereby degrading context protection. 

\subsection{Defend against Diffusion–based Deepfakes}\label{subsec:diffusion_model}

Diffusion-based face-swapping models generally produce higher-quality results than GAN-based methods and have gained increasing popularity in recent years. As discussed in Section~\ref{subsec:exp_simswap}, \name\ achieves image protection by manipulating identity and context feature extractors in the latent space, enabling unified protection across different model architectures without relying on specific downstream face-swapping models. In this section, we evaluate \name\ against DiffFace~\cite{kim2025diffface} to assess its effectiveness on diffusion-based face-swapping systems.

\noindent\textbf{Dataset and Settings.} {The default test set for DiffFace is the video-based FaceForensics++ dataset, consistent with FaceShifter. To ensure sufficient identity diversity, and because DiffFace is trained using FFHQ as the conditional DDPM dataset, we instead adopt FFHQ as the test set.
The identity-distance threshold for baseline cloak selection is determined on the validation set and set to 1.31, and is increased to 1.55 under the protection-prioritized setting.}

\noindent\textbf{Results.} Table~\ref{tab:protection_performance} summarizes the protection performance of \name\ against DiffFace.
\textbf{(1)} \name\ requires substantially smaller perturbations when defending against DiffFace, achieving an MSE of 57 and an SSIM of 0.77. We attribute this to differences in context manipulation between GAN- and diffusion-based face-swapping models: while GAN-based models typically require larger perturbations to induce significant context deviation, diffusion models can be disrupted by steering facial landmarks toward non-facial categories, which demands smaller perturbations.
\textbf{(2)} The protected images preserve the original identity, with over 99.5\% of $\tilde{\mathbf{x}}$ recognized as $\mathbf{x}$.
\textbf{(3)} Identity-stealing attacks are effectively mitigated, with $ASR_{id}<0.5\%$. 
\textbf{(4)} \name~maintains strong tracing effectiveness, achieving a $TSR$ of 68.97\% on FaceNet-512 and 68.93\% on Face++. Although the $TSR$ values on DiffFace appear lower than those on SimSwap, they are already within 1.2\% of the corresponding $ASR_{o}$ values. This is because DM-based face-swapping pipelines generally produce higher-quality results, but often require image-specific parameter tuning to achieve optimal performance. To ensure a fair and large-scale evaluation, we use the default DiffFace parameters across all 3,000 image pairs, resulting in lower baseline face-swapping success rates of 69.67\% on FaceNet-512 and 70.13\% on Face++.
\textbf{(5)} Context protection remains effective, with $ASR_{ctx}$ limited to 1.00\% and 0.50\%.

\noindent\textbf{Cloak Image Selection Strategies.} We report the cloak selection experiments for DiffFace in Table~\ref{tab:cloak_image_selection}. Overall, trends are consistent with those observed for the GAN-based deepfake.
\textbf{(1)} Using AI-generated cloaks leads to lower tracing success rates, which we attribute to the domain gap between synthetic and real faces.
\textbf{(2)} Selecting opposite-gender cloak images improves identity protection (lower $ASR_{id}$) at the cost of reduced tracing effectiveness.
\textbf{(3)} When protection is prioritized, \name\ achieves $ASR_{id}=0\%$ while maintaining a comparable $ASR_{ctx}$.
\textbf{(4)} As the identity distance used in cloak selection increases from 1.04 to 2.00, $ASR_{id}$ consistently decreases: from 8.30\% to 0\% (FaceNet-512) and from 1.10\% to 0.00\% (Megvii Face++), respectively. In contrast, $TSR$ declines from 69.73\% and 69.90\% to 64.33\% and 64.00\%. Notably, under different cloak image selection strategies, $ASR_{ctx}$ remains largely stable rather than increasing. Visually, across all cloak image selection strategies, identity-stealing results resemble the cloak identity, while context-stealing attempts fail to insert the swapped face into the victim’s context. Examples are provided in Appendix Table~\ref{tab:diffface_anchor_sample}.

\begin{table}[t]
  \centering
  \small
  \captionsetup{skip=3 pt}
  \caption{Unified identity protection performance of \name\ against SimSwap, FaceShifter, HifiFace and DiffFace.}
  \label{tab:unified_model_evaluation}

  \setlength\tabcolsep{6.15 pt}
  \begin{tabular}{c}
    \hline
    \name~Utility: MSE 108, PSNR 27.82, SSIM 0.69, and LPIPS 0.31. \\
    \hline
  \end{tabular}
  
  \vspace{0.3 em}
  \hrule
  \vspace{0.3 em}
  
  \setlength\tabcolsep{2.85 pt}
  \renewcommand{\arraystretch}{1.1}
  \begin{tabular}{c | c  c  c  c | c  c  c  c} 
    \hline
    \multirow{2}*{\makecell{Models}} & \multicolumn{4}{c}{FaceNet-512} \vline & \multicolumn{4}{c}{Megvii Face++} \\
    \cline{2-9}
    ~ & $ASR_o$ & $ASR_{id}{\downarrow}$ & $TSR{\uparrow}$ & $\mathcal{S}{\uparrow}$ & $ASR_o$ & $ASR_{id}{\downarrow}$ & $TSR{\uparrow}$ & $\mathcal{S}{\uparrow}$ \\
    \hline
    SimSwap & 94.13 & 2.14 & 87.83 & 1.86 & 98.20 & 0.62 & 89.86 & 1.89 \\
    FaceShifter & 70.17 & 2.03 & 63.58 & 1.62 & 72.33 & 0.21 & 71.72 & 1.72 \\
    HifiFace & 93.12 & 2.47 & 95.95 & 1.94 & 95.67 & 1.23 & 95.98 & 1.95 \\
    \hline
     DiffFace &  69.67 &  10.97 &  47.33 &  1.36 &  70.13 &  2.73 &  38.63 &  1.36 \\
    \hline
  \end{tabular}
  \vspace{-4 mm}
\end{table}

\subsection{Unified Defense and Black-box Defense}\label{subsec:black_box}

\noindent\textbf{Unified Defense. }
For efficient identity manipulation and contextual consistency, face-swapping models typically operate in the latent space instead of directly editing pixels \cite{zhao2023diffswap,kim2025diffface,zhou2023uniface,gao2021information,li2023e4s,shiohara2023blendface}.
Leveraging this shared architecture, \name~can generalize across diverse face-swapping models by jointly targeting multiple feature extractors. We implement a unified defense to jointly target the identity extractors of three face-swapping methods, SimSwap, FaceShifter, and HifiFace, to generate a \textit{single protected image}, and then evaluate its protection effectiveness against each model.

As shown in Table~\ref{tab:unified_model_evaluation}, unified identity protection exhibits trends consistent with single-model defense:
\textbf{(1)} Only minor perturbations are added to the protected images, e.g, the mean MSE is 108.
\textbf{(2)} \name~significantly reduces identity-stealing $ASR$ across all models ($ASR_{id}{<}2.5\%$).
\textbf{(3)} The cloak identities are highly traceable under unified protection, with performance comparable to single-model defense, e.g., for HifiFace, the cloak-matching rates reach 95.95\% and 95.98\% on FaceNet-512 and Face++, respectively. \textbf{(4)} We further evaluate the protected images on an unseen diffusion-based face-swapping model, DiffFace, and observe strong transferability, with significantly reduced $ASR_{id}$ while maintaining $TSR$.

\noindent\textbf{Black-box Defense. }
In real-world defense, the attacker’s face-swapping model is typically unknown, making it impractical to tailor protection to a specific model/extractor. Building upon the unified defense, we extend \name~to black-box setting, where the target model’s identity extractor is unknown/inaccessible. We follow NullSwap~\cite{wang2025nullswap} and optimize perturbations against standard face recognition models that commonly serve as the backbone of identity extractors. All experiments are conducted on 3000 randomly sampled image pairs from the FFHQ dataset for consistency.

In Table~\ref{tab:blackbox_protection}, we report the black-box identity protection performance across nine face-swapping models with different architectures and processing pipelines:
\textbf{(1)} Only minor perturbations are needed to achieve effective black-box defense (mean MSE of 70 and SSIM of 0.85).
\textbf{(2)} Across all models, \name~significantly reduces the face-swapping success rate. For example, UniFace's ASR is reduced from 89.10\% to 2.07\% and from 87.83\% to 0.77\%, as evaluated by FaceNet and Face++. The cloak image plays a significant role in defense. Please refer to the ablation study in Section~\ref{sec:ablation_studies} for details.
\textbf{(3)} However, under the black-box setting, \name's tracing performance ($TSR$) decreased. 
\textbf{(4)} We further evaluate the protected images on the commercial face-swapping tool Deep-Live-Cam \cite{deep_live_cam}. \name\ still achieves strong defense as shown in Table~\ref{tab:blackbox_protection}. The slight performance decrease is likely due to additional optimizations in the commercial system.

\begin{table}[t]
  \centering
  \small
  \captionsetup{skip=3 pt}
  \caption{Black-box identity protection of \name~and NullSwap~\cite{wang2025nullswap}. $F^1$: FaceNet-512, $F^2$: Megvii Face++. Bold indicates better $ASR_{id}$ between \name~and NullSwap.} \label{tab:blackbox_protection}
  \setlength\tabcolsep{3.28 pt}
  \renewcommand{\arraystretch}{1.1}
    
  \begin{tabular}{c  c  c  c | c  c  c  c}
    \hline
    \multicolumn{4}{c}{\name~Utility} \vline & \multicolumn{4}{c}{NullSwap Utility} \\
    \hline
    MSE $\downarrow$ & PSNR $\uparrow$ & SSIM $\uparrow$ & LPIPS $\downarrow$ & MSE $\downarrow$ & PSNR $\uparrow$ & SSIM $\uparrow$ & LPIPS $\downarrow$ \\
    \hline
    70 & 29.81 & 0.85 & 0.24 & 70 & 29.82 & 0.85 & 0.24 \\
    \hline
  \end{tabular}
  
  \vspace{0.3 em}
  \hrule
  \vspace{0.3 em}
  
  \setlength\tabcolsep{2.04 pt}
  \begin{tabular}{c | c  c | c  c  c  c | c  c}
    \hline
    \multirow{3}*{Model} & \multicolumn{2}{c}{\multirow{2}{*}{$ASR_{o}$}} \vline & \multicolumn{4}{c}{\name} \vline & \multicolumn{2}{c}{NullSwap} \\
    \cline{4-9}
    ~ & ~ & ~ & \multicolumn{2}{c}{$TSR{\uparrow}$} & \multicolumn{2}{c}{$ASR_{id}{\downarrow}$}  \vline & \multicolumn{2}{c}{$ASR_{id}{\downarrow}$} \\
    \cline{2-9}
    ~ & $F^1$ & $F^2$ & $F^1$ & $F^2$ & $F^1$ & $F^2$ & $F^1$ & $F^2$ \\
    \hline
    SimSwap~\cite{chen2020simswap} & 91.03 & 89.47 & 15.27 & 3.20 & \textbf{8.33} & \textbf{2.97} & 11.83 & 3.00 \\
    FaceShifter~\cite{li2019faceshifter} & 84.03 & 86.67 & 26.03 & 13.40 & \textbf{1.27} & \textbf{0.40} & 8.57 & 2.47 \\
    HifiFace~\cite{wang2021hififace} & 82.63 & 73.73 & 12.27 & 1.97 & \textbf{9.33} & 4.73 & 14.63 & \textbf{2.47} \\
    DiffFace~\cite{kim2025diffface} & 69.67 & 70.13 & 10.30 & 3.13 & \textbf{4.67} & 1.83 & 6.17 & \textbf{1.43} \\
    UniFace~\cite{zhou2023uniface} & 89.10 & 87.83 & 13.83 & 3.73 & \textbf{2.07} & \textbf{0.77} & 12.53 & 3.00 \\
    InfoSwap~\cite{gao2021information} & 87.53 & 84.17 & 20.10 & 5.87 & \textbf{2.47} & \textbf{1.00} & 11.40 & 3.53 \\
    E4S~\cite{li2023e4s} & 76.47 & 67.50 & 5.17 & 0.73 & \textbf{7.77} & \textbf{3.63} & 14.43 & 5.30 \\
    DiffSwap~\cite{zhao2023diffswap} & 80.80 & 84.67 & 1.70 & 0.93 & 8.70 & \textbf{2.87} & \textbf{7.57} & 3.13 \\
    \hline
    Deep-Live-Cam~\cite{deep_live_cam} & 90.90 & 95.30 & 11.87 & 3.14 & \textbf{13.67} & \textbf{15.70} & 23.03 & 20.80 \\
    \hline
  \end{tabular}
  \vspace{-3 mm}
\end{table}

\begin{table}[t]
  \centering
  \small
  \captionsetup{skip=3 pt}
  \caption{White-box identity protection of \name\ and LowKey~\cite{cherepanova2021lowkey}. Bold indicates better $ASR_{id}$. Model index: 1 SimSwap, 2 FaceShifter, 3 HifiFace, 4 DiffFace.
  }\label{tab:protection_comparison_with_lowkey}
  \setlength\tabcolsep{1.1 pt}
  \renewcommand{\arraystretch}{1.1}
  
  \begin{tabular}{c | c  c | c  c | c  c | c  c  c  c | c  c}
    \hline
    \multirow{3}{*}{\rotatebox[origin=c]{90}{Model}} & \multicolumn{2}{c}{\multirow{2}{*}{$ASR_{o}$}} \vline & \multicolumn{2}{c}{\multirow{2}{*}{\name}} \vline & \multicolumn{2}{c}{\multirow{2}{*}{LowKey}} \vline & \multicolumn{4}{c}{\name} \vline & \multicolumn{2}{c}{LowKey} \\
    \cline{8-13}
    ~ & ~ & ~ & ~ & ~ & ~ & ~ & \multicolumn{2}{c}{$TSR{\uparrow}$} & \multicolumn{2}{c}{$ASR_{id}{\downarrow}$} \vline & \multicolumn{2}{c}{$ASR_{id}{\downarrow}$} \\
    \cline{2-13}
     ~ & $F^1$ & $F^2$ & PSNR & SSIM & PSNR & SSIM & $F^1$ & $F^2$ & $F^1$ & $F^2$  & $F^1$ & $F^2$ \\
    \hline
    1 & 94.13 & 98.20 & 36.20 & 0.92 & 36.73 & 0.83 & 96.10 & 98.93 & \textbf{0.53} & \textbf{0.20} & 25.33 & 27.67 \\
    2 & 70.17 & 72.33 & 29.43 & 0.80 & 30.44 & 0.77 & 66.67 & 71.03 & \textbf{1.63} & \textbf{0.23} & 9.50 & 10.33 \\
    3 & 93.12 & 95.67 & 29.23 & 0.74  & 28.24 & 0.70 & 96.00 & 96.23 & \textbf{1.03} & \textbf{0.40} & 11.40 & 11.03 \\
    4 & 69.67 & 70.13 & 31.91 & 0.82 & 30.94 & 0.77 & 70.13 & 70.73 & \textbf{0.23} & \textbf{0.09} & 6.56 & 5.43 \\
    \hline
  \end{tabular}
  
  \setlength\tabcolsep{1.2 pt}
  \vspace{-3mm}
\end{table}

\begin{table}[t]
  \centering
  \small
  \captionsetup{skip=3 pt}
  \caption{Protection performance evaluated with additional FR models and human evaluation.}
  \label{tab:more_FR_and_human_evaluation}
  \setlength\tabcolsep{1.18 pt}
  \renewcommand{\arraystretch}{1.1}
  \begin{tabular}{c  c  c | c  c  c | c  c  c | c  c  c}
    \hline
    \multicolumn{12}{c}{Additional FR models evaluation} \\
    \hline
    \multicolumn{6}{c}{Face Recognition~\cite{face_recognition}} \vline & \multicolumn{6}{c}{AWS Rekognition~\cite{aws_rekognition}} \\
    \hline
    \multicolumn{3}{c}{$\mathcal{T}_{identity}$} \vline & \multicolumn{2}{c}{$\mathcal{T}_{context}$} & \multirow{2}*{$\mathcal{S}$} & \multicolumn{3}{c}{$\mathcal{T}_{identity}$} \vline & \multicolumn{2}{c}{$\mathcal{T}_{context}$} & \multirow{2}*{$\mathcal{S}$} \\
    \cline{1-5} \cline{7-11}
    $ASR_o$ & $ASR_p$ & $TSR$ & $ASR_o$ & $ASR_p$ & ~ & $ASR_o$ & $ASR_p$ & $TSR$ & $ASR_o$ & $ASR_p$ & ~ \\
    \hline
    92.86 & 1.13 & 92.63 & 93.20 & 2.00 & 2.90 & 95.33 & 0.00 & 94.67 & 96.07 & 4.33 & 2.90 \\
    \hline
  \end{tabular}

  \vspace{0.3 em}
  \hrule
  \vspace{0.3 em}
  
  \setlength\tabcolsep{4.26 pt}
  \begin{tabular}{c | c  c  c | c  c  c }
    \hline
    \multicolumn{7}{c}{\makecell{Human evaluation Q2: Does the deepfake result better  resemble \\ the victim identity or the cloak identity?}} \\
    \hline
    & Victim ID & Cloak ID & Skip & Victim ID & Cloak ID & Skip \\
    \hline
    $\mathcal{T}_{identity}$ & \multicolumn{3}{c|}{ w/ Pt. VGGFace2/FFHQ} & \multicolumn{3}{c}{w/ Pt. AI-Generate} \\
    \hline
    SimSwap & 26 & 132 & 102 & 41 & 123 & 111 \\
    \hline
    DiffFace & 36 & 137 & 102 & 21 & 130 & 79 \\
    \hline
  \end{tabular}
  \vspace{-6 mm}
\end{table}

\begin{table}[t]
  \centering
  \small
  \captionsetup{skip=3 pt}
  \caption{Ablation study: removing the RGB channel bound $\delta_{RGB}$, identity deviation bound $\delta_{id}$ and context deviation bound $\delta_{ctx}$. Here, $A_{i}$ and $A_{c}$ denotes $ASR_{id}$ and $ASR_{ctx}$.}
  \label{tab:ablation_study}
  \setlength\tabcolsep{1.05 pt}
  \begin{tabular}{c | c  c  c  c | c  c  c  c | c  c  c  c}
    \hline
    \multirow{3}*{\makecell{No \\ Bound}} & \multicolumn{4}{c}{Utility} \vline & \multicolumn{8}{c}{Matching Rate ($\%$)} \\
    \cline{2-13}
    ~ & \multicolumn{4}{c}{$\mathbf{x}$ and $\tilde{\mathbf{x}}$ } \vline & \multicolumn{4}{c}{FaceNet-512} \vline & \multicolumn{4}{c}{Megvii Face++} \\
    \cline{2-13}
    ~ & MSE & PSNR & SSIM & LPIPS & $A_{i}{\downarrow}$ & $TSR{\uparrow}$ & $A_{c}{\downarrow}$ & $\mathcal{S}{\uparrow}$ & $A_{i}{\downarrow}$& $TSR{\uparrow}$ & $A_{c}{\downarrow}$ & $\mathcal{S}{\uparrow}$ \\
    \hline
    None & 106 & 27.88 & 0.68 & 0.25 & 0.73 & 94.93 & 1.90 & 2.92 & 0.30 & 97.97 & 3.20 & 2.95 \\
    $\delta_{RGB}$ & 128 & 27.23 & 0.65 & 0.28 & 0.55 & 95.00 & 0.30 & 2.94 & 0.23 & 99.43 & 0.10 & 2.99 \\
    $\delta_{id}$ & 112 & 27.64 & 0.67 & 0.27 & 0.27 & 0.00 & 19.30 & 1.80 & 0.20 & 0.00 & 35.20 & 1.65 \\
    $\delta_{ctx}$ & 110 & 27.71 & 0.67 & 0.25 & 1.13 & 93.53 & 1.90 & 2.91 & 0.23 & 97.90 & 3.13 & 2.95 \\
    All & 415 & 21.98 & 0.44 & 0.49 & 0.07 & 0.00 & 0.00 & 2.00 & 0.00 & 0.00 & 0.00 & 2.00 \\
    \hline
  \end{tabular}
  \vspace{-3mm}
\end{table}

\subsection{Comparison with Previous Works}\label{subsec:comparison}

We compare the protection performance of \name~with representative prior works, a white-box identity protection method, LowKey\cite{cherepanova2021lowkey}, and a black-box identity protection method, NullSwap~\cite{wang2025nullswap}. For a fair comparison, we configure \name~to focus solely on identity protection by setting $\delta_{ctx}$ to zero.
 
As \name~optimizes perturbations using MSE but LowKey uses LPIPS, we align PSNR and SSIM to ensure a fair comparison. As shown in Table~\ref{tab:protection_comparison_with_lowkey},
\textbf{(1)} With similar perturbation levels (PSNR and SSIM), \name~outperforms LowKey across all face-swapping models. For example, on SimSwap, \name~achieves an $ASR_{id}$ of 0.53\% and 0.20\% with FaceNet and Face++, respectively, whereas LowKey achieves 25.33\% and 27.67\%.
\textbf{(2)} As an additional benefit, \name~achieves $TSR$ at a level comparable to the inherent face-swapping capability of the model, i.e., all ``successful'' swappings are swapped to the cloak identity. For example, on HifiFace, the $ASR_o$ is 93.12\% (FaceNet) and 95.67\% (Face++), while \name~achieves a $TSR$ of 96.00\% and 96.23\%, respectively.

Table~\ref{tab:blackbox_protection} compares the protection performance of \name~and NullSwap under similar perturbation levels. \name~consistently outperforms NullSwap across most face-swapping models. For example, on UniFace, \name~reduces $ASR_{id}$ to 2.07\% and 0.77\% when evaluated with FaceNet and Face++, respectively, compared with 12.53\% and 3.00\% achieved by NullSwap.

\name~protects both identity and context against face-swapping deepfakes. To our best knowledge, there are no existing proactive defenses specifically designed to protect image context against deepfake, i.e., to prevent an external identity from being inserted into a protected image. In Appendix~\ref{app:context_comparison}, we compare the context protection performance of \name~with representative proactive defenses that comprehensively protect facial images against other types of deepfake manipulation.

\subsection{Additional FR and Human Evaluations}\label{subsec:more_FR_and_human_evaluation}

\noindent\textbf{Additional FR Model Evaluations.} {The design of \name\ is entirely independent of face recognition models, and FaceNet-512 and Face++ are used solely for \textit{ex post facto} evaluation. To avoid potential bias, we additionally evaluate its performance using the Face Recognition library~\cite{face_recognition} and AWS Rekognition~\cite{aws_rekognition}.
We repeat the experiments in Section~\ref{subsec:exp_simswap} following the same settings. As shown in Table~\ref{tab:more_FR_and_human_evaluation}, evaluations with Face Recognition and AWS Rekognition yield results consistent with those obtained using FaceNet-512 and Face++. These results demonstrate that \name\ provides face recognition–agnostic protection and confirm that the use of FaceNet-512 and Face++ does not introduce evaluation bias.} 

\noindent\textbf{Human Evaluations.} Since the ultimate audience of deepfakes is human, we conduct a user study to evaluate human perception of \name’s usability and effectiveness. The survey begins with an IRB information statement, followed by background questions (e.g., whether participants are concerned about deepfakes). Participants then complete three tasks of five image batches each, designed to answer the following questions:
\textbf{Q1:} Is the perturbation visually acceptable to human viewers?
\textbf{Q2:} Do deepfakes generated from \textit{protected} images resemble the victim identity or the cloak identity?
\textbf{Q3:} Are context-stealing face-swapping results perceived as successful deepfakes by human observers?
For all tasks, participants can skip questions if they are not confident in their judgment.

The user study was distributed to CS/engineering students and faculty, yielding 107 SimSwap responses (52 VGGFace2 cloaks, 55 AI-generated cloaks) and
101 DiffFace responses (55 FFHQ cloaks, 46 AI cloaks), with over 3,000 annotations. Among all participants, only 43.11\% reported that they post images online (e.g., on social media or LinkedIn). In contrast, 81.30\% expressed concern about the protection of their shared images against malicious applications, indicating the real-world relevance of \name.

\noindent$\bullet~$ \textbf{Q1:} In a side-by-side comparison of original and protected images, perturbations in 85.08\% of SimSwap images and 79.43\% of DiffFace images were labeled visually acceptable, indicating that the protection is mostly negligible for human observers. However, the utility metric for DiffFace outperforms that of SimSwap (Table~\ref{tab:protection_performance}). These metrics are inconsistent with the human evaluation results, which could be caused by differences in perturbation patterns: SimSwap produces more uniformly distributed perturbations, whereas DiffFace introduces more irregularly shaped perturbations that are more noticeable to human participants.

\noindent$\bullet~$ \textbf{Q2:} As shown in Table~\ref{tab:more_FR_and_human_evaluation}, the majority of the participants in Q2 perceived deepfakes generated from protected images as resembling the cloak identity, not the victim (79.19\% for SimSwap and 82.41\% for DiffFace). 
The human evaluation results are consistent with the results from the automated face recognition models reported in Table~\ref{tab:cloak_image_selection}.
The results confirm \name's effectiveness in identity protection against deepfakes.
In addition, about 40\% of the images (both tasks/models)
were skipped by the participants. This is expected, as humans are known to be less effective in face recognition than state-of-the-art deep learning models \cite{ranjan2018deep,guo2019survey}.

\noindent$\bullet~$ \textbf{Q3:} 94.21\% of the SimSwap context-stealing attacks and 97.36\% of DiffFace attacks were labeled as unsuccessful by the participants, which confirms the effectiveness of \name’s context protection. Protection performance remained comparable when real or AI-generated images were used as cloaks (95.38\% vs. 93.09\% for SimSwap, and 97.73\% vs. 96.93\% for DiffFace). This trend aligns with the results in Sections~\ref{subsec:cloak_image} and~\ref{subsec:diffusion_model}, suggesting that the type of the cloak images has a limited effect on context protection.
\section{Ablation Study}\label{sec:ablation_studies}

\noindent\textbf{Bounds on Noise, Identity Deviation, and Context Deviation.} We impose explicit noise bounds to constrain the perturbation magnitude and set upper bounds on $|\Delta \mathbf{z}_s|$ and $|\Delta \mathbf{z}_t|$ within a prioritized constrained optimization framework (Section~\ref{subsec:methodology}).
In this section, we analyze the effects of these bounds on SimSwap as a representative face-swapping model.

As shown in Table~\ref{tab:ablation_study}, removing the noise bounds improves both identity and context protection. $ASR_{id}$ decreases from 0.73\% to 0.55\% (FaceNet-512) and from 0.30\% to 0.23\% (Face++).
$ASR_{ctx}$ drops substantially, from 1.90\% to 0.30\% (FaceNet-512) and from 3.20\% to 0.10\% (Face++).
These gains come at the cost of increased perturbation, with MSE rising from 106 to 128.
The results indicate that, for GAN-based face-swapping models, context protection is more sensitive to the noise level than identity protection: allowing larger perturbations yields substantially stronger context protection.

When either the identity or context deviation bound is removed individually, the perturbation magnitude increases only marginally, indicating that both deviations are implicitly constrained by the noise bound $\delta_{RGB}$.
Removing the identity deviation bound yields stronger identity protection: the FaceNet-512–based $ASR_{id}$ drops to 0.27\%, and the Face++–based $ASR_{id}$ decreases to 0.20\%.
However, this improvement comes at the expense of complete tracing failure and a severe degradation in context protection.
In contrast, removing the context deviation bound alone has minimal impact, and the overall protection performance remains largely unchanged.

However, when all bounds are removed, the perturbation magnitude increases dramatically, with the MSE rising from 106 to 415. Under this unconstrained setting, both $ASR_{id}$ and $ASR_{ctx}$ drop to zero, indicating maximal protection. However, forensic tracing becomes entirely ineffective, with $TSR$ also dropping to zero.

\begin{figure*}[htbp]
    \centering
    \begin{minipage}[t]{0.246\textwidth}
        \centering
        \includegraphics[width=\linewidth]{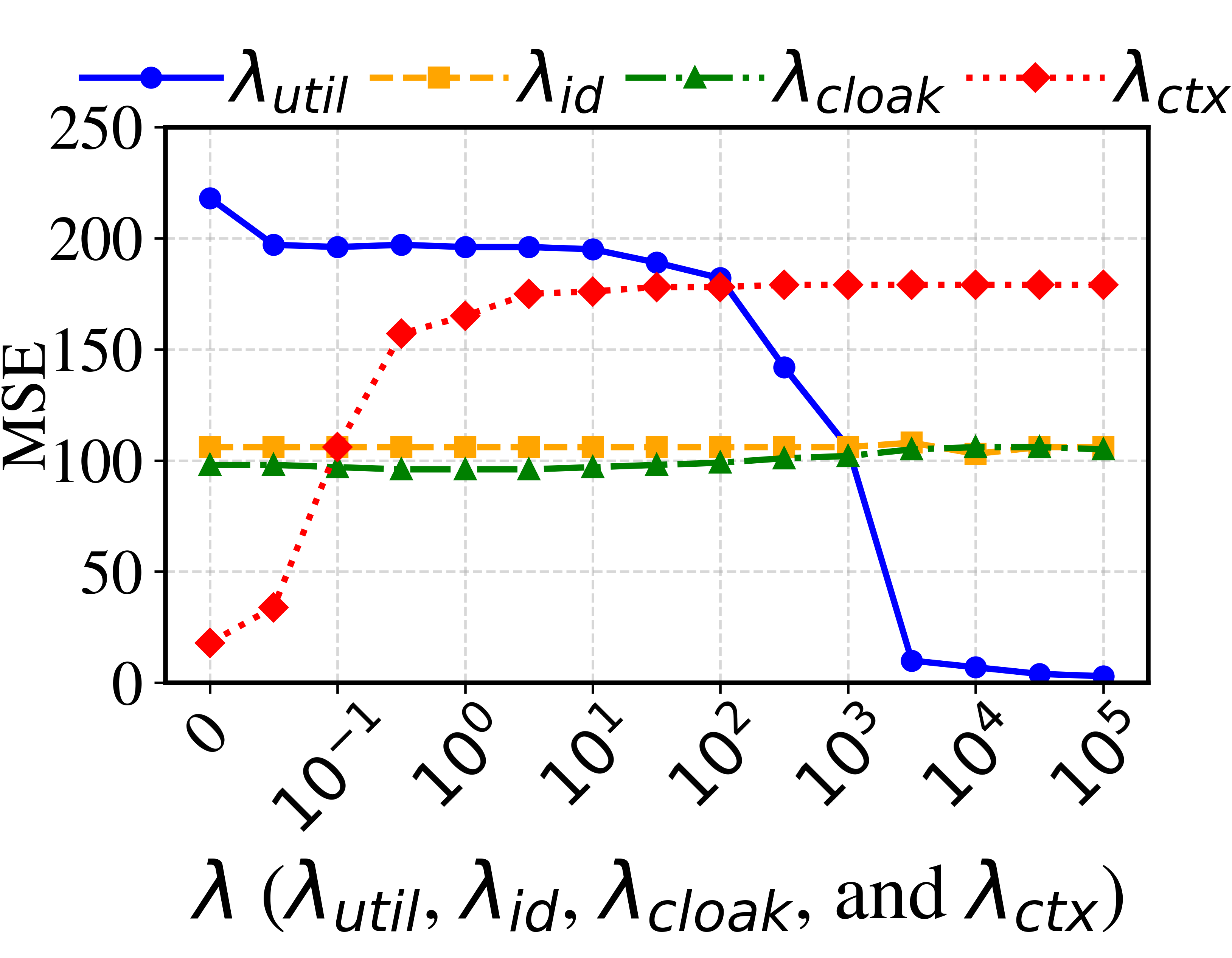}
    \end{minipage}
    \begin{minipage}[t]{0.246\textwidth}
        \centering
        \includegraphics[width=\linewidth]{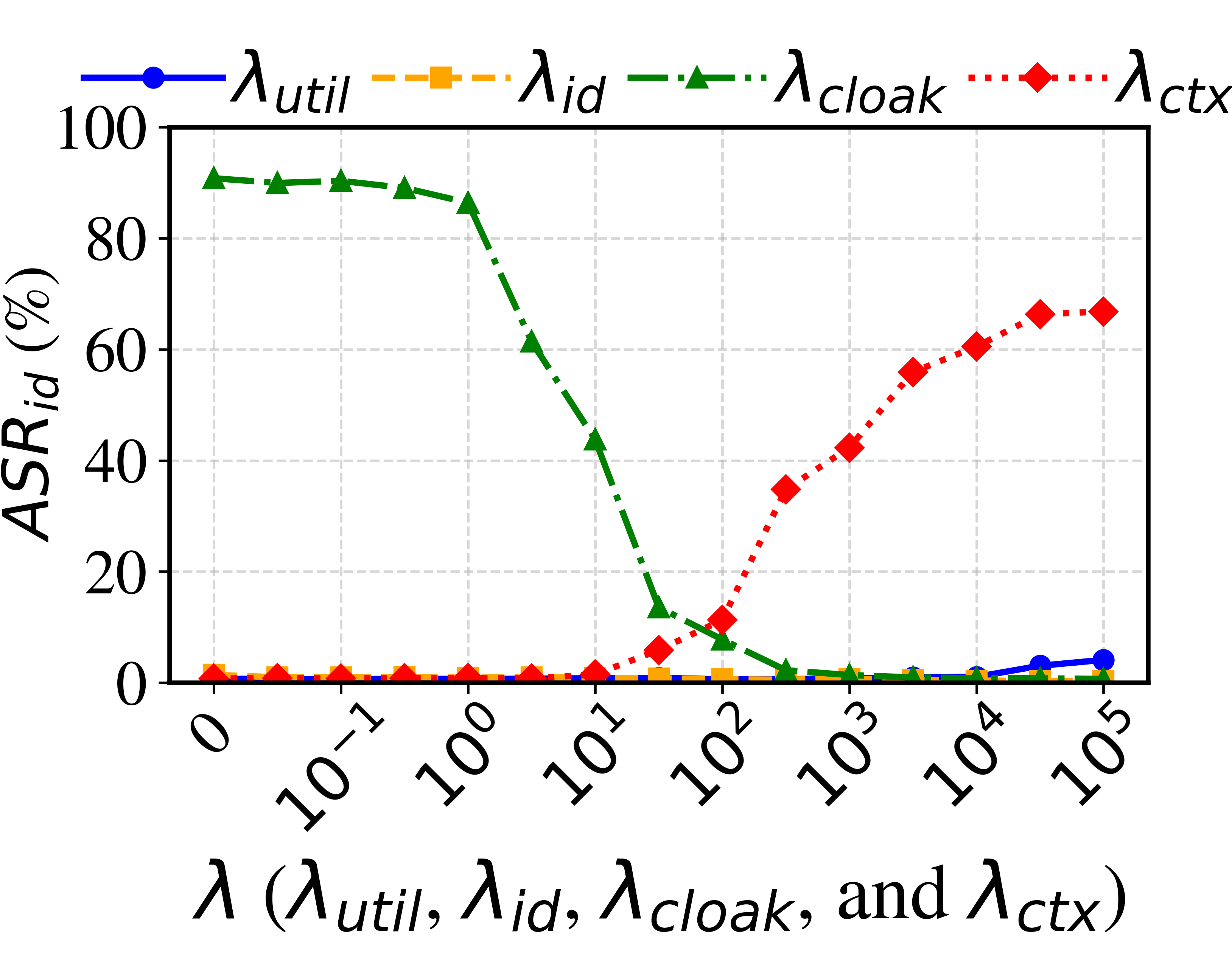}
    \end{minipage}
    \begin{minipage}[t]{0.246\textwidth}
        \centering
        \includegraphics[width=\linewidth]{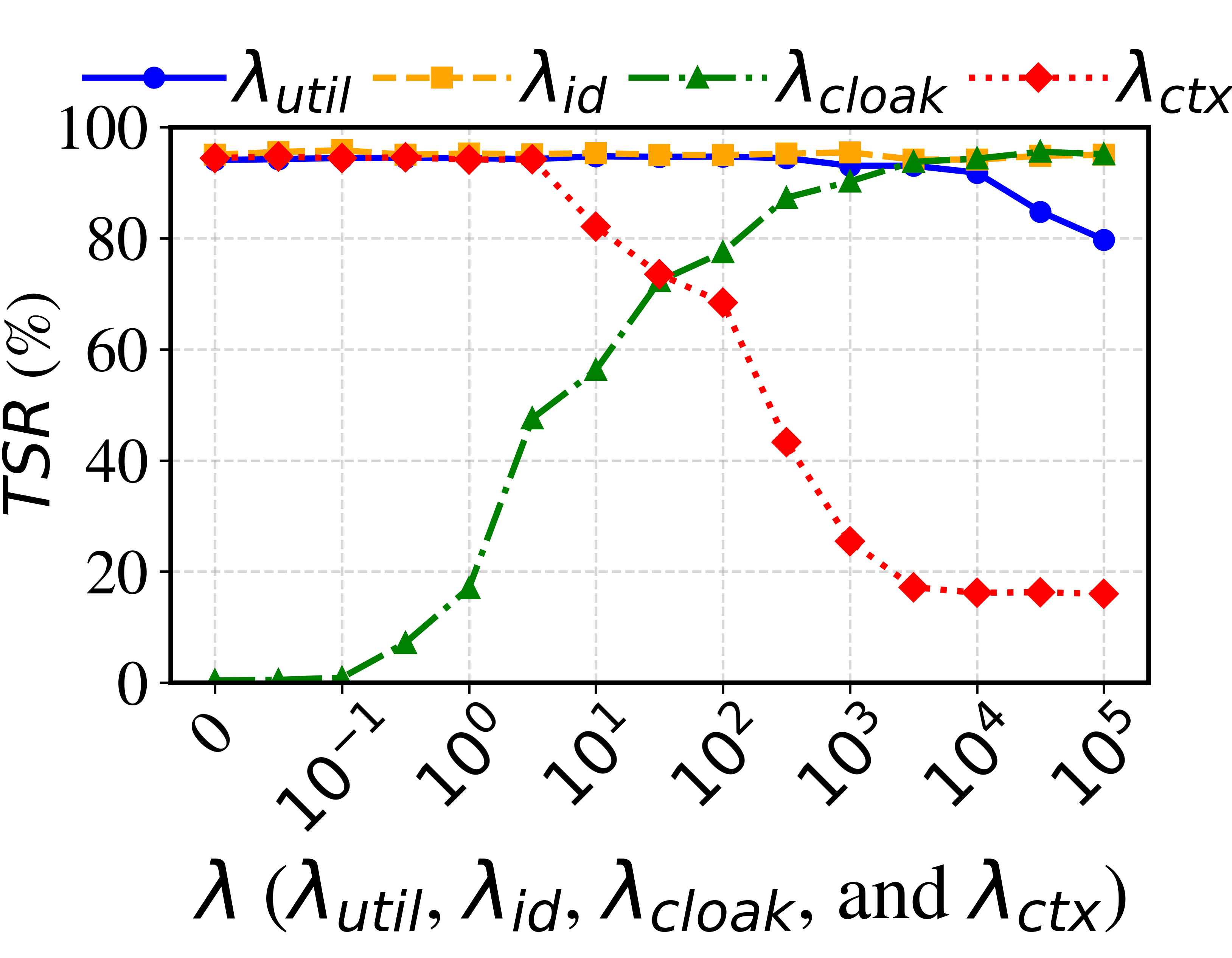}
    \end{minipage}
    \begin{minipage}[t]{0.246\textwidth}
        \centering
        \includegraphics[width=\linewidth]{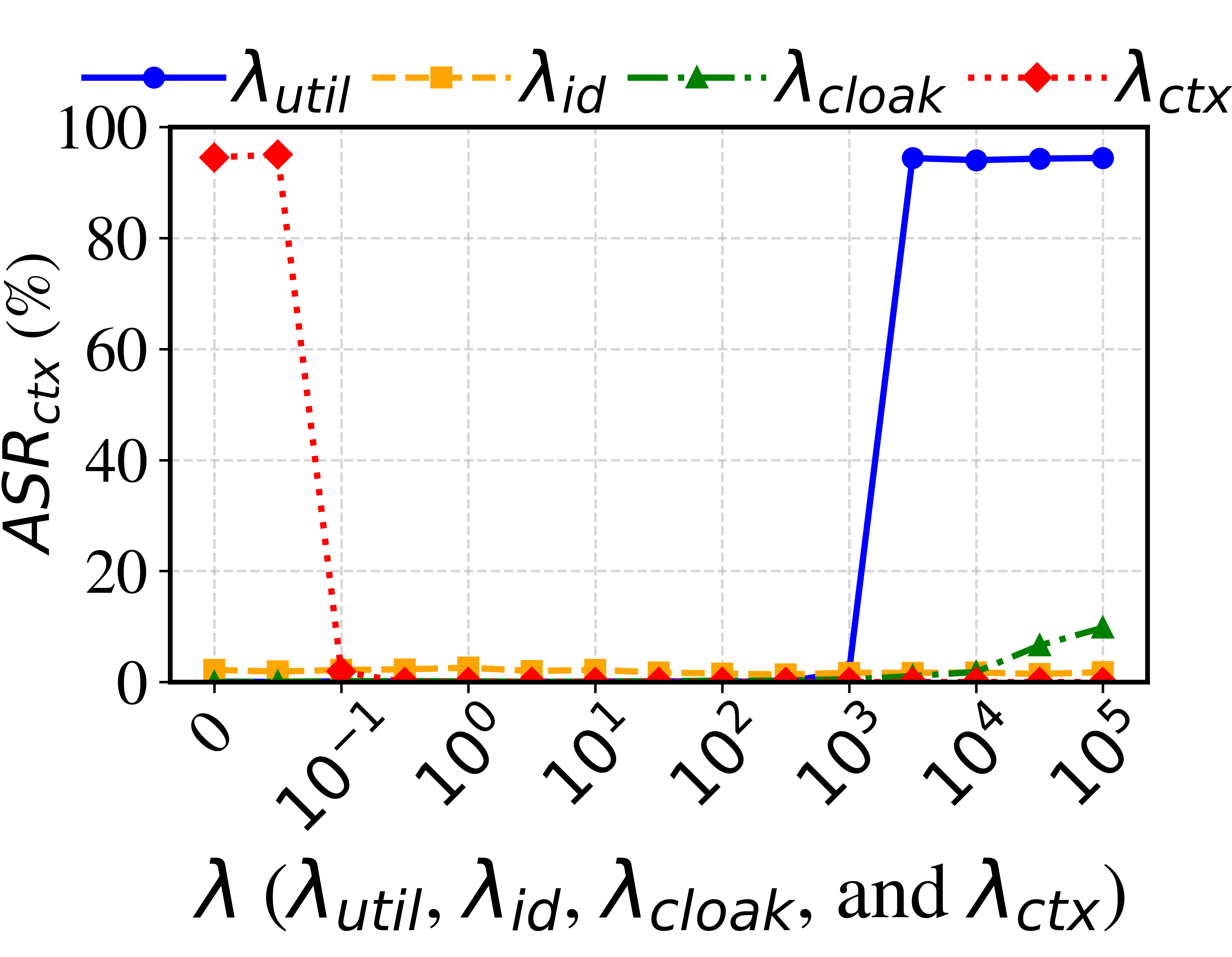}
    \end{minipage}
    \captionsetup{skip=-3 pt}
    \vspace{1 mm}
    \caption{Sensitivity analysis of coefficient weights. The x-axis represents the weight values ranging from 0 to $10^5$ for $\lambda$ ($\lambda_{util}$, $\lambda_{id}$, $\lambda_{cloak}$, and $\lambda_{ctx}$). The y-axis shows MSE, $ASR_{id}$ (\%), $TSR$ (\%), and $ASR_{ctx}$ (\%) from left to right.}
    \Description{}
    \label{fig:ablation_study}
    \vspace{-3 mm}
\end{figure*}

\vspace{1 mm}
\noindent\textbf{The Coefficients in the Loss Function.}
In Section~\ref{subsec:exp_simswap}, we adopt the default loss weight coefficients $\lambda_{util}=1000$, $\lambda_{id}=10000$, $\lambda_{cloak}=10000$, and $\lambda_{ctx}=0.1$ in Eq.~\ref{eqa:pgd_simswap_all}.
These heterogeneous weights compensate for scale differences among the perturbation magnitude, latent features, and identity embeddings.
To examine the effects of imbalanced coefficients, we vary each weight over a wide range, from small values (0, 0.05, 0.1), through intermediate values (10, 50, 100, \dots), to large values (50{,}000 and 100{,}000).
Figure~\ref{fig:ablation_study} summarizes the results. For clarity and space considerations, all effectiveness metrics in the figure are evaluated with FaceNet-512.

\noindent$\bullet~$\textbf{$\lambda_{util}$ (default: 1000).} When $\lambda_{util}$ ranges from 0 to 100, the MSE decreases marginally (from 218 to 182), while $ASR_{id}$ and $ASR_{ctx}$ remain low and $TSR$ remains high.
In contrast, increasing $\lambda_{util}$ from 500 to 5{,}000 leads to a sharp reduction in MSE (from 182 to 10) and a rapid increase in $ASR_{ctx}$, ultimately resulting in complete failure of context protection.
This trend is consistent with the results in Table~\ref{tab:ablation_study}, indicating that context protection is substantially more constrained by perturbation magnitude than identity protection.

\noindent$\bullet~$\textbf{$\lambda_{id}$ (default: 10000).} $ASR_{id}$ increases to 1.43\% when $\lambda_{id}{=}0$. $ASR_{id}$ decreases to 0.80\% when $\lambda_{id}{=}100$. The results indicate that $\lambda_{id}$ provides only a modest impact in identity protection.
That is, the dominant protection effect is driven by guidance from the cloak identity rather than by the explicit identity loss weight.

\noindent$\bullet~$\textbf{$\lambda_{cloak}$ (default: 10000).} $\lambda_{cloak}$ plays an important role in \name’s performance.
As $\lambda_{cloak}$ increases from 0 to 100{,}000, $ASR_{id}$ steadily decreases, while $TSR$ consistently increases.
Across this range, the utility MSE remains essentially unchanged, and $ASR_{ctx}$ shows only a slight increase once $\lambda_{cloak}$ exceeds its default value of 10{,}000.
These results indicate that $\lambda_{cloak}$ is central to achieving both identity protection and traceability.

\noindent$\bullet~$\textbf{$\lambda_{ctx}$ (default: 0.1).} The context deviation magnitude is substantially larger than that of other components, making protection performance highly sensitive to $\lambda_{ctx}$.
When $\lambda_{ctx}$ is set to 0 or 0.05, $ASR_{ctx}$ remains high, with the original attack success rate exceeding 90\%.
In contrast, increasing $\lambda_{ctx}$ to 0.1 (the default setting) causes $ASR_{ctx}$ to drop sharply to 1.90\%.
As $\lambda_{ctx}$ is further increased to 0.5, $ASR_{ctx}$ consistently drops to zero.
However, this improvement comes at a significant cost: the MSE surges from 18 to 178, while identity protection and tracing performance deteriorate markedly as context deviation begins to dominate the perturbation.

In summary, minimizing perturbation while maximizing identity and context protection are inherently conflicting objectives.
The four loss components in Eq.~\ref{eqa:pgd_simswap_all} are essential and complementary in navigating this trade-off.
Our default weight configuration achieves a balanced level of protection across all objectives.
In practice, defenders can flexibly adjust these weights to accommodate different priorities and deployment requirements.

\section{Robustness}\label{sec:robustness}

In real-world deployment, an image owner (Alice) may apply AI-based image beautification before sharing an image, while the image-sharing pipeline itself often involves common pre-processing such as compression, cropping, or platform logo insertion. These transformations may degrade the protection performance of \name. In addition, adversaries (Bob and Charlie) may employ pre-processing techniques or adaptive attacks to attenuate or remove the injected perturbations. To evaluate the robustness of \name\ under such practical conditions, we conduct experiments using SimSwap as a representative face-swapping model.

In this section, we evaluate the robustness of \name\ under six common image transformations: Gaussian noise addition (mean 0, standard deviation 0.1), WebP compression with a quality factor of 80, cropping 20 pixels from each image border, insertion of an ACM CCS logo at the bottom-right corner, and brightness adjustment by ±25\%. We adopt two cloak image selection strategies, baseline and prioritizing-protection, as described in Section~\ref{key:strategy}.

\subsection{Protection Robustness}
\label{subsec:protection_robustness}

\noindent\textbf{AI-based Beautification Robustness} We evaluate \name’s robustness to AI-based beautification using two commonly used services: the AILab Tools Smart Beauty API~\cite{ailab} and the Tencent Cloud BeautifyPic API~\cite{tencentcloud}. We apply beautification to the original images and then protect the resulting images with \name. As shown in Table~\ref{tab:robustness_experiments} (beautification robustness), beautification introduces only a negligible change in \name's protection effectiveness. These results suggest that \name\ is compatible with practical beautification pipelines.

\begin{table}[t]
  \centering
  \small
  \captionsetup{skip=3 pt}
  \caption{Robustness of \name\ against AI-beautifiers, image transformations, and adaptive attacks.}
  \label{tab:robustness_experiments}
  \setlength\tabcolsep{1.25 pt}
  \begin{tabular}{c | c  c  c  c  c  c}
    \hline
    \multirow{2}*{} & \multicolumn{3}{c|}{FaceNet-512} & \multicolumn{3}{c}{Megvii Face++}\\
    \cline{2-7} 
    & $ASR_{id}{\downarrow}$ & $TSR{\uparrow}$ & \multicolumn{1}{c|}{$ASR_{ctx}{\downarrow}$} & $ASR_{id}{\downarrow}$ & $TSR{\uparrow}$ & $ASR_{ctx}{\downarrow}$ \\
    \hline
    Baseline & 0.73 & 94.93 & 1.90 & 0.30 & 97.97 & 3.20 \\
    \hline
    \hline
    Services & \multicolumn{6}{c}{AI-based beautification robustness} \\
    \hline
    AILab Tools   & 0.77 & 94.76 & 2.35 & 0.23 & 97.83 & 3.37 \\
    Tencent Cloud & 0.70 & 94.60 & 2.20 & 0.10 & 96.80 & 3.40 \\
    \hline
    \hline
    Operation & \multicolumn{6}{c}{Image transformations robustness} \\
    \hline
    Compress & 0.63 & 94.77 & 1.90 & 0.33 & 98.60 & 3.83 \\
    Crop     & 1.13 & 93.47 & 4.60 & 0.20 & 97.20 & 5.20 \\
    Logo     & 0.73 & 94.03 & 2.10 & 0.10 & 97.97 & 3.10 \\
    \hline
    \hline
    Scenario & \multicolumn{6}{c}{Adaptive attack robustness} \\
    \hline
    Adaptive 1 & 34.07 & 54.43 & 2.80 & 24.70 & 52.53 & 3.00 \\
    Adaptive 2 & 36.50 & 48.67 & 0.00 & 21.63 & 45.50 & 0.00 \\
    \hline
    \hline
    Training Samples & \multicolumn{6}{c}{Adaptive denoiser robustness} \\
    \hline
    10  & 0.95 & 95.50 & 4.00  & 0.40 & 98.10 & 7.10  \\
    50  & 1.30 & 95.10 & 56.70 & 0.50 & 98.85 & 77.20 \\
    100 & 1.60 & 95.20 & 97.10 & 0.50 & 98.30 & 99.20 \\
    500 & 2.10 & 94.25 & 94.20 & 0.25 & 97.35 & 98.95 \\
    \hline
  \end{tabular}
  \vspace{-5 mm}
\end{table}

\noindent\textbf{Image Transformations Robustness.} In the real-world deployment, \name\ can be integrated into online image-sharing platforms to automatically protect uploaded images using platform-generated cloaks. During the sharing process, images may undergo pre-processing operations like compression, cropping, or logo insertion. In a practical setting, we evaluate \name's protection effectiveness when such operations are applied \textit{before} protection. As shown in Table~\ref{tab:robustness_experiments} (Image transformation robustness), applying \name\ after these pre-processing steps does not noticeably degrade protection performance.

\noindent\textbf{Adaptive Attack Robustness.} An adaptive attacker with knowledge of \name’s perturbation generation pipeline may attempt to estimate and remove the injected noise, as considered in prior work~\cite{fu2025diffpad, zhou2020manifold, deb2023faceguard}.
To evaluate this threat, we consider a strong adaptive attacker with access to the protected image ($\tilde{\mathbf{x}}$) and the exact cloak image ($\mathbf{x}_c$). The attacker aims to approximate the perturbation $\Delta\mathbf{x}$ and recover the original image by computing $\mathbf{x} \approx \tilde{\mathbf{x}} - \Delta\mathbf{x}$.
To achieve this, the attacker first identifies a seed image $\mathbf{x}'$ that is similar to $\mathbf{x}$. The attacker then invokes \name\ on $\mathbf{x}'$ with the same cloak image $\mathbf{x}_c$ to obtain an estimated perturbation $\Delta\mathbf{x}'$. The reconstructed image is obtained as $\tilde{\mathbf{x}} - \Delta\mathbf{x}'$. 
We consider two adaptive strategies for selecting $\mathbf{x}'$: (A1) using an image from the same dataset as $\mathbf{x}$, and (A2) directly using the protected image $\tilde{\mathbf{x}}$.
As shown in Table~\ref{tab:robustness_experiments} (Adaptive attack robustness), both adaptive attacks result in a moderate degradation of protection and tracing effectiveness. For A1, $ASR_{id}$ increases to 34.07\% and 24.70\% when evaluated with FaceNet-512 and Face++, respectively. Nevertheless, context protection remains effective, with $ASR_{ctx}$ values of 2.80\% and 3.00\%. A similar trend is observed for A2, which yields a slightly higher $ASR_{id}$ while achieving improved context protection. We further visualize and compare the perturbations generated by both adaptive attack strategies in Appendix~\ref{app:extra_robustness}.

\noindent\textbf{Adaptive Denoiser Robustness.}
An adaptive attacker may attempt to collect pairs of clean and protected images to train a denoiser to remove \name's perturbation from new protected images. To mimic this adaptive denoiser attack, we train a residual U-Net~\cite{ronneberger2015u, ho2020denoising} with $n$ random pairs of known clean and protected images, and test it with 2,000 new protected images.
As shown in Table~\ref{tab:robustness_experiments} (Adaptive denoiser robustness): (1) with the denoised images, the identity protection and tracing are still highly effective (low $ASR_{id}$ and high $TSR$). (2) $ASR_{id}$ remains low and $TSR$ remains high even when the training set size increases to 500. (3) Context protection performance decreases: $ASR_{ctx}$ started low but significantly increases from 4.00\% to 97.10\% (FaceNet-512) when the training set size increases to 100 (pairs). Detailed results with training set sizes ranging from 10 to 500 are provided in Appendix~\ref{app:denoiser}.

\noindent\textbf{Robustness against Pre-processing.} 
Attackers may \textit{pre-process} protected images \textit{before} performing face swapping, attempting to remove the protection. In watermarking literature, there is a standard strategy to expose the model to typical distortions during training to improve robustness of generated watermarks \cite{wu2023sepmark}. Inspired by this, we design a compression-in-the-loop approach to incorporate compression in the 1{,}000-epoch perturbation process: we compress the intermediate image after 335 and 670 epochs, and continue the optimization process using the compressed image.

In Table~\ref{tab:robustness_metric}, we report \name's defense robustness when the protected images are pre-processed before face swapping. We have the following key observations:
\textbf{(1)} Adding Gaussian noise to protected images substantially degrades protection effectiveness. This is expected, as the noise magnitude (0.1) exceeds the perturbation magnitude of \name, making such strong distortions difficult to withstand.
\textbf{(2)} Under image compression and brightness reduction, \name’s context protection degrades, with $ASR_{ctx}$ increasing to over 95\% and 50\%, respectively. However, as shown in Table~\ref{tab:robustness_metric} (context protection samples), even when face-swapped outputs are classified as successful attacks, they exhibit clearly malicious visual artifacts and remain highly noticeable to human eyes. To prove this, we conducted a user study that recruited 101 participants with 505 annotated images for both compression and brightness reduction. With compression, participants were able to identify 66.93\% (baseline) and 65.94\% (prioritizing protection) of the images as malicious. With decreased brightness, participants identified 99.21\% (baseline) and 99.01\% (prioritizing protection) of the images as malicious.
\textbf{(3)} For other pre-processing operations, both $ASR_{id}$ and $ASR_{ctx}$ remain consistently low, indicating robustness of \name\ to these modifications.
\textbf{(4)} Prioritizing protection effectiveness during cloak image selection further improves the robustness of identity protection. (5) The compression-in-the-loop design significantly enhances \name's robustness against pre-processing (please see Appendix~\ref{app:robustness} for details).

\begin{table}[t]
\small
    \captionsetup{skip=0 pt}
    \caption{\name\ robustness against pre-processing. Process: 0 No operation, 1 Gaussian noise, 2 compress, 3 crop, 4 logo, 5 increase brightness, 6 decrease brightness. $F^1$: FaceNet-512, and $F^2$: Megvii Face++.}
    \label{tab:robustness_metric}
    
    \setlength\tabcolsep{5.65 pt}
    \begin{tabular}{c | c  c  c  c | c  c  c  c}
    \hline
    \multirow{3}*{\raisebox{-1.5 ex}{\rotatebox{90}{Process}}} & \multicolumn{4}{c}{Baseline} \vline & \multicolumn{4}{c}{Prioritizing protection}  \\
    \cline{2-9}
    ~ & \multicolumn{2}{c}{$ASR_{id}$} & \multicolumn{2}{c}{$ASR_{ctx}$} \vline & \multicolumn{2}{c}{$ASR_{id}$} & \multicolumn{2}{c}{$ASR_{ctx}$} \\
    \cline{2-9}
    ~ & $F^1{\downarrow}$ & $F^2{\downarrow}$ & $F^1{\downarrow}$ & $F^2{\downarrow}$ & $F^1{\downarrow}$ & $F^2{\downarrow}$ & $F^1{\downarrow}$ & $F^2{\downarrow}$ \\
    \hline
    0 & 0.20  & 0.07  & 3.53  & 4.73  & 0.13  & 0.00  & 4.00  & 6.43  \\
    1 & 27.07 & 13.30 & 0.80  & 1.63  & 19.40 & 10.90 & 1.13  & 2.00  \\
    2 & 5.50  & 1.60  & 95.50 & 98.50 & 2.37  & 1.13  & 96.03 & 98.77 \\
    3 & 0.63  & 0.27  & 10.07 & 13.90 & 0.20  & 0.17  & 12.17 & 15.37 \\
    4 & 0.23  & 0.10  & 2.97  & 4.73  & 0.10  & 0.00  & 3.50  & 5.57  \\
    5 & 0.57  & 0.33  & 7.73  & 10.20 & 0.27  & 0.20  & 8.30  & 10.40 \\
    6 & 0.27  & 0.13  & 52.17 & 59.97 & 0.10  & 0.00  & 53.57 & 65.27 \\
    \hline
    
    \addlinespace[0.3 em]
    \hline
    \addlinespace[0.3 em]
    
    \hline
    \multicolumn{9}{c}{Context protection samples} \\
    \end{tabular}

    \setlength\tabcolsep{2.62 pt}
    \newcommand{\wI}{1.00 em}
    \newcommand{\wN}{4.10 em}
    \newcolumntype{C}[1]{>{\centering\arraybackslash}m{#1}}
    
    \begin{tabular}{C{\wI} C{\wN} C{\wN} C{\wN} C{\wN} C{\wN} C{\wN}}
      \toprule
      2 &
      \raisebox{-0.6 ex}{\includegraphics[width=\linewidth]{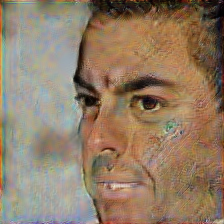}} &
      \raisebox{-0.6 ex}{\includegraphics[width=\linewidth]{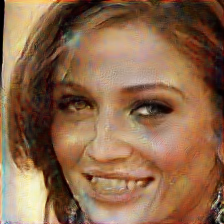}} &
      \raisebox{-0.6 ex}{\includegraphics[width=\linewidth]{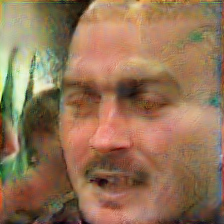}} &
      \raisebox{-0.6 ex}{\includegraphics[width=\linewidth]{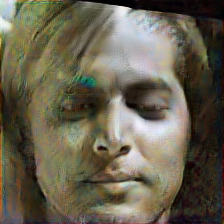}} &
      \raisebox{-0.6 ex}{\includegraphics[width=\linewidth]{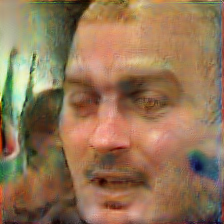}} &
      \raisebox{-0.6 ex}{\includegraphics[width=\linewidth]{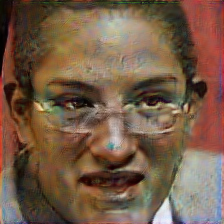}} \\
      \midrule
      6 &
      \raisebox{-0.6 ex}{\includegraphics[width=\linewidth]{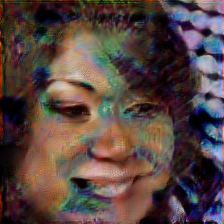}} &
      \raisebox{-0.6 ex}{\includegraphics[width=\linewidth]{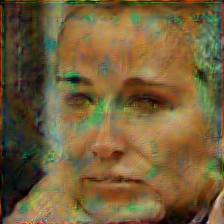}} &
      \raisebox{-0.6 ex}{\includegraphics[width=\linewidth]{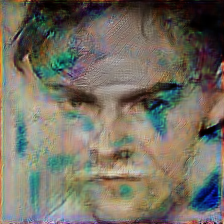}} &
      \raisebox{-0.6 ex}{\includegraphics[width=\linewidth]{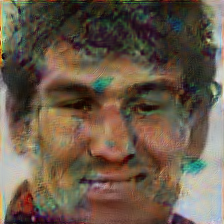}} &
      \raisebox{-0.6 ex}{\includegraphics[width=\linewidth]{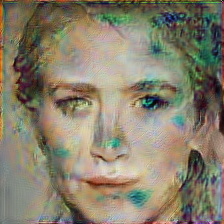}} &
      \raisebox{-0.6 ex}{\includegraphics[width=\linewidth]{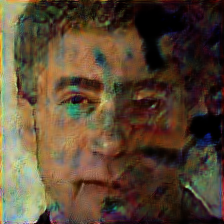}} \\
      \bottomrule
    \end{tabular}

    \setlength\tabcolsep{4.22 pt}
    \begin{tabular}{c | c  c  c | c  c  c}
      \multicolumn{7}{c}{User Study: ``failed'' means the user did not label the image as malicious.} \\\hline
      \multirow{2}*{Process} & \multicolumn{3}{c}{Baseline} \vline & \multicolumn{3}{c}{Prioritizing protection} \\
      \cline{2-7}
      ~ & Failed &  Identified & $ASR_{ctx}$ & Failed &  Identified & $ASR_{ctx}$ \\
      \hline
      2 & 167 & 338 & 33.07 & 172 & 333 & 34.06 \\
      6 & 4   & 501 & 0.79  & 5   & 500 & 0.99  \\
      \hline
    \end{tabular}
    \vspace{-4 mm}
\end{table}

\subsection{Forensics Robustness}\label{subsec:forensics_robustness}

To prevent defenders from tracing deepfake images, attackers may apply post-processing operations \textit{after} face swapping.
As discussed in Section~\ref{key:strategy}, the tracing success rate ($TSR$) decreases as the identity distance between the cloak and protected image increases, and the minimum cloak distance of 1.04 yields the highest $TSR$. Accordingly, we evaluate forensic tracing robustness under two cloak image selection settings: the baseline strategy and a tracing-prioritized setting with the cloak distance fixed at 1.04.
We compare \name\ with two representative forensic watermarking methods: Artificial Fingerprinting (AF)~\cite{yu2021artificial} and SepMark~\cite{wu2023sepmark}. AF embeds fingerprints into training samples via steganographic techniques, while SepMark employs a single encoder to embed watermarks and two decoders to extract them at different robustness levels.

Both AF and SepMark use the Binary Error Rate (BER) as the evaluation metric. For a fair comparison, we convert BER to identification accuracy using Eq.~(\ref{eqa:ber_to_accuracy}): 
\begin{equation}
\begin{split}
\label{eqa:ber_to_accuracy}
\text{Accuracy} = \sum_{i=0}^{\mathbf{k}} \binom{\mathbf{N}}{i} (\text{BER})^i (1 - \text{BER})^{\mathbf{N}-i}
\end{split}
\end{equation}
where $\mathbf{N}$ denotes the fingerprint length (100 for AF and 128 for SepMark), and $\mathbf{k}$ denotes the maximum number of tolerable bit errors for successful tracing. Following~\cite{mannam2024feasibility}, we adopt a BER threshold of 10\%, corresponding to $\mathbf{k}=10$ for AF and $\mathbf{k}=13$ for SepMark.

From Table \ref{tab:robustness_comparison}, we can observe the following:
\textbf{(1)} Both \name\ and SepMark achieve high tracing accuracy on raw SimSwap outputs without postprocessing, while AF exhibits noticeably lower accuracy.
\textbf{(2)} For \name, Gaussian noise is the only operation that causes a substantial degradation in tracing performance, while other perturbations result in only minor decreases in matching rates.
\textbf{(3)} Gaussian noise also significantly degrades the tracing performance of SepMark, with a much larger impact than that for \name. In addition, other operations, including compression, crop overlay, and brightness adjustment, further reduce SepMark’s tracing effectiveness.
\textbf{(4)} When tracing is explicitly prioritized in cloak selection, \name\ exhibits reduced tracing robustness. This trade-off arises because selecting cloak images closer to Alice in feature space improves tracing under clean conditions, but increases vulnerability to post-processing distortions.

\begin{table}[t]
  \centering
  \small
  \captionsetup{skip=3 pt}
  \caption{Forensic tracing robustness of \name, Artificial Fingerprints (AF) \cite{yu2021artificial}, and SepMark \cite{wu2023sepmark}.}
  \label{tab:robustness_comparison}
  \setlength\tabcolsep{4 pt}
  \begin{tabular}{c | c  c  c  c | c  c | c  c}
    \hline
    \multirow{3}*{\raisebox{-1.5 ex}{\rotatebox{90}{Process}}} & \multicolumn{4}{c}{\name\ (ours)} \vline & \multicolumn{2}{c}{AF} \vline & \multicolumn{2}{c}{SepMark} \\
    \cline{2-9}
    ~ & \multicolumn{2}{c}{Baseline} & \multicolumn{2}{c}{Pr. tracing} \vline & \multirow{2}*{BER$\downarrow$} & \multirow{2}*{ACC$\uparrow$} & \multirow{2}*{BER$\downarrow$} & \multirow{2}*{ACC$\uparrow$} \\ 
    \cline{2-5}
    ~ & $F^1{\uparrow}$ & $F^2{\uparrow}$ & $F^1{\uparrow}$ & $F^2{\uparrow}$ & ~ & ~ & ~ & ~ \\
    \hline
    0 & 82.77 & 88.33 & 81.47 & 85.13 & 12.86 & 25.26 & 7.91  & 86.43 \\
    1 & 70.40 & 64.67 & 67.47 & 58.53 & 40.33 & 0.00  & 31.99 & 0.00  \\
    2 & 82.27 & 87.80 & 81.10 & 84.50 & 38.03 & 0.00  & 23.99 & 0.01  \\
    3 & 80.90 & 86.10 & 79.20 & 83.47 & 20.02 & 0.63  & 15.07 & 7.36  \\
    4 & 82.37 & 87.53 & 80.53 & 84.50 & 20.17 & 0.55  & 8.73  & 77.12 \\
    5 & 81.93 & 87.30 & 80.53 & 83.60 & 16.32 & 5.64  & 10.28 & 55.47 \\
    6 & 81.93 & 88.47 & 81.43 & 85.37 & 13.13 & 22.90 & 7.99  & 85.40 \\
    \hline
  \end{tabular}
  \vspace{-3mm}
\end{table}

We emphasize that forensic tracing is a secondary objective of \name, whereas AF and SepMark are explicitly designed for fingerprint-based forensic tracing.
Despite this difference in design objectives, \name\ achieves higher tracing accuracy than both methods under most post-processing operations.

\section{Discussions and Limitations}\label{sec:limitations}

\noindent\textbf{Decoupled Identity and Context Extractors.}
In the white-box protection (Sections~\ref{subsec:gan_exp} and \ref{subsec:diffusion_model}), \name~relies on decoupled identity and context extractors to guide the perturbation. Such decoupled extractors are widely adopted in face-swapping pipelines, including autoencoder-, GAN-, and diffusion-based models. Adapting to this structure allows \name~to cover a broad range of existing face-swapping attacks. 
Furthermore, as shown in Section~\ref{subsec:black_box}, \name\ remains effective against unseen face-swapping models in black-box settings.

\noindent\textbf{Identity, Context Protection and Tracing under Different Threat Models.}
\name~protects image identity and context, and provides forensic tracing in a white-box setting. In the black-box setting, while achieving state-of-the-art identity protection performance, its context protection and tracing capabilities drop significantly. This limitation is caused by the fact that context extractors are more architecturally diverse than identity extractors across face-swapping models. Hence, \name\ trained for one specific context extractor is less transferable and generalizable to heterogeneous context extractors adopted by other attacks.

\noindent\textbf{Robustness against Adaptive Attacks.}
In Section~\ref{subsec:protection_robustness}, \name's protection performance degrades under adaptive attacks.
We note that the adaptive attack requires access to the raw cloak image, which is considered impractical in the real world, as assumed in Section~\ref{sec:threat_model}. Therefore, this setting represents the upper bound of the attacker's capability rather than a practical attack scenario.

\noindent\textbf{Robustness against Adaptive Denoiser.}
We acknowledge that \name's context protection significantly degrades with adaptive denoising. Nevertheless, it still maintains strong identity protection and tracing performance even when the denoiser is trained with 500 pairs of clean/protected images. Moreover, under our recommended deployment setting, where \name~is integrated into online image-sharing platforms, it would be highly suspicious for a user/attacker to upload images from many different identities to collect paired clean and protected images as training data. This practical barrier prevents the attacker from obtaining sufficient image pairs to train a high-performance adaptive denoiser.

\section{Conclusion}\label{sec:conclusion}

In this paper, we present \name, a proactive defense against face-swapping deepfakes that simultaneously protects users’ images from identity-stealing and context-stealing attacks.
By leveraging a novel ``cloak'' concept, \name\ prevents the swapped face from resembling the victim's identity while enabling forensic traceability.
Through extensive experiments, we demonstrate the utility of \name\ and its effectiveness in identity protection, forensic tracing, context protection, as well as its robustness under various practical conditions.

\section*{Acknowledgments}\label{app:acknowledgements}

Liangqin Ren, Fengjun Li, and Bo Luo were supported in part by US NSF IIS-2014552, DGE-1565570, NSF ACCESS program \cite{10.1145/3569951.3597559}, and the Ripple University Blockchain Research Initiative. The authors would like to thank the anonymous reviewers and shepherd for their valuable comments and suggestions.

\bibliographystyle{ACM-Reference-Format}
\bibliography{reference}

@misc{faceplusplus,
  author = {Megvii},
  title = {Face++},
  howpublished = {\url{https://console.faceplusplus.com/documents/5679308}},
  year = {2026}
}

@inproceedings{cui2023face,
  title = {Face transformer: Towards high fidelity and accurate face swapping},
  author = {Cui, Kaiwen and Wu, Rongliang and Zhan, Fangneng and Lu, Shijian},
  booktitle = {CVPRW},
  year = {2023}
}

@article{li2019faceshifter,
  title={FaceShifter: Towards High Fidelity and Occlusion Aware Face Swapping},
  author = {Li, L. and Bao, J. and Yang, H. and Chen, D. and Wen, F.},
  journal = {arXiv:1912.13457},
  year = {2019}
}

@article{hsieh2014combining,
  title={Combining digital watermarking and fingerprinting techniques to identify copyrights for color images},
  author={Hsieh, Shang-Lin and Chen, Chun-Che and Shen, Wen-Shan},
  journal={The Scientific World Journal},
  volume={2014},
  number={1},
  pages={454867},
  year={2014}
}

@inproceedings{liu2023bifpro,
  title = {BiFPro: A Bidirectional Facial-data Protection Framework against DeepFake},
  author = {Liu, Honggu and Li, Xiaodan and Zhou, Wenbo and Fang, Han and Bestagini, P. and Zhang, Weiming and Chen, Y. and Tubaro, S. and Yu, Nenghai and He, Yuan and Xue, Hui},
  booktitle = {ACM MM},
  year = {2023}
}

@inproceedings{wang2024lampmark,
  title = {LampMark: Proactive Deepfake Detection via Training-Free Landmark Perceptual Watermarks},
  author = {Wang, Tianyi and Huang, Mengxiao and Cheng, Harry and Zhang, Xiao and Shen, Zhiqi},
  booktitle = {ACM MM},
  year = {2024}
}

@inproceedings{tan2024frequency,
  title = {Frequency-Aware Deepfake Detection: Improving Generalizability through Frequency Space Domain Learning},
  author = {Tan, Chuangchuang and Zhao, Yao and Wei, Shikui and Gu, Guanghua and Liu, Ping and Wei, Yunchao},
  booktitle = {AAAI},
  year = {2024}
}

@article{kingma2019introduction,
  title = {An introduction to variational autoencoders},
  author={Diederik, P Kingma and Max, Welling},
  journal = {Foundations and Trends in Machine Learning},
  volume = {12},
  number = {},
  pages = {307--392},
  year = {2019}
}

@inproceedings{thakkar2024deepfakes,
  title = {From Deepfakes to Digital Truths: The Role of Watermarking in AI-Generated Image Verification},
  author = {Thakkar, J. J. and Kaur, A.},
  booktitle = {IEEE TSP},
  year = {2024}
}

@article{li2019hiding,
  title = {Hiding faces in plain sight: Disrupting ai face synthesis with adversarial perturbations},
  author = {Li, Yuezun and Yang, Xin and Wu, Baoyuan and Lyu, Siwei},
  journal = {arXiv:1906.09288},
  year = {2019}
}

@inproceedings{liu2024disrupting,
  title = {Disrupting diffusion: Token-level attention erasure attack against diffusion-based customization},
  author = {Liu, Yisu and An, Jinyang and Zhang, Wanqian and Wu, Dayan and Gu, Jingzi and Lin, Zheng and Wang, Weiping},
  booktitle = {ACM MM},
  year = {2024}
}

@article{li2021obstructing,
  title = {Obstructing deepfakes by disrupting face detection and facial landmarks extraction},
  author = {Li, Yuezun and Lyu, Siwei},
  journal = {Deep Learning-Based Face Analytics},
  pages = {247--267},
  volume = {},
  number = {},
  year = {2021}
}

@article{tang2024feature,
  title = {Feature Extraction Matters More: An Effective and Efficient Universal Deepfake Disruptor},
  author = {Tang, Long and Ye, Dengpan and Lu, Zhenhao and Zhang, Yunming and Chen, Chuanxi},
  journal = {ACM TOMM},
  volume = {},
  number = {},
  pages = {},
  year = {2024}
}

@article{zhang2023disrupting,
  title={Disrupting deepfakes via union-saliency adversarial attack},
  author={Zhang, Guisheng and Gao, Mingliang and Li, Qilei and Zhai, Wenzhe and Zou, Guofeng and Jeon, Gwanggil},
  journal={IEEE TCE},
  volume={70},
  number={1},
  pages={2018--2026},
  year={2023},
}

@article{segalis2020ogan,
  title = {OGAN: Disrupting deepfakes with an adversarial attack that survives training},
  author = {Segalis, Eran and Galili, Eran},
  journal = {arXiv:2006.12247},
  year = {2020}
}

@inproceedings{asnani2022proactive,
  title = {Proactive image manipulation detection},
  author = {Asnani, Vishal and Yin, Xi and Hassner, Tal and Liu, Sijia and Liu, Xiaoming},
  booktitle = {CVPR},
  year = {2022}
}

@article{lan2024facial,
  title = {Facial Features Matter: a Dynamic Watermark based Proactive Deepfake Detection Approach},
  author = {Lan, Shulin and Liu, Kanlin and Zhao, Yazhou and Yang, Chen and Wang, Yingchao and Yao, Xingshan and Zhu, Liehuang},
  journal = {arXiv:2411.14798},
  year = {2024}
}

@article{yang2021faceguard,
  title = {Faceguard: Proactive deepfake detection},
  author = {Yang, Yuankun and Liang, Chenyue and He, Hongyu and Cao, Xiaoyu and Gong, Neil Zhenqiang},
  journal = {arXiv:2109.05673},
  year = {2021}
}

@article{zhang2021restore,
  title = {Restore DeepFakes video frames via identifying individual motion styles},
  author = {Zhang, Haichao and Lu, Zhe-Ming and Luo, Hao and Feng, Ya-Pei},
  journal = {Electronics Letters},
  volume = {57},
  number = {4},
  pages = {183--186},
  year = {2021}
}

@inproceedings{niudiffusion,
  title={A diffusion-based approach for restoring face-swapped images},
  author={Niu, Yuanchen and Li, Yuanman and Zhang, Guijia and Li, Xia},
  booktitle={APSIPA ASC},
  year={2024},
}

@inproceedings{chen2021magdr,
  title = {Magdr: Mask-guided detection and reconstruction for defending deepfakes},
  author = {Chen, Zhikai and Xie, Lingxi and Pang, Shanmin and He, Yong and Zhang, Bo},
  booktitle = {CVPR},
  year = {2021}
}

@article{temir2020deepfake,
  title = {Deepfake: new era in the age of disinformation \& end of reliable journalism},
  author = {Temir, Erkam},
  journal = {Sel{\c{c}}uk {\.I}leti{\c{s}}im},
  volume = {13},
  number = {2},
  pages = {1009--1024},
  year = {2020}
}

@article{jeong2024faceshield,
  title = {FaceShield: Defending Facial Image against Deepfake Threats},
  author = {Jeong, Jaehwan and In, Sumin and Kim, Sieun and Shin, Hannie and Jeong, Jongheon and Yoon, Sang Ho and Chung, Jaewook and Kim, Sangpil},
  journal = {arXiv:2412.09921},
  year = {2024}
}

@article{zhu2024hiding,
  title = {Hiding Faces in Plain Sight: Defending DeepFakes by Disrupting Face Detection},
  author = {Zhu, Delong and Li, Yuezun and Wu, Baoyuan and Zhou, Jiaran and Wang, Zhibo and Lyu, Siwei},
  journal = {arXiv:2412.01101},
  year = {2024}
}

@article{li2024landmarkbreaker,
  title = {LandmarkBreaker: A proactive method to obstruct DeepFakes via disrupting facial landmark extraction},
  author = {Li, Yuezun and Sun, Pu and Qi, Honggang and Lyu, Siwei},
  journal = {CVIU},
  volume = {240},
  pages = {103935},
  number = {},
  year = {2024}
}

@article{shim2023leat,
  title = {Leat: Towards robust deepfake disruption in real-world scenarios via latent ensemble attack},
  author = {Shim, J. and Yoon, H.},
  journal = {arXiv:2307.01520},
  year = {2023}
}

@inproceedings{guan2022defending,
  title = {Defending against deepfakes with ensemble adversarial perturbation},
  author = {Guan, Weinan and He, Ziwen and Wang, Wei and Dong, Jing and Peng, Bo},
  booktitle = {ICPR},
  year = {2022}
}

@inproceedings{dong2021visually,
  title = {Visually maintained image disturbance against deepfake face swapping},
  author = {Dong, Junhao and Xie, Xiaohua},
  booktitle = {ICME},
  year = {2021}
}

@article{zhai2023defending,
  title = {Defending fake via warning: Universal proactive defense against face manipulation},
  author = {Zhai, Rui and Ni, Rongrong and Chen, Yu and Yu, Yang and Zhao, Yao},
  journal = {IEEE SPL},
  volume={30},
  pages={1072--1076},
  number = {},
  year = {2023}
}

@inproceedings{dai2024idguard,
  title = {IDGuard: Robust, General, Identity-Centric POI Proactive Defense Against Face Editing Abuse},
  author = {Dai, Yunshu and Fei, Jianwei and Huang, Fangjun},
  booktitle = {CVPR},
  year = {2024}
}

@inproceedings{lin2022source,
  title = {Source-ID-Tracker: Source Face Identity Protection in Face Swapping},
  author = {Lin, Yuzhen and Chen, Han and Maiorana, Emanuele and Campisi, Patrizio and Li, Bin},
  booktitle = {ICME},
  year = {2022}
}

@article{wu2024watermarks,
  title = {Are Watermarks Bugs for Deepfake Detectors? Rethinking Proactive Forensics},
  author = {Wu, Xiaoshuai and Liao, Xin and Ou, Bo and Liu, Yuling and Qin, Zheng},
  journal = {arXiv:2404.17867},
  year = {2024}
}

@inproceedings{hu2023draw,
  title = {Draw: Defending camera-shooted raw against image manipulation},
  author = {Hu, Xiaoxiao and Ying, Qichao and Qian, Zhenxing and Li, Sheng and Zhang, Xinpeng},
  booktitle = {ICCV},
  year = {2023}
}

@article{li2024dual,
  title = {Dual Protection for Image Privacy and Copyright via Traceable Adversarial Examples},
  author = {Li, Ming and Yang, Zhaoli and Wang, Tao and Zhang, Yushu and Wen, Wenying},
  journal = {IEEE TCSVT},
  volume={34},
  number={12},
  pages={13401--13412},
  year = {2024}
}

@article{sun2023faketracer,
  title={Faketracer: Catching face-swap deepfakes via implanting traces in training},
  author={Sun, Pu and Qi, Honggang and Li, Yuezun and Lyu, Siwei},
  journal={TETC},
  volume={13},
  number={},
  pages={134--146},
  year={2024},
}

@article{guarnera2020fighting,
  title={Fighting Deepfake by Exposing the Convolutional Traces on Images}, 
  author={Guarnera, Luca and Giudice, Oliver and Battiato, Sebastiano},
  journal={IEEE Access},
  volume={8},
  number={},
  pages={},
  year = {2020}
}

@article{yu2024dfrec,
  title = {DFREC: DeepFake Identity Recovery Based on Identity-aware Masked Autoencoder},
  author={Yu, Peipeng and Gao, Hui and Fei, Jianwei and Huang, Zhitao and Xia, Zhihua and Chang, Chip-Hong},
  journal = {TIFS},
  year = {2026},
  volume={21},
  number = {},
  pages={4319-4332},
}

@inproceedings{ai2023deepreversion,
  title = {DeepReversion: Reversely Inferring the Original Face from the DeepFake Face},
  author = {Ai, Jiaxin and Wang, Zhongyuan and Huang, Baojin and Han, Zhen},
  booktitle = {IJCNN},
  year = {2023}
}

@inproceedings{ai2023deepfake,
  title = {Deepfake Face Provenance for Proactive Forensics},
  author = {Ai, Jiaxin and Wang, Zhongyuan and Huang, Baojin and Han, Zhen and Zou, Qin},
  booktitle = {ICIP},
  year = {2023}
}

@article{shen2024hiding,
  title = {Hiding Face Into Background: A Proactive Countermeasure Against Malicious Face Swapping},
  author = {Shen, Xiaofeng and Yao, Heng and Tan, Shunquan and Qin, Chuan},
  journal = {IEEE TII},
  volume = {},
  number = {},
  pages = {},
  year = {2024}
}

@inproceedings{beuve2023waterlo,
  title = {Waterlo: Protect images from deepfakes using localized semi-fragile watermark},
  author = {Beuve, Nicolas and Hamidouche, Wassim and D{\'e}forges, Olivier},
  booktitle = {ICCVW},
  year = {2023}
}

@inproceedings{wang2020deepsonar,
  title = {Deepsonar: Towards effective and robust detection of ai-synthesized fake voices},
  author = {Wang, Run and Juefei-Xu, Felix and Huang, Yihao and Guo, Qing and Xie, Xiaofei and Ma, Lei and Liu, Yang},
  booktitle = {ACM MM},
  year = {2020}
}

@inproceedings{barni2020cnn,
  title = {CNN detection of GAN-generated face images based on cross-band co-occurrences analysis},
  author = {Barni, Mauro and Kallas, Kassem and Nowroozi, Ehsan and Tondi, Benedetta},
  booktitle = {WIFS},
  year = {2020}
}

@inproceedings{liu2021spatial,
  title = {Spatial-phase shallow learning: rethinking face forgery detection in frequency domain},
  author = {Liu, Honggu and Li, Xiaodan and Zhou, Wenbo and Chen, Yuefeng and He, Yuan and Xue, Hui and Zhang, Weiming and Yu, Nenghai},
  booktitle = {CVPR},
  year = {2021}
}

@inproceedings{huang2020fakepolisher,
  title = {Fakepolisher: Making deepfakes more detection-evasive by shallow reconstruction},
  author={Huang, Yihao and Juefei-Xu, Felix and Wang, Run and Guo, Qing and Ma, Lei and Xie, Xiaofei and Li, Jianwen and Miao, Weikai and Liu, Yang and Pu, Geguang},
  booktitle = {ACM MM},
  year = {2020}
}

@article{huang2022fakelocator,
  title = {Fakelocator: Robust localization of gan-based face manipulations},
  author = {Huang, Y. and Juefei-Xu, F. and Guo, Q. and Liu, Yang and Pu, G.},
  journal = {IEEE TIFS},
  volume = {17},
  year = {2022}
}

@inproceedings{hu2021exposing,
  title = {Exposing GAN-generated faces using inconsistent corneal specular highlights},
  author = {Hu, Shu and Li, Yuezun and Lyu, Siwei},
  booktitle = {ICASSP},
  year = {2021}
}

@inproceedings{chai2020makes,
  title = {What makes fake images detectable? understanding properties that generalize},
  author = {Chai, Lucy and Bau, David and Lim, Ser-Nam and Isola, Phillip},
  booktitle = {ECCV},
  year = {2020}
}

@inproceedings{carlini2020evading,
  title = {Evading deepfake-image detectors with white-and black-box attacks},
  author = {Carlini, Nicholas and Farid, Hany},
  booktitle = {CVPR Workshops},
  year = {2020}
}

@article{li2020identification,
  title = {Identification of deep network generated images using disparities in color components},
  author = {Li, Haodong and Li, Bin and Tan, Shunquan and Huang, Jiwu},
  journal = {Signal Processing},
  volume = {174},
  pages = {107616},
  number = {},
  year = {2020}
}

@article{nataraj2019detecting,
  title = {Detecting GAN generated fake images using co-occurrence matrices},
  author = {Nataraj, Lakshmanan and Mohammed, Tajuddin Manhar and Chandrasekaran, Shivkumar and Flenner, Arjuna and Bappy, Jawadul H and Roy-Chowdhury, Amit K and Manjunath, BS},
  journal = {arXiv:1903.06836},
  year = {2019}
}

@inproceedings{frank2020leveraging,
  title = {Leveraging frequency analysis for deep fake image recognition},
  author = {Frank, Joel and Eisenhofer, Thorsten and Sch{\"o}nherr, Lea and Fischer, Asja and Kolossa, Dorothea and Holz, Thorsten},
  booktitle = {ICML},
  year = {2020}
}

@inproceedings{zhang2018unreasonable,
  title = {The unreasonable effectiveness of deep features as a perceptual metric},
  author = {Zhang, R. and Isola, P. and Efros, A. A and Shechtman, E. and Wang, O.},
  booktitle = {CVPR},
  year = {2018}
}

@inproceedings{li2024adversarial,
  title = {The Adversarial AI-Art: Understanding, Generation, Detection, and Benchmarking},
  author = {Li, Yuying and Liu, Zeyan and Zhao, Junyi and Ren, Liangqin and Li, Fengjun and Luo, Jiebo and Luo, Bo},
  booktitle = {ESORICS},
  year = {2024}
}

@inproceedings{yu2021artificial,
  title = {Artificial fingerprinting for generative models: Rooting deepfake attribution in training data},
  author = {Yu, N. and Skripniuk, V. and Abdelnabi, S. and Fritz, M.},
  booktitle = {ICCV},
  year = {2021}
}

@article{nguyen2022deep,
  title = {Deep learning for deepfakes creation and detection: A survey},
  author={Nguyen, Thanh Thi and Nguyen, Quoc Viet Hung and Nguyen, Dung Tien and Nguyen, Duc Thanh and Huynh-The, Thien and Nahavandi, Saeid and Nguyen, Thanh Tam and Pham, Quoc-Viet and Nguyen, Cuong M},
  journal = {CVIU},
  volume = {223},
  pages = {103525},
  number = {},
  year = {2022}
}

@article{lee2019deepfake,
  title = {Deepfake Salvador Dal{\'\i} takes selfies with museum visitors},
  author = {Lee, Dami},
  journal = {The Verge},
  volume = {10},
  number = {},
  pages = {},
  year = {2019}
}

@article{9115874,
  author = {Verdoliva, Luisa},
  title = {Media Forensics and DeepFakes: An Overview},
  journal = {IEEE JSTSP},
  year = {2020},
  volume = {14},
  number = {5},
  pages = {910--932},
}

@article{itzkoff2016rogue,
  title = {How ‘Rogue One’ brought back familiar faces},
  author = {Itzkoff, Dave},
  journal = {New York Times},
  volume = {27},
  number = {},
  pages = {},
  year = {2016}
}

@article{cole2017ai,
  title = {AI-assisted fake porn is here and we’re all fucked},
  author = {Cole, Samantha},
  journal = {Motherboard Tech by Vice},
  year = {2017},
  volume = {},
  number = {},
  pages = {},
}

@article{madry2017towards,
  title = {Towards deep learning models resistant to adversarial attacks},
  author = {Madry, Aleksander and Makelov, Aleksandar and Schmidt, Ludwig and Tsipras, Dimitris and Vladu, Adrian},
  journal = {arXiv:1706.06083},
  year = {2017}
}

@article{goodfellow2014explaining,
  title = {Explaining and harnessing adversarial examples},
  author = {Goodfellow, Ian J and Shlens, Jonathon and Szegedy, Christian},
  journal = {arXiv:1412.6572},
  year = {2014}
}

@article{kurakin2016adversarial,
  title = {Adversarial machine learning at scale},
  author = {Kurakin, Alexey and Goodfellow, Ian and Bengio, Samy},
  journal = {arXiv:1611.01236},
  year = {2016}
}

@inproceedings{karras2019style,
  title = {A style-based generator architecture for generative adversarial networks},
  author = {Karras, Tero and Laine, Samuli and Aila, Timo},
  booktitle = {CVPR},
  year = {2019}
}

@inproceedings{chen2020simswap,
  title = {Simswap: An efficient framework for high fidelity face swapping},
  author = {Chen, Renwang and Chen, Xuanhong and Ni, Bingbing and Ge, Yanhao},
  booktitle = {ACM MM},
  year = {2020}
}

@article{mirza2014conditional,
  title = {Conditional generative adversarial nets},
  author = {Mirza, Mehdi and Osindero, Simon},
  journal = {arXiv:1411.1784},
  year = {2014}
}

@inproceedings{8638330,
  author = {Matern, Falko and Riess, Christian and Stamminger, Marc},
  title = {Exploiting Visual Artifacts to Expose Deepfakes and Face Manipulations},
  booktitle = {WACV Workshops},
  year = {2019}
}

@inproceedings{zhao2021multi,
  title = {Multi-attentional deepfake detection},
  author = {Zhao, Hanqing and Zhou, Wenbo and Chen, Dongdong and Wei, Tianyi and Zhang, Weiming and Yu, Nenghai},
  booktitle = {CVPR},
  year = {2021}
}

@inproceedings{huang2021initiative,
  title = {Initiative defense against facial manipulation},
  author = {Huang, Qidong and Zhang, Jie and Zhou, Wenbo and Zhang, Weiming and Yu, Nenghai},
  booktitle = {AAAI},
  year = {2021}
}

@inproceedings{wu2023sepmark,
  title = {Sepmark: Deep separable watermarking for unified source tracing and deepfake detection},
  author = {Wu, Xiaoshuai and Liao, Xin and Ou, Bo},
  booktitle = {ACM MM},
  year = {2023}
}

@article{zhang2019adversarial,
  title = {Adversarial examples: Opportunities and challenges},
  author = {Zhang, Jiliang and Li, Chen},
  journal = {TNNLS},
  volume = {31},
  number = {7},
  pages = {2578--2593},
  year = {2019}
}

@inproceedings{li2023unganable,
  title = {{UnGANable}: Defending Against {GAN-based} Face Manipulation},
  author = {Li, Zheng and Yu, Ning and Salem, Ahmed and Backes, Michael and Fritz, Mario and Zhang, Yang},
  booktitle = {USENIX Security},
  year = {2023}
}

@inproceedings{schroff2015facenet,
  title = {Facenet: A unified embedding for face recognition and clustering},
  author = {Schroff, Florian and Kalenichenko, Dmitry and Philbin, James},
  booktitle = {CVPR},
  year = {2015}
}

@article{wang2004image,
  title = {Image quality assessment: from error visibility to structural similarity},
  author = {Wang, Zhou and Bovik, Alan C and Sheikh, Hamid R and Simoncelli, Eero P},
  journal = {TIP},
  year = {2004}
}

@article{wang2022anti,
  title = {Anti-forgery: Towards a stealthy and robust deepfake disruption attack via adversarial perceptual-aware perturbations},
  author = {Wang, Run and Huang, Ziheng and Chen, Zhikai and Liu, Li and Chen, Jing and Wang, Lina},
  journal = {IJCAI},
  year = {2022},
  volume = {},
  number = {},
  pages = {},
}

@inproceedings{cao2018vggface2,
  title = {VGGFace2: A dataset for recognising faces across pose and age},
  author = {Cao, Qiong and Shen, Li and Xie, Weidi and Parkhi, Omkar M and Zisserman, Andrew},
  booktitle = {FG},
  year = {2018}
}

@article{serengil2024lightface,
  title = {A Benchmark of Facial Recognition Pipelines and Co-Usability Performances of Modules},
  author={Serengil, Sefik and {\"O}zp{\i}nar, Alper},
  journal = {JIT},
  volume = {17},
  number = {2},
  pages = {95--107},
  year = {2024}
}

@inproceedings{serengil2020lightface,
  title={Lightface: A hybrid deep face recognition framework},
  author={Serengil, Sefik Ilkin and Ozpinar, Alper},
  booktitle = {ASYU},
  pages = {23--27},
  year = {2020}
}

@article{mannam2024feasibility,
  title = {On Feasibility of Transferring Watermarks from Training Data to GAN-Generated Fingerprint Images},
  author = {Mannam, Venkata Srinath and Makrushin, Andrey and Dittmann, Jana},
  journal = {VISIGRAPP},
  volume = {4},
  pages = {435-445},
  number = {},
  year = {2024}
}

@inproceedings{he2022defeating,
  title = {Defeating DeepFakes via Adversarial Visual Reconstruction},
  author = {He, Ziwen and Wang, Wei and Guan, Weinan and Dong, Jing and Tan, Tieniu},
  booktitle = {ACM MM},
  pages = {2464--2472},
  year = {2022}
}

@misc{ailab,
  title = {AILab Smart Beauty API},
  author = {AI Innovate Technology Limited},
  year = {2026},
  howpublished = {\url{https://www.ailabtools.com/portrait-intelligent-beautification-example}}
}

@misc{tencentcloud,
  title = {Tencent Cloud BeautifyPic API},
  author = {Tencent Technology Limited},
  year = {2026},
  howpublished = {\url{https://cloud.tencent.com/document/product/1172/40715}}
}

@article{wang2021hififace,
  title = {HifiFace: 3D shape and semantic prior guided high fidelity face swapping},
  author = {Wang, Yuhan and Chen, Xu and Zhu, Junwei and Chu, Wenqing and Tai, Ying and Wang, Chengjie and Li, Jilin and Wu, Yongjian and Huang, Feiyue and Ji, Rongrong},
  journal = {arXiv:2106.09965},
  year = {2021}
}

@inproceedings{zhao2023diffswap,
  title = {DiffSwap: High-fidelity and controllable face swapping via 3D-aware masked diffusion},
  author = {Zhao, Wenliang and Rao, Yongming and Shi, Weikang and Liu, Zuyan and Zhou, Jie and Lu, Jiwen},
  booktitle = {CVPR},
  year = {2023}
}

@article{kim2025diffface,
  title = {DiffFace: Diffusion-based face swapping with facial guidance},
  author = {Kim, Kihong and Kim, Yunho and Cho, Seokju and Seo, Junyoung and Nam, Jisu and Lee, Kychul and Kim, Seungryong and Lee, KwangHee},
  journal = {PR},
  volume = {163},
  pages = {111451},
  number = {},
  year = {2025}
}

@inproceedings{zhou2023uniface,
  title = {UniFace: Unified cross-entropy loss for deep face recognition},
  author = {Zhou, Jiancan and Jia, Xi and Li, Qiufu and Shen, Linlin and Duan, Jinming},
  booktitle = {ICCV},
  year = {2023}
}

@inproceedings{gao2021information,
  title = {Information bottleneck disentanglement for identity swapping},
  author = {Gao, Gege and Huang, Huaibo and Fu, Chaoyou and Li, Zhaoyang and He, Ran},
  booktitle = {CVPR},
  year = {2021}
}

@article{li2023e4s,
  title = {E4S: Fine-grained Face Swapping via Editing With Regional GAN Inversion},
  author = {Li, Maomao and Yuan, Ge and Wang, Cairong and Liu, Zhian and Zhang, Yong and Nie, Yongwei and Wang, Jue and Xu, Dong},
  journal = {arXiv:2310.15081},
  year = {2023}
}

@inproceedings{shiohara2023blendface,
  title = {BlendFace: Re-designing identity encoders for face-swapping},
  author = {Shiohara, Kaede and Yang, Xingchao and Taketomi, Takafumi},
  booktitle = {ICCV},
  year = {2023}
}

@misc{face_recognition,
  author = {ageitgey},
  title = {Face Recognition},
  year = {2026},
  howpublished = {\url{https://github.com/ageitgey/face_recognition}}
}

@misc{aws_rekognition,
  author = {{Amazon Web Services}},
  title = {Amazon Rekognition Documentation},
  year = {2026},
  howpublished = {\url{https://docs.aws.amazon.com/rekognition/}}
}

@article{ranjan2018deep,
  title = {Deep learning for understanding faces: Machines may be just as good, or better, than humans},
  author = {Ranjan, Rajeev and Sankaranarayanan, Swami and Bansal, Ankan and Bodla, Navaneeth and Chen, Jun-Cheng and Patel, Vishal M and Castillo, Carlos D and Chellappa, Rama},
  journal = {SPM},
  volume = {35},
  number = {1},
  pages = {66--83},
  year = {2018}
}

@article{guo2019survey,
  title = {A survey on deep learning based face recognition},
  author = {Guo, Guodong and Zhang, Na},
  journal = {CVIU},
  volume = {189},
  pages = {102805},
  number = {},
  year = {2019}
}

@misc{deep_live_cam,
  author = {hacksider},
  title = {Deep-Live-Cam},
  howpublished = {\url{https://github.com/hacksider/Deep-Live-Cam}},
  year = {2026}
}

@misc{faceswap,
  author = {deepfakes},
  title = {faceswap},
  howpublished = {\url{https://github.com/deepfakes/faceswap}},
  year = {2026}
}

@misc{faceswap_deepfake_pytorch,
  author = {Oldpan},
  title = {Faceswap-Deepfake-Pytorch},
  howpublished = {\url{https://github.com/Oldpan/Faceswap-Deepfake-Pytorch}},
  year = {2026}
}

@misc{deep_face_live,
  author = {iperov},
  title = {DeepFaceLive},
  howpublished = {\url{https://github.com/iperov/DeepFaceLive}},
  year = {2026}
}

@misc{herahaven,
  title = {HeraHaven},
  author = {HeraHaven},
  howpublished = {\url{https://herahaven.com/}},
  year = {2026}
}

@misc{cathy_newman,
  title = {`Haunting' to see deepfake pornography of myself, says journalist Cathy Newman},
  author = {Hannah Roberts},
  howpublished = {\url{https://www.standard.co.uk/showbiz/celebrity-news/cathy-newman-good-morning-britain-british-ai-technology-government-b1203312.html}},
  year = {2025}
}

@inproceedings{fu2025diffpad,
  title = {DiffPAD: Denoising Diffusion-Based Adversarial Patch Decontamination},
  author = {Fu, Jia and Zhang, Xiao and Pashami, S. and Rahimian, F. and Holst, Anders},
  booktitle = {WACV},
  year = {2025}
}

@inproceedings{zhou2020manifold,
  title = {Manifold projection for adversarial defense on face recognition},
  author = {Zhou, Jianli and Liang, Chao and Chen, Jun},
  booktitle = {ECCV},

  year = {2020}
}

@inproceedings{deb2023faceguard,
  title = {FaceGuard: A self-supervised defense against adversarial face images},
  author = {Deb, Debayan and Liu, Xiaoming and Jain, Anil K},
  booktitle = {FG},
  year = {2023}
}

@inproceedings{walker2024merging,
  title = {Merging AI incidents research with political misinformation research: introducing the political DeepFakes incidents database},
  author={Walker, Christina P and Schiff, Daniel S and Schiff, Kaylyn Jackson},
  booktitle = {AAAI},
  year = {2024}
}

@misc{wikimediaFileDrKlaus,
  title = {File: Dr. Klaus Schwab Founder of the WEF},
  author = {John Kerry},
  howpublished = {\url{https://commons.wikimedia.org/wiki/File:Dr._Klaus_Schwab_Founder_of_the_WEF_(24511853176).jpg}},
  year = {2026}
}

@article{wang2024instantid,
  title={Instantid: Zero-shot identity-preserving generation in seconds},
  author={Wang, Qixun and Bai, Xu and Wang, Haofan and Qin, Zekui and Chen, Anthony and Li, Huaxia and Tang, Xu and Hu, Yao},
  journal={arXiv:2401.07519},
  year={2024}
}

@inproceedings{cherepanova2021lowkey,
  title={LowKey: Protecting Social Media Users from Facial Recognition},
  author={Cherepanova, Valeriia and Goldblum, Micah and Foley, Harrison and Duan, Shiyuan and Dickerson, John and Taylor, Gavin and Goldstein, Tom},
  booktitle={ICLR},
  year={2021}
}

@inproceedings{wang2025nullswap,
  title={NullSwap: Proactive Identity Cloaking Against Deepfake Face Swapping},
  author={Wang, Tianyi and Niu, Shuaicheng and Cheng, Harry and Zhang, Xiao and Wang, Yinglong},
  booktitle={ICCV},
  year={2025}
}

@misc{brandstudio,
  title = {Brand Studio},
  author = {brandstudio.ai},
  howpublished = {\url{https://brandstudio.com/}},
  year={2026}
}

@inproceedings{ronneberger2015u,
  title={U-net: Convolutional networks for biomedical image segmentation},
  author={Ronneberger, Olaf and Fischer, Philipp and Brox, Thomas},
  booktitle={MICCAI},
  year={2015},
}

@inproceedings{huang2022cmua,
  title={Cmua-watermark: A cross-model universal adversarial watermark for combating deepfakes},
  author={Huang, Hao and Wang, Yongtao and Chen, Zhaoyu and Zhang, Yuze and Li, Yuheng and Tang, Zhi and Chu, Wei and Chen, J. and Lin, W. and Ma, Kai-Kuang},
  booktitle={AAAI},
  year={2022}
}

@article{li2026inject,
  title={Inject Where It Matters: Training-Free Spatially-Adaptive Identity Preservation for Text-to-Image Personalization},
  author={Li, Guandong and Ye, Mengxia},
  journal={arXiv preprint arXiv:2602.13994},
  year={2026}
}

@inproceedings{ho2020denoising,
  title={Denoising diffusion probabilistic models},
  author={Ho, Jonathan and Jain, Ajay and Abbeel, Pieter},
  booktitle = {NeurIPS},
  year={2020}
}

@inproceedings{10.1145/3569951.3597559,
author = {Boerner, Timothy J. and Deems, Stephen and Furlani, Thomas R. and Knuth, Shelley L. and Towns, John},
title = {ACCESS: Advancing Innovation: NSF’s Advanced Cyberinfrastructure Coordination Ecosystem: Services \& Support},
year = {2023},
pages = {173–176}
}

\appendix

\section*{Ethics Considerations}\label{app:ethics}

\name~focuses on defending against deepfake attacks rather than creating or advancing them. All data and models used in this study are publicly available, and we do not introduce any new attack methods or conduct real-world attacks.

\noindent\textbf{User Study.} We conducted a user study involving human participants to evaluate the perceptual performance of \name. The study was reviewed and approved by the Human Research Protection Program (HRPP) of University of Kansas. All participants were informed of the study’s purpose, procedure, risks, and outcome. They were provided with an information statement in lieu of signed consent forms. They were free to withdraw at any time during the study without penalty. No personally identifiable information was collected or stored. The study was designed to pose no greater risk or discomfort than participants would ordinarily encounter in daily life.

\noindent\textbf{Cloak Image Selection.}\label{ethic:cloak_image_selection}
\name\ utilizes a cloak image to guide the face-swapping process such that the deepfake outputs resemble the cloak identity rather than the original victim. In our experiments, we use subjects from the VGGFace2~\cite{cao2018vggface2} and FFHQ~\cite{karras2019style}, the same dataset used for training and evaluating face-swapping models, only to reflect the general performance of \name. That said, the increasing availability of high-quality AI-generated faces makes it easy for benign users to select appropriate synthetic identities as cloak images. In practice, \textit{AI-generated cloaks should always be used}. With a small performance decrease in tracing accuracy, using AI cloaks eliminates the ethical concerns with cloak selection. 
In its real-world deployment, \name\ can be integrated into online image-sharing platforms, where the platform itself generates, selects, and applies AI-generated cloak images automatically. This eliminates the need for users to make cloak selection decisions and prevents inadvertent involvement of real individuals.

\noindent\textbf{Non-consensual Use of Cloak Images.}
The cloak image scheme may raise ethical concerns that a malicious user could use a victim’s image as the cloak without the victim's consent. However, such behavior will not bring any additional benefit to the malicious user.
\textbf{(1)} For normal/benign users, \name~recommends selecting appropriate AI-generated images as the cloak image. Because AI-generated images are virtually unlimited, \name\ can select the most suitable cloak image based on the user's facial identity and the user's protection priorities to achieve optimal protection performance. As a result, benign users have no practical advantage in manually selecting real-world images.
\textbf{(2)} If a malicious user obtains an unprotected victim image and intends to perform non-consensual face-swapping, they can directly execute the face-swapping using this unprotected image. Therefore, \name~does not offer additional advantages to adversaries.
\textbf{(3)} If a protected image is reused as the cloak, the resulting identity follows the original cloak rather than the victim. That is, if Alice posts a protected image with an AI cloak, and Chuck uses the protected image as the cloak for his own image, an identity-stealing face swap will produce an image that resembles the AI cloak, not Alice, not Chuck. An experiment of 3000 random image pairs indicates that $ASR_{id}$ of the victim identity is 1.03\% and 0.13\% by FaceNet and Face++, respectively. However, the $TSR$ of the cloak identity of the original protected victim image is 90.17\% (FaceNet) and 96.20\% (Face++). Please refer to Appendix~\ref{app:cloak_transferability} for more details.
As a result, with \name, the victim image cannot be used as the source, target, or cloak image.

\section{Glossary}\label{app:glossary}

{Table \ref{tab:notations} summarizes the notations used in the paper.}

\begin{table}[t]
    \centering
    \small
    \caption{Notations used in the paper.}
    \label{tab:notations}
    \renewcommand{\arraystretch}{1.25}
    \begin{tabular}{
        >{\raggedright\arraybackslash}m{0.14\linewidth} |
        >{\raggedright\arraybackslash}m{0.76\linewidth}
    }
    
    \hline
    Notation & Description \\
    \hline
    $\mathbf{x}$ & The image to protect.  \\
    \hline
    $\mathbf{x}_{s}$, $\mathbf{x}_{t}$ & The face-swapping source and target image. \\
    \hline
    $\tilde{\mathbf{x}}$ & The protected image. \\
    \hline
    $\mathcal{E}_{id}$, $\mathcal{E}_{ctx}$ & The identity and context extractors. \\
    \hline
    $\mathbf{z}_{s}$, $\mathbf{z}_{t}$ & The image's identity- and context-encoded features. \\
    \hline
    $\tilde{\mathbf{z}}_{s}$, $\tilde{\mathbf{z}}_{t}$ & The protected image's identity- and context-encoded features. \\
    \hline
    $\hat{\mathbf{x}}$ & The face-swapping result. \\
    \hline
    $\tilde{I}_c$ & The cloak identity. \\
    \hline
    $I_s$ & The source or original identity. \\
    \hline
    $\mathcal{T}_{identity}$, $\mathcal{T}_{context}$ & The identity and context protections. \\
    \hline
    $\mathcal{L}_{identity}$, $\mathcal{L}_{context}$ & The identity and context protection loss functions. \\
    \hline
    $\lambda_{util}$, $\lambda_{id}$, $\lambda_{cloak}$, and $\lambda_{ctx}$  & The loss function weights for utility, identity, cloak, and context. \\
    \hline
    $\delta_{RGB}$, $\delta_{id}$, and $\delta_{ctx}$ & The deviation limits for RGB-channel perturbation, identity, and context. \\
    \hline
    $\Delta \mathbf{x}$ & The difference between the protected image and the original image, i.e., the noise perturbation. \\
    \hline
    $\Delta \mathbf{z_s}$ & The identity feature difference between the protected image and the original image. \\
    \hline
    $\Delta \mathbf{z_t}$ & The context feature difference between the protected image and the original image. \\
    \hline
    $\Delta \mathbf{z_c}$ & The identity feature difference between the protected image and the cloak image. \\
    \hline
    $\mathcal{S}$ & The composite protection score. \\
    \hline
    $w_{id}$, $w_{ctx}$, and $w_{cloak}$ & Weights for identity protection, context protection, and tracing in the composite protection score. \\
    \hline
    $ASR_o$   & Original attack success rate (i.e., the face-swapping model's baseline performance). \\
    \hline
    $ASR_p$   & Overall attack success rate under protection. \\
    \hline
    $ASR_{id}$  & Identity attack success rate under protection. \\
    \hline
    $ASR_{ctx}$ & Context attack success rate under protection. \\
    \hline
    $TSR$ & Tracing success rate. \\
    \hline
    $\mathcal{X}_c$ & Cloak image set with N cloak images. \\
    \hline
    $\mathbf{x}_c^{(i)}$ & Cloak image indexed by i. \\
    \hline
    \end{tabular}
\end{table}

\section{DiffFace Protection Samples with Different Cloak Image Selection Strategies}\label{app:diffface}

\begin{table}[t]
  \centering
  \captionsetup{skip=3pt}
  \caption{DiffFace face-swapping results with different cloak image selections. \textbf{w/o Pt.}: without protection.}
  \label{tab:diffface_anchor_sample}\vspace{-1mm}
  \setlength\tabcolsep{3.6pt}
  \begin{tabular}{c  c  c  c  c}
    \toprule
    $\mathbf{x}$ (Alice) & Bob & w/o Pt. & Charlie & w/o Pt. \\
    \midrule
    \raisebox{-0.5\height}{\includegraphics[scale=0.16]{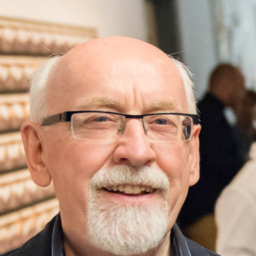}} &
    \raisebox{-0.5\height}{\includegraphics[scale=0.16]{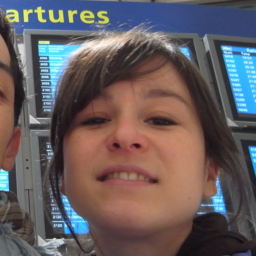}} &
    \raisebox{-0.5\height}{\includegraphics[scale=0.16]{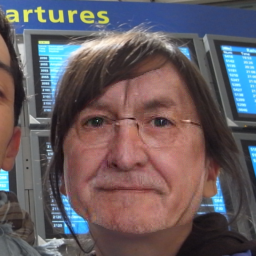}} &
    \raisebox{-0.5\height}{\includegraphics[scale=0.16]{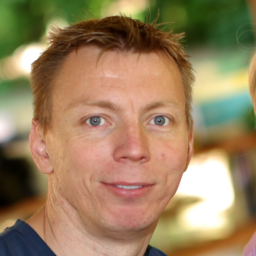}} &
    \raisebox{-0.5\height}{\includegraphics[scale=0.16]{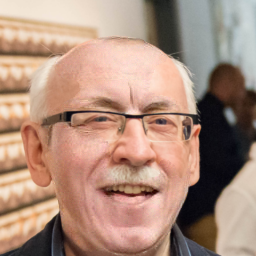}} \\
    \bottomrule
  \end{tabular}

  \setlength\tabcolsep{0.5pt}
  \begin{tabular}{ c  c  c  c  c  c  c  c}
    & \makecell{Baseline}
    & \makecell{Style\\GAN3}
    & \makecell{Gender\\Aware}
    & \makecell{Pr. Pro-\\tection}
    & \makecell{Dist\\1.04}
    & \makecell{Dist\\1.50} 
    & \makecell{Dist\\2.00} \\
    \midrule
    \raisebox{-0.5\height}{\rotatebox{90}{Cloak}} &
    \raisebox{-0.5\height}{\includegraphics[scale=0.125]{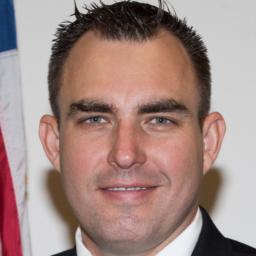}} &
    \raisebox{-0.5\height}{\includegraphics[scale=0.125]{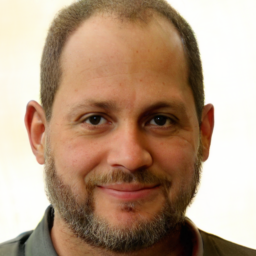}} &
    \raisebox{-0.5\height}{\includegraphics[scale=0.125]{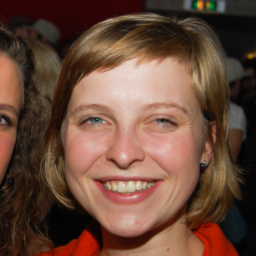}} &
    \raisebox{-0.5\height}{\includegraphics[scale=0.125]{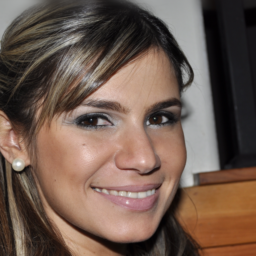}} &
    \raisebox{-0.5\height}{\includegraphics[scale=0.125]{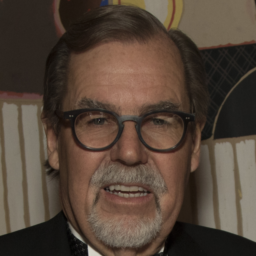}} &
    \raisebox{-0.5\height}{\includegraphics[scale=0.125]{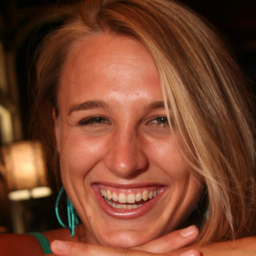}} &
    \raisebox{-0.5\height}{\includegraphics[scale=0.125]{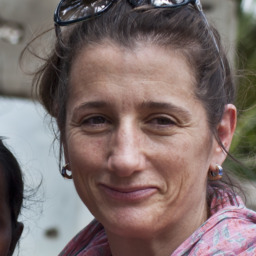}} \\
    \midrule
    \raisebox{-0.5\height}{\rotatebox{90}{$\mathcal{T}_{identity}$}} &  \raisebox{-0.5\height}{\includegraphics[scale=0.125]{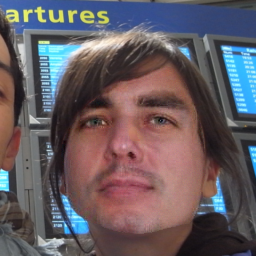}} &
    \raisebox{-0.5\height}{\includegraphics[scale=0.125]{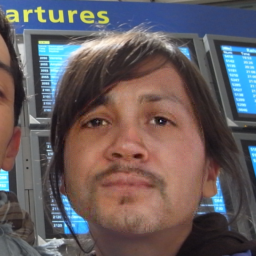}} &
    \raisebox{-0.5\height}{\includegraphics[scale=0.125]{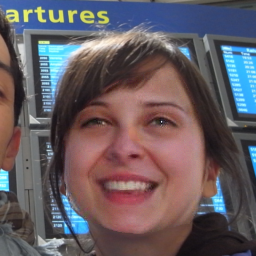}} &    
    \raisebox{-0.5\height}{\includegraphics[scale=0.125]{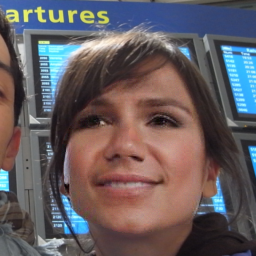}} &    
    \raisebox{-0.5\height}{\includegraphics[scale=0.125]{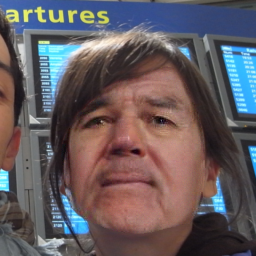}} &    
    \raisebox{-0.5\height}{\includegraphics[scale=0.125]{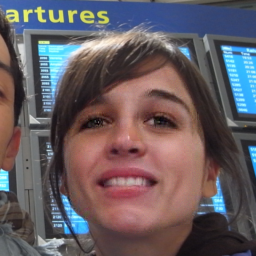}} & 
    \raisebox{-0.5\height}{\includegraphics[scale=0.125]{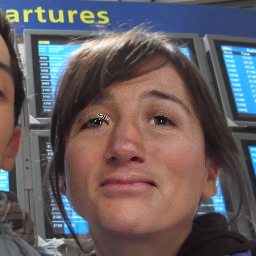}} \\
    \midrule
    \raisebox{-0.5\height}{\rotatebox{90}{$\mathcal{T}_{context}$}} &   
    \raisebox{-0.5\height}{\includegraphics[scale=0.125]{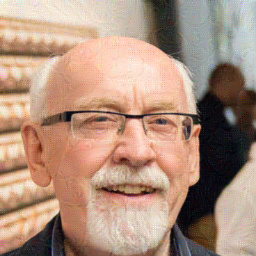}} &
    \raisebox{-0.5\height}{\includegraphics[scale=0.125]{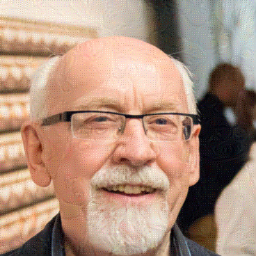}} &
    \raisebox{-0.5\height}{\includegraphics[scale=0.125]{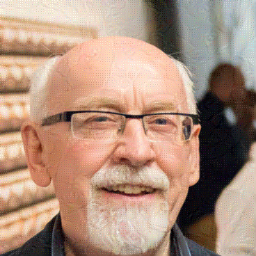}} &    
    \raisebox{-0.5\height}{\includegraphics[scale=0.125]{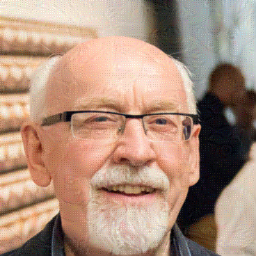}} &    
    \raisebox{-0.5\height}{\includegraphics[scale=0.125]{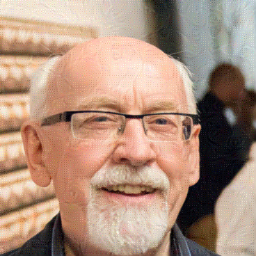}} &    
    \raisebox{-0.5\height}{\includegraphics[scale=0.125]{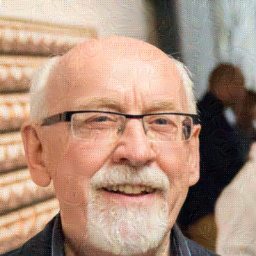}} & 
    \raisebox{-0.5\height}{\includegraphics[scale=0.125]{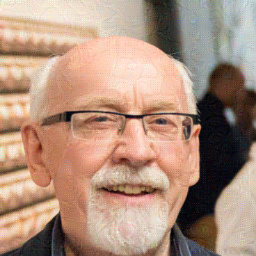}} \\
    \bottomrule
  \end{tabular}
  \vspace{-3mm}
\end{table}

{Table~\ref{tab:diffface_anchor_sample} illustrates the cloak images and the corresponding face-swapping results for identity and context protection. Visually, across all cloak images, identity-stealing face swaps more closely resemble the cloak identity rather than the victim Alice, while context-stealing attempts fail to alter the original image, indicating that Charlie cannot successfully swap his face into Alice’s image.}

\section{SimSwap Robustness Experiment Samples}\label{app:extra_robustness}

We analyze the adaptive attack in Section~\ref{subsec:protection_robustness}, where the attacker attempts to approximate the added perturbation $\Delta\mathbf{x}$ and recover the original image via $\mathbf{x} = \tilde{\mathbf{x}} - \Delta\mathbf{x}$. Representative examples of such attacks are shown in Table~\ref{tab:adaptive_samples}. For visualization clarity, the perturbations are magnified by a factor of 10 to better expose their structural characteristics. When a random image is used to generate the perturbation, the resulting patterns vary considerably across protected images. In contrast, when the protected image itself is used, the perturbations exhibit high similarity, which explains the observed drop in identity protection effectiveness but improved traceability performance. Complete experimental details are provided in Section~\ref{subsec:protection_robustness}.

\begin{table}[t]
  \centering
  \captionsetup{skip=3 pt}
  \caption{\name~adaptive attack samples.}
  \label{tab:adaptive_samples}
  \setlength\tabcolsep{1.7pt}
  \begin{tabular}{c  c  c  c  c}    
    \toprule
    \multicolumn{5}{c}{Adaptive Attack 1 (random image)} \\
    \midrule
    $\mathbf{x_A}$ & $\mathbf{x_B}$ & $\mathbf{x_C}$ & $\tilde{\mathbf{x}}_A$ - $\mathbf{x}_A$ & $\tilde{\mathbf{x}}_B$ - $\mathbf{x_B}$ \\
    \midrule
    \raisebox{-0.5\height}{\includegraphics[scale=0.20]{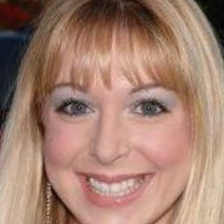}} &
    \raisebox{-0.5\height}{\includegraphics[scale=0.20]{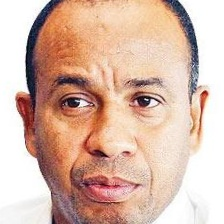}} &
    \raisebox{-0.5\height}{\includegraphics[scale=0.20]{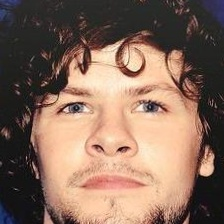}} &
    \raisebox{-0.5\height}{\includegraphics[scale=0.20]{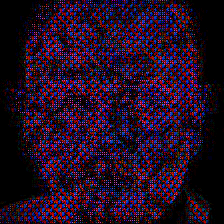}} &
    \raisebox{-0.5\height}{\includegraphics[scale=0.20]{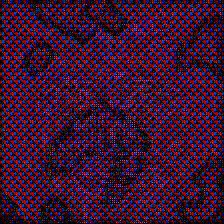}} \\
    \midrule
    \raisebox{-0.5\height}{\includegraphics[scale=0.20]{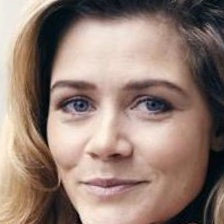}} &
    \raisebox{-0.5\height}{\includegraphics[scale=0.20]{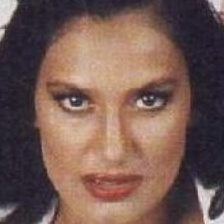}} &
    \raisebox{-0.5\height}{\includegraphics[scale=0.20]{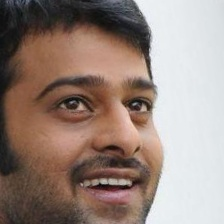}} &
    \raisebox{-0.5\height}{\includegraphics[scale=0.20]{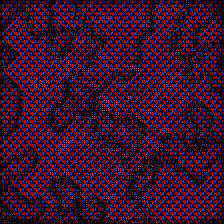}} &
    \raisebox{-0.5\height}{\includegraphics[scale=0.20]{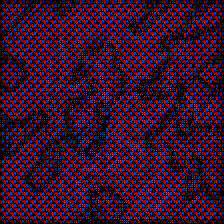}} \\
    \midrule
    \raisebox{-0.5\height}{\includegraphics[scale=0.20]{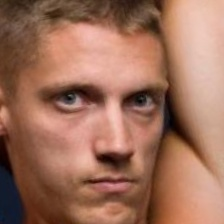}} &
    \raisebox{-0.5\height}{\includegraphics[scale=0.20]{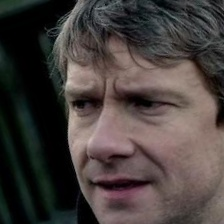}} &
    \raisebox{-0.5\height}{\includegraphics[scale=0.20]{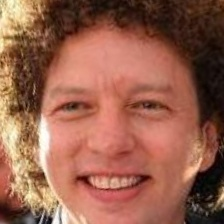}} &
    \raisebox{-0.5\height}{\includegraphics[scale=0.20]{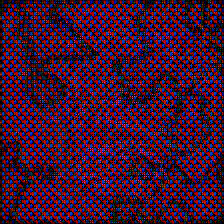}} &
    \raisebox{-0.5\height}{\includegraphics[scale=0.20]{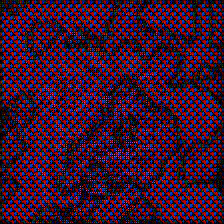}} \\
    \midrule
    \multicolumn{5}{c}{Adaptive Attack 2 (protected image)}\\
    \midrule
    $\mathbf{x_A}$ & $\mathbf{x_B}$ & Cloak & $\tilde{\mathbf{x}}_A$ - $\mathbf{x}_A$ & $\tilde{\mathbf{x}}_B$ - $\mathbf{x_B}$ \\
    \midrule
    \raisebox{-0.5\height}{\includegraphics[scale=0.20]{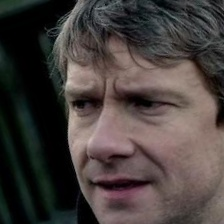}} &
    \raisebox{-0.5\height}{\includegraphics[scale=0.20]{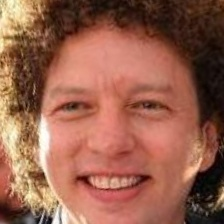}} &
    \raisebox{-0.5\height}{\includegraphics[scale=0.20]{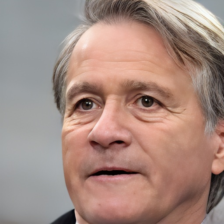}} &
    \raisebox{-0.5\height}{\includegraphics[scale=0.20]{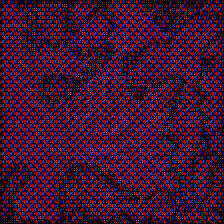}} &
    \raisebox{-0.5\height}{\includegraphics[scale=0.20]{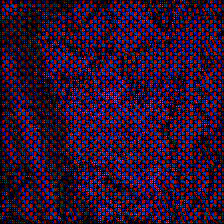}} \\
    \midrule
    \raisebox{-0.5\height}{\includegraphics[scale=0.20]{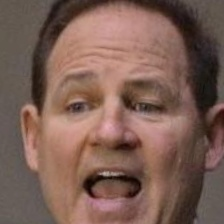}} &
    \raisebox{-0.5\height}{\includegraphics[scale=0.20]{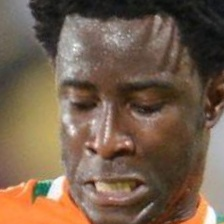}} &
    \raisebox{-0.5\height}{\includegraphics[scale=0.20]{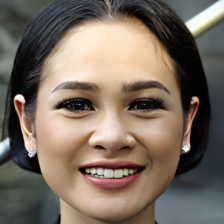}} &
    \raisebox{-0.5\height}{\includegraphics[scale=0.20]{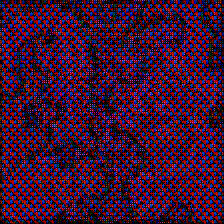}} &
    \raisebox{-0.5\height}{\includegraphics[scale=0.20]{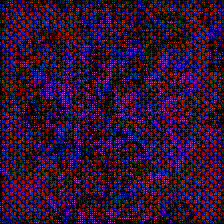}} \\
    \midrule
    \raisebox{-0.5\height}{\includegraphics[scale=0.20]{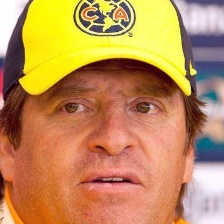}} &
    \raisebox{-0.5\height}{\includegraphics[scale=0.20]{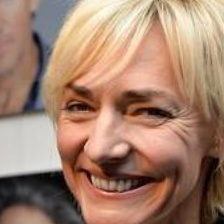}} &
    \raisebox{-0.5\height}{\includegraphics[scale=0.20]{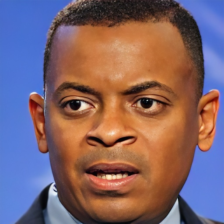}} &
    \raisebox{-0.5\height}{\includegraphics[scale=0.20]{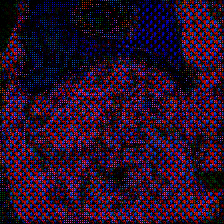}} &
    \raisebox{-0.5\height}{\includegraphics[scale=0.20]{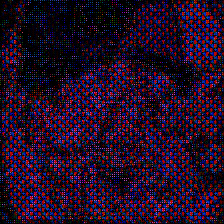}} \\
    \bottomrule
  \end{tabular}
  \vspace{-1mm}
\end{table}

\section{Comparison with Previous Works on Context Protection}\label{app:context_comparison}

\name~provides comprehensive protection against face swapping deepfakes by simultaneously protecting both image identity and context. To the best of our knowledge, there are no existing proactive defenses specifically designed to protect image context against face swapping, i.e., to prevent a protected image from being manipulated by replacing its original identity with another person’s face. Therefore, for context protection comparison, we evaluate \name~against representative proactive defenses that aim to comprehensively protect facial images from other types of deepfake manipulation. Specifically, we compare with Initiative Defense~\cite{huang2021initiative}, which is designed to defend against face attribute editing and face reenactment, and CMUA-Watermark~\cite{huang2022cmua}, which targets facial manipulation and attribute editing. For consistency, all comparisons are conducted using SimSwap as the representative face-swapping model.

To ensure a fair comparison, we adjust the perturbation magnitude of all methods to achieve similar utility, with MSE values ranging from 106 to 112. As shown in Table~\ref{tab:protection_comparison_context}, under comparable perturbation levels, both Initiative Defense~\cite{huang2021initiative} and CMUA-Watermark~\cite{huang2022cmua} provide almost no protection against SimSwap-based face-swapping deepfakes. Specifically, their $ASR_{ctx}$ values remain close to the original face-swapping success rates ($ASR_{o}$). For example, Initiative Defense achieves $ASR_{ctx}$ values of 94.50\% and 97.90\%, compared to $ASR_{o}$ values of 94.87\% and 98.80\%, while CMUA-Watermark achieves 93.40\% and 97.73\%. In contrast, \name~reduces $ASR_{ctx}$ to 1.90\% and 3.20\%, demonstrating effective context protection against SimSwap-based face-swapping deepfakes.

\begin{table}[t]
  \centering
  \small
  \captionsetup{skip=3 pt}
  
  \caption{Context protection of \name~, Initiative Defense~\cite{huang2021initiative}, and CMUA-Watermark~\cite{huang2022cmua}. $F^1$: FaceNet-512; $F^2$: Megvii Face++. Bold indicates the best $ASR_{ctx}$ across all protection methods.
  }\label{tab:protection_comparison_context}
  \setlength\tabcolsep{1.00 pt}
  \renewcommand{\arraystretch}{1.1}
  
  \begin{tabular}{c | c  c | c  c  c  c | c  c}
    \hline
    \multirow{2}*{} & \multicolumn{2}{c}{$ASR_{o}$} \vline & \multicolumn{4}{c}{Utility} \vline & \multicolumn{2}{c}{$ASR_{ctx}{\downarrow}$} \\
    \cline{2-9}
    ~ & $F^1$ & $F^2$ & MSE $\downarrow$ & PSNR $\uparrow$ & SSIM $\uparrow$ & LPIPS $\downarrow$ & $F^1$ & $F^2$\\
    \hline
    \name & \multirow{3}*{94.87} & \multirow{3}*{98.80} & 106 & 27.88 & 0.68 & 0.25 & 1.90 & 3.20 \\
    Initiative Defense & ~ & ~ & 112 & 27.61 & 0.67 & 0.36 & 94.50 & 97.90 \\
    CMUA-Watermark & ~ & ~ & 109 & 27.90 & 0.74 & 0.31 & 93.40 & 97.73 \\
    \hline
  \end{tabular}
  
  \setlength\tabcolsep{1.2 pt}
  \vspace{-1mm}
\end{table}

\section{Transferability of Cloak Identity Protection under Non-consensual Use}\label{app:cloak_transferability}

We further analyze a potential non-consensual usage scenario involving cloak images. Consider user Alice who posted her own images, and an attacker Chuck who uses Alice's protected images as the cloak without consent, hoping the face-swapping output to present Alice’s identity instead of Chuck’s.

However, when Alice already employed \name\ to protect her images using a cloak identity $\mathcal{A}$, Chuck's attack becomes ineffective. Specifically, if Chuck uses Alice's protected image as the cloak, the resulting face-swapping output is dominated by Alice's cloak identity $\mathcal{A}$, not Alice, not Chuck. This indicates that the cloak identity protection is \emph{transferable}, i.e., the protection effect propagates through the protected image.

Table~\ref{tab:cloak_transferability} reports the quantitative results. The identity of Alice remains well protected, with $ASR_{id}$ close to zero, while the identity of Chuck is also effectively suppressed ($ASR_{id}$ of 1.03\% and 0.13\% under FaceNet and Face++, respectively). Meanwhile, the cloak identity $\mathcal{A}$ remains highly traceable, with $TSR$ exceeding 96\% across both evaluation models.

These results demonstrate that even under non-consensual reuse of protected images, \name~does not expose the victim’s identity. Instead, the protection consistently redirects the output toward the originally embedded cloak identity.

\begin{table}[t]
  \centering
  \small
  \captionsetup{skip=3pt}
  \caption{Cloak identity protection transferability.}
  \label{tab:cloak_transferability}
  
  \setlength\tabcolsep{7pt}
  \renewcommand{\arraystretch}{1.1}
  
  \begin{tabular}{c  c  c | c  c  c} 
    \hline
    \multicolumn{3}{c}{FaceNet-512} \vline & \multicolumn{3}{c}{Megvii Face++} \\
    \hline
    Alice & Chuck & $\mathcal{A}$ & Alice & Chuck & $\mathcal{A}$ \\
    $ASR_{id}{\downarrow}$ & $ASR_{id}{\downarrow}$ & $TSR{\uparrow}$ & $ASR_{id}{\downarrow}$ & $ASR_{id}{\downarrow}$ & $TSR{\uparrow}$ \\    
    \hline
    0.17 & 1.03 & 96.17 & 0.00 & 0.13 & 96.20 \\
    \hline
  \end{tabular}
  
\end{table}

\section{Qualitative Results under Different Perturbation Levels}

In this section, we provide additional qualitative examples of protected images under different perturbation strengths. Specifically, we present visual comparisons across a range of MSE/PSNR levels to illustrate the perceptual impact of the perturbations.

Since \name~is designed to introduce minimal yet effective perturbations, these examples demonstrate that the protection can substantially degrade face-swapping performance while preserving the overall visual fidelity of the original images. By examining cases with varying perturbation magnitudes, we provide an intuitive understanding of the trade-off between protection effectiveness and perceptual quality.

\begin{table}[t]
  \centering
  \small
  \captionsetup{skip=1.0 pt}
  \caption{Qualitative samples of \name\ protection.}
  \label{tab:qualitative_samples}
  \setlength\tabcolsep{1.0 pt}
  
  \begin{tabular}{c  c  c  c  c  c}
    \hline
    \multicolumn{5}{c}{SimSwap protected image samples in Table~\ref{tab:protection_performance}} \\
    \toprule
    \raisebox{-0.5\height}{\includegraphics[scale=0.2]{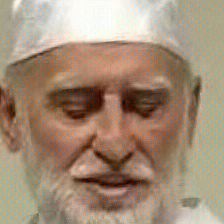}} & 
    \raisebox{-0.5\height}{\includegraphics[scale=0.2]{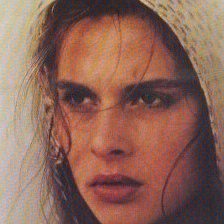}} & 
    \raisebox{-0.5\height}{\includegraphics[scale=0.2]{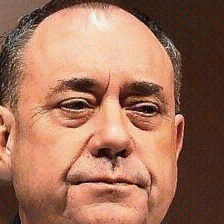}} &
    \raisebox{-0.5\height}{\includegraphics[scale=0.2]{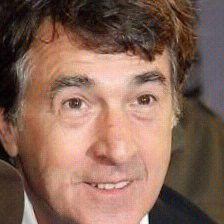}} &
    \raisebox{-0.5\height}{\includegraphics[scale=0.2]{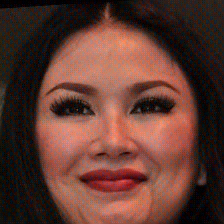}} \\
    \midrule
    \raisebox{-0.5\height}{\includegraphics[scale=0.2]{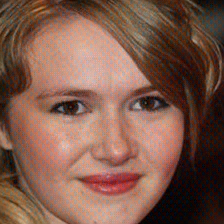}} & 
    \raisebox{-0.5\height}{\includegraphics[scale=0.2]{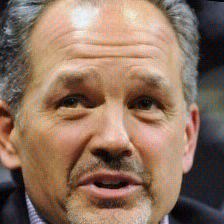}} &
    \raisebox{-0.5\height}{\includegraphics[scale=0.2]{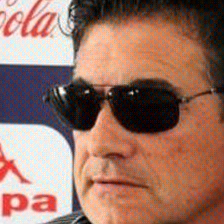}} &
    \raisebox{-0.5\height}{\includegraphics[scale=0.2]{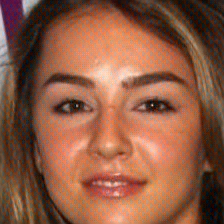}} &
    \raisebox{-0.5\height}{\includegraphics[scale=0.2]{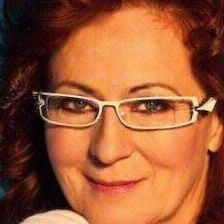}} \\
    \bottomrule

    \multicolumn{5}{c}{DiffFace protected image samples in Table~\ref{tab:protection_performance}} \\
    \toprule
    \raisebox{-0.5\height}{\includegraphics[scale=0.175]{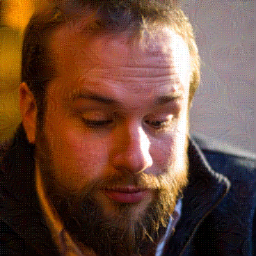}} & 
    \raisebox{-0.5\height}{\includegraphics[scale=0.175]{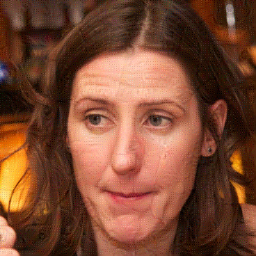}} & 
    \raisebox{-0.5\height}{\includegraphics[scale=0.175]{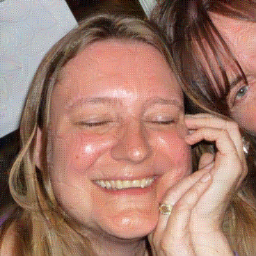}} &
    \raisebox{-0.5\height}{\includegraphics[scale=0.175]{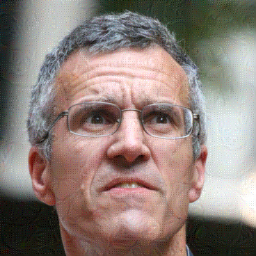}} &
    \raisebox{-0.5\height}{\includegraphics[scale=0.175]{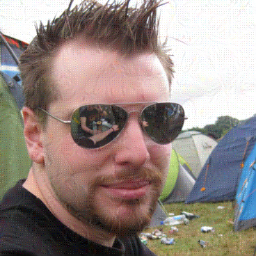}} \\
    \midrule
    \raisebox{-0.5\height}{\includegraphics[scale=0.175]{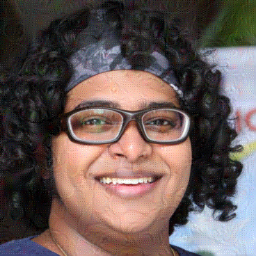}} & 
    \raisebox{-0.5\height}{\includegraphics[scale=0.175]{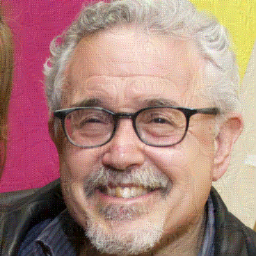}} &
    \raisebox{-0.5\height}{\includegraphics[scale=0.175]{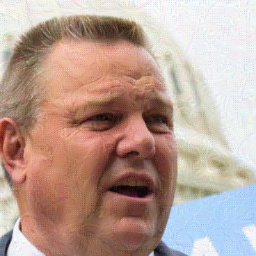}} &
    \raisebox{-0.5\height}{\includegraphics[scale=0.175]{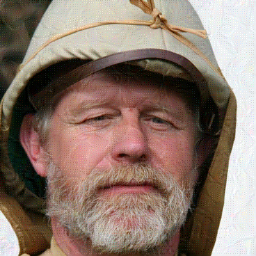}} &
    \raisebox{-0.5\height}{\includegraphics[scale=0.175]{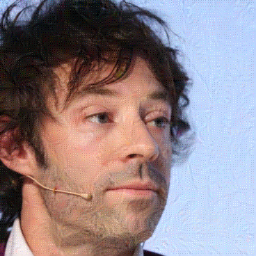}} \\
    \bottomrule

    \multicolumn{5}{c}{Black-box protected image samples in Table~\ref{tab:blackbox_protection}} \\
    \toprule
    \raisebox{-0.5\height}{\includegraphics[scale=0.175]{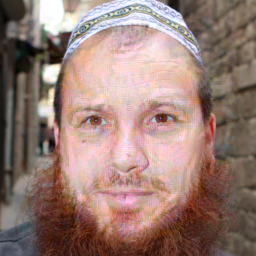}} & 
    \raisebox{-0.5\height}{\includegraphics[scale=0.175]{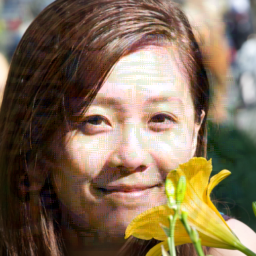}} & 
    \raisebox{-0.5\height}{\includegraphics[scale=0.175]{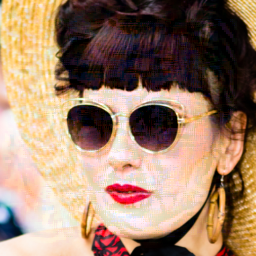}} &
    \raisebox{-0.5\height}{\includegraphics[scale=0.175]{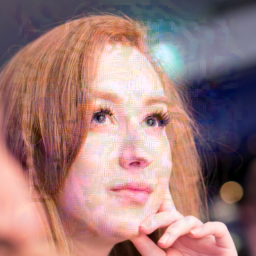}} &
    \raisebox{-0.5\height}{\includegraphics[scale=0.175]{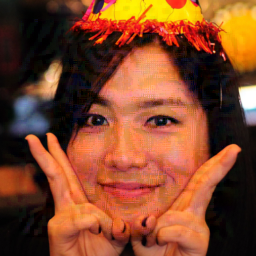}} \\
    \midrule
    \raisebox{-0.5\height}{\includegraphics[scale=0.175]{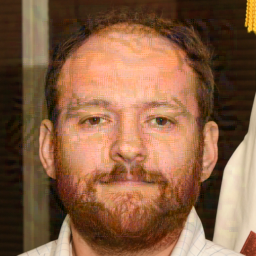}} & 
    \raisebox{-0.5\height}{\includegraphics[scale=0.175]{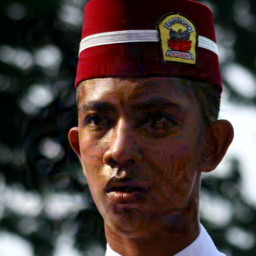}} &
    \raisebox{-0.5\height}{\includegraphics[scale=0.175]{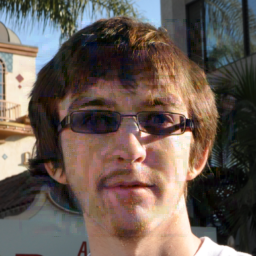}} &
    \raisebox{-0.5\height}{\includegraphics[scale=0.175]{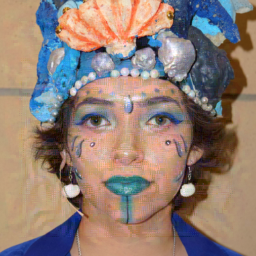}} &
    \raisebox{-0.5\height}{\includegraphics[scale=0.175]{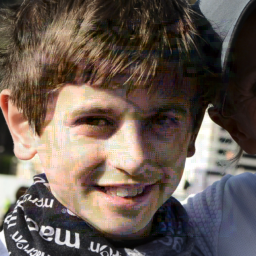}} \\
    \bottomrule
    \end{tabular}
    
    \vspace{-4mm}
\end{table}

\section{Effect of Compression-in-the-Loop Optimization on Robustness to Pre-processing}\label{app:robustness}

In Section~\ref{subsec:protection_robustness}, we introduced the compression-in-the-loop approach to the perturbation process to improve robustness of the generated images. Here, we provide a quantitative evaluation of this design by comparing the robustness of \name\ against common pre-processing transformations when the perturbations are generated with and without compression-in-the-loop. The results are summarized in Table~\ref{tab:robustness_metric_comparison}.

As shown in Table~\ref{tab:robustness_metric_comparison}, incorporating two compression operations in the 1,000-epoch optimization process substantially improves identity protection robustness. For example, when the attacker pre-processes the protected images with compression before face swapping, the attack success rate ($ASR_{id}$) of the baseline drops to 5.50\% (FaceNet) and 1.60\% (Face++), compared to 67.80\% and 62.97\% without compression-in-the-loop. The consistent improvement of robustness across all settings indicates that the proposed compression-in-the-loop design effectively enhances robustness against pre-processing transformations.

In contrast, the improvement in context protection is limited. For example, under the same pre-processing of compression, the $ASR_{ctx}$ with compression-in-the-loop is 95.50\% and 98.50\%, while the corresponding $ASR_{ctx}$ without compression-in-the-loop are 95.70\% and 98.30\%, respectively. This suggests that compression-in-the-loop primarily benefits identity protection, with limited impact on context protection robustness.

\begin{table}[t]
\small
    \captionsetup{skip=0 pt}
    \caption{\name~robustness against pre-processing. Process: 0 No operation, 1 Gaussian noise, 2 compress, 3 crop, 4 logo, 5 increase brightness, 6 decrease brightness. $F^1$: FaceNet-512, and $F^2$: Megvii Face++.}
    \label{tab:robustness_metric_comparison}
    
    \setlength\tabcolsep{5.65 pt}
    \begin{tabular}{c | c  c  c  c | c  c  c  c}
    \hline
    \multirow{3}*{\raisebox{-1.5 ex}{\rotatebox{90}{Process}}} & \multicolumn{4}{c}{Baseline} \vline & \multicolumn{4}{c}{Prioritizing protection}  \\
    \cline{2-9}
    ~ & \multicolumn{2}{c}{$ASR_{id}$} & \multicolumn{2}{c}{$ASR_{ctx}$} \vline & \multicolumn{2}{c}{$ASR_{id}$} & \multicolumn{2}{c}{$ASR_{ctx}$} \\
    \cline{2-9}
    ~ & $F^1{\downarrow}$ & $F^2{\downarrow}$ & $F^1{\downarrow}$ & $F^2{\downarrow}$ & $F^1{\downarrow}$ & $F^2{\downarrow}$ & $F^1{\downarrow}$ & $F^2{\downarrow}$ \\
    \hline
    0 & 0.20  & 0.07  & 3.53  & 4.73  & 0.13  & 0.00  & 4.00  & 6.43  \\
    \hline
    \multicolumn{9}{c}{ \name~protection robustness \textbf{without} compression-in-the-loop} \\
    \hline
     1 &   63.63 &   64.30 &   2.03  &   2.97  &   58.63 &   59.70 &   2.53  &   4.10  \\
     2 &   67.80 &   62.97 &   95.70 &   98.30 &   62.23 &   61.20 &   96.77 &   97.30 \\
     3 &   6.83  &   1.80  &   5.90  &   6.57  &   3.93  &   2.27  &   7.70  &   9.93  \\
     4 &   1.60  &   0.57  &   2.50  &   2.33  &   0.47  &   0.77  &   2.70  &   4.83  \\
     5 &   4.63  &   2.90  &   5.53  &   6.20  &   3.10  &   2.40  &   7.40  &   8.30  \\
     6 &   1.80  &   0.60  &   41.47 &   61.20 &   0.60  &   0.10  &   48.67 &   68.13 \\
    \hline
    \hline
    \multicolumn{9}{c}{\name~protection robustness \textbf{with} compression-in-the-loop} \\
    \hline
    1 &   27.07 &  13.30 &  0.80  &  1.63  &  19.40 &  10.90 &  1.13  &  2.00  \\
    2 &  5.50  &  1.60  &  95.50 &  98.50 &  2.37  &  1.13  &  96.03 &  98.77 \\
    3 &  0.63  &  0.27  &  10.07 &  13.90 &  0.20  &  0.17  &  12.17 &  15.37 \\
    4 &  0.23  &  0.10  &  2.97  &  4.73  &  0.10  &  0.00  &  3.50  &  5.57  \\
    5 &  0.57  &  0.33  &  7.73  &  10.20 &  0.27  &  0.20  &  8.30  &  10.40 \\
    6 &  0.27  &  0.13  &  52.17 &  59.97 &  0.10  &  0.00  &  53.57 &  65.27 \\
    \hline
    \end{tabular}

    \vspace{-4 mm}
\end{table}

\section{Robustness against Adaptive Denoisers}\label{app:denoiser}

\begin{figure}[t]
    \centerline{\includegraphics[width=0.49\textwidth]{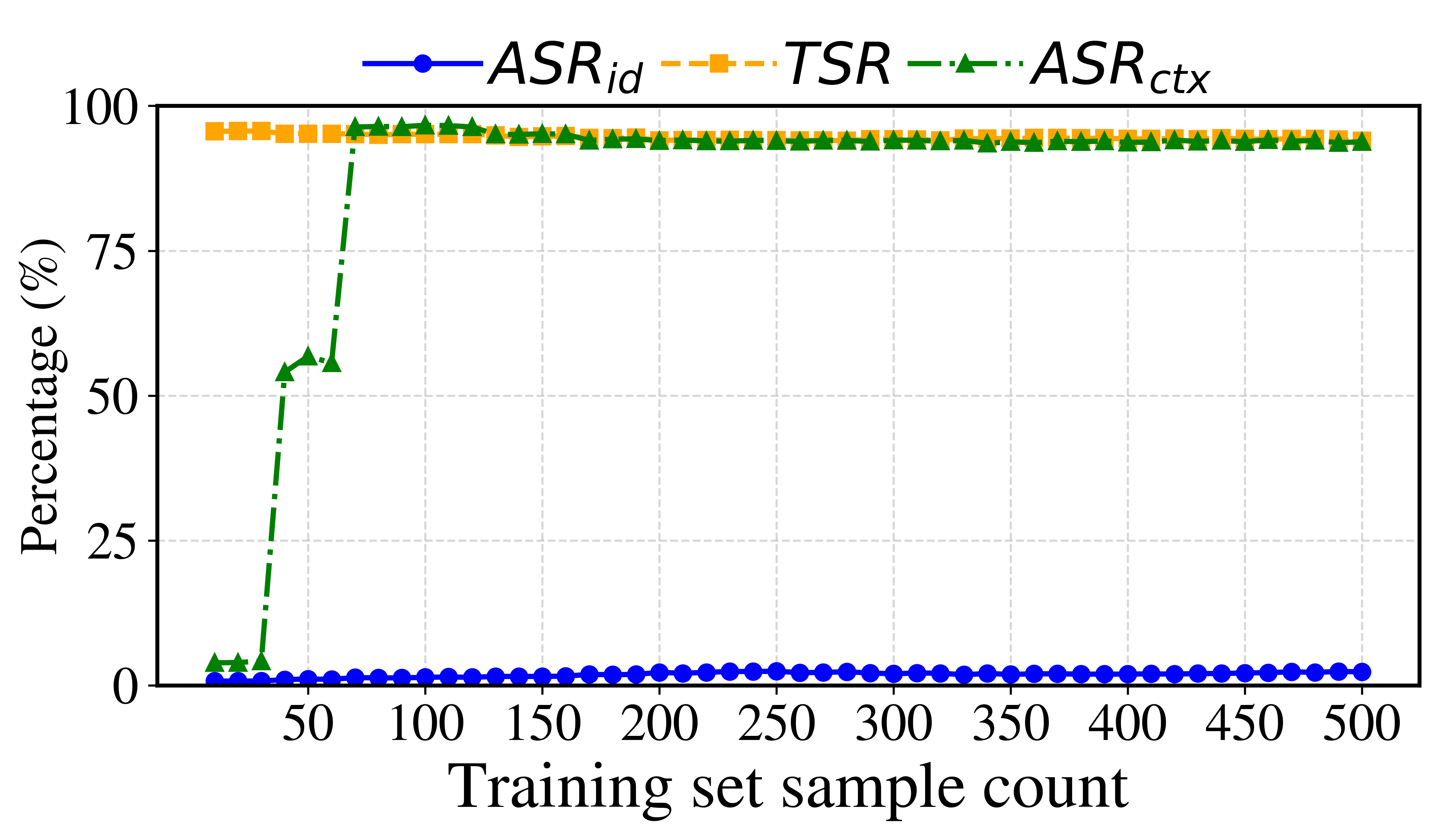}}
    \captionsetup{skip=0 pt}
    \caption{\name~robustness against denoisers trained with different numbers of training samples.}
    \Description{}
    \label{fig:denoiser}
    \vspace{-3mm}
\end{figure}

In this section, we further analyze the robustness of \name~against adaptive denoisers trained with different numbers of image pairs. To reduce API cost, all metrics are evaluated using FaceNet-512 only. As shown in Figure \ref{fig:denoiser}, $ASR_{id}$ increases only slightly as the number of training samples increases from 10 to 500, while $TSR$ remains stable throughout. When the number of training samples increases from 10 to 30, $ASR_{ctx}$ remains low and nearly unchanged. However, when the training set size reaches 40, $ASR_{ctx}$ increases sharply to more than 50\%. When the number of training samples reaches 70, the context protection fails, and additional training samples do not further improve the attack significantly. Overall, \name~maintains effective identity protection and tracing even when the adaptive denoiser is trained with 500 image pairs, while context protection remains effective only when the denoiser is trained with a relatively small number of image pairs, such as 30.

\section{Defending Autoencoder-based Deepfakes}\label{app:autoencoder}

\subsection{Methodology}\label{subsec:faceswap_methodologies}

In autoencoder-based deepfake, each identity has its own decoder. During inference, a shared encoder $\mathcal{E}$ extracts attributes from the target image, and the decoder corresponding to the source identity reconstructs the swapped face:
\begin{equation*}
\begin{aligned}
\mathbf{z}_{t} = \mathcal{E}\left(\mathbf{x}_{t}\right) \qquad
\hat{\mathbf{x}} = \mathcal{D}_{s}\left(\mathbf{z}_{t}\right)
\label{eqa:aeswap}
\end{aligned}
\end{equation*}
where $\mathcal{D}_{s}$ is the decoder trained on the source identity.

To protect Alice’s image $\mathbf{x}$, \name\ perturbs it into a protected version $\tilde{\mathbf{x}}$. For \textit{Context Protection}, our goal is to prevent accurate extraction of attributes with minimal perturbations. The objective function is similar to the one adopted by \name\ against GAN-based deepfakes:
\begin{equation}
\mathcal{L}_{context} = \alpha \|\Delta \mathbf{x}\| - \beta \|\Delta \mathbf{z}_t\| 
\label{eqa:faceswap_loss_context}
\end{equation}

However, the scope of \textit{Identity Protection} in autoencoder-based face-swapping is fundamentally different from the GAN-based case. In this setting, the attacker trains an identity-specific decoder using Alice's images. Therefore, \name~poisons this training process by generating \textit{hard-to-learn} examples.
\begin{equation}
\mathcal{L}_{identity} = \alpha \|\Delta \mathbf{x}\| - \beta \|\Delta \mathbf{z}_s\| 
\label{eqa:faceswap_loss_identity}
\end{equation}

Since the decoder is tightly coupled to the source identity, redirecting it to a completely different identity as the cloak is highly difficult, if not infeasible. Therefore, we focus on non-cloaking identity protection for autoencoder-based deepfakes.
 
Again, by combining both objectives, \name\ simultaneously protects context and identity:
\begin{equation}
\mathcal{L} = \mathcal{L}_{context} \oplus \mathcal{L}_{identity} = \alpha \|\Delta \mathbf{x}\| - \beta \|\Delta \mathbf{z}\|
\label{eqa:pgd_faceswap_all}
\end{equation}

\subsection{Experiments on FaceSwap-Deepfake-Pytorch}\label{subsec:first_autoencoder}

\noindent\textbf{Model and Dataset.} We use FaceSwap-Deepfake-Pytorch \cite{faceswap_deepfake_pytorch}, a widely used autoencoder-based face-swapping model. To reduce identity bias, we evaluate the model on five randomly chosen identities, denoted as $I_A$ to $I_E$, from the VGGFace2 dataset \cite{cao2018vggface2}, which contains 248, 237, 279, 287, and 250 images, respectively. This setup yields 20 unique source-target identity pairs.

\noindent\textbf{Parameter Settings.} For the loss function, we set $\alpha$ = 750 and $\beta$ = 0.1. The maximum number of epochs is 1,000. Losses are calculated using the L2 norm, and the upper bound for $\|\Delta \mathbf{z}\|$ is set as 75. The Human-Perception-Aware Noise Bounds for RGB channels are set to $\epsilon = [0.08, 0.03, 0.1]$. 

\noindent\textbf{Results.} The protected images show slight color shifts. Such subtle perturbations significantly disrupt face-swapping. The resulting deepfake images display unnatural distortions, gaze shifts, altered angles, blurring, and degraded quality.

\noindent$\bullet$ Utility. Table~\ref{tab:faceswap_protection_performance} presents utility measurements of all the protected images (1301 images from 5 identities). MSE and LPIPS are low, while PSNR and SSIM remain high. FaceNet-512 matching rates stay above 99.4\% and Face++ matching rate is 94.9\%, confirming that the protected images still retain high utility and resemble the original person. 

\begin{table}[t]
  \centering
  \captionsetup{skip=3pt}
  \caption{Utility of \name\ on FaceSwap-Deepfake-Pytorch}.
  \label{tab:faceswap_protection_performance}
  \setlength\tabcolsep{7pt}
  \renewcommand{\arraystretch}{1.1}
  \begin{tabular}{c | c | c | c | c }
    \hline
    Input & MSE$\downarrow$ & PSNR$\uparrow$ & SSIM$\uparrow$ & LPIPS$\downarrow$ \\
    \hline
    $I_A$ & 128 & 27.07 & 0.86 & 0.08 \\
    $I_B$ & 112 & 27.65 & 0.87 & 0.08 \\
    $I_C$ & 108 & 27.81 & 0.87 & 0.07 \\
    $I_D$ & 107 & 27.83 & 0.87 & 0.07 \\
    $I_E$ & 111 & 27.68 & 0.86 & 0.08 \\
    \hline
  \end{tabular}
  \vspace{-3mm}
\end{table}

\begin{table}[]
  \centering
  \captionsetup{skip=3pt}
  \setlength\tabcolsep{3.0 pt}
  \caption{Protect the identity over FaceSwap-Deepfake-Pytorch}: Deepfake is trained with original images of $I_A$ and 10\% \name\ perturbed images of $I_E$ (early stop at lowest training loss)
  \label{tab:faceswap_identity}
  
  \begin{tabular}{ c | c | c | c | c | c }
    \hline
    \multicolumn{3}{c}{$I_A$ (unprotected)} \vline & \multicolumn{3}{c}{$I_E$ (with 10\% protected images)} \\
    \hline
    Alice ($\mathbf{x}$) & Rebuild & w/ Pt. & Alice ($\mathbf{x}$) & Rebuild & w/ Pt. \\
    \toprule
    \raisebox{-0.5\height}{\includegraphics[scale=0.5]{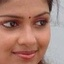}} &
    \raisebox{-0.5\height}{\includegraphics[scale=0.5]{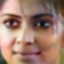}} &
    \raisebox{-0.5\height}{\includegraphics[scale=0.5]{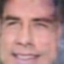}} &
    \raisebox{-0.5\height}{\includegraphics[scale=0.5]{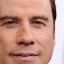}} &
    \raisebox{-0.5\height}{\includegraphics[scale=0.5]{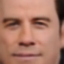}} &
    \raisebox{-0.5\height}{\includegraphics[scale=0.5]{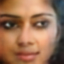}} \\
    \midrule
    \raisebox{-0.5\height}{\includegraphics[scale=0.5]{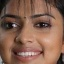}} &
    \raisebox{-0.5\height}{\includegraphics[scale=0.5]{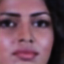}} &
    \raisebox{-0.5\height}{\includegraphics[scale=0.5]{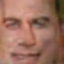}} &
    \raisebox{-0.5\height}{\includegraphics[scale=0.5]{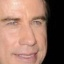}} &
    \raisebox{-0.5\height}{\includegraphics[scale=0.5]{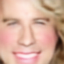}} &
    \raisebox{-0.5\height}{\includegraphics[scale=0.5]{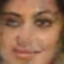}} \\
    \midrule
    \raisebox{-0.5\height}{\includegraphics[scale=0.5]{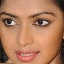}} &
    \raisebox{-0.5\height}{\includegraphics[scale=0.5]{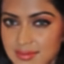}} &
    \raisebox{-0.5\height}{\includegraphics[scale=0.5]{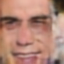}} &
    \raisebox{-0.5\height}{\includegraphics[scale=0.5]{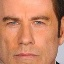}} &
    \raisebox{-0.5\height}{\includegraphics[scale=0.5]{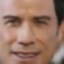}} &
    \raisebox{-0.5\height}{\includegraphics[scale=0.5]{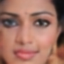}} \\
    \bottomrule
  \end{tabular}
  \vspace{-3mm}
\end{table}

\begin{figure}[]
    \centerline{\includegraphics[width=0.47\textwidth]{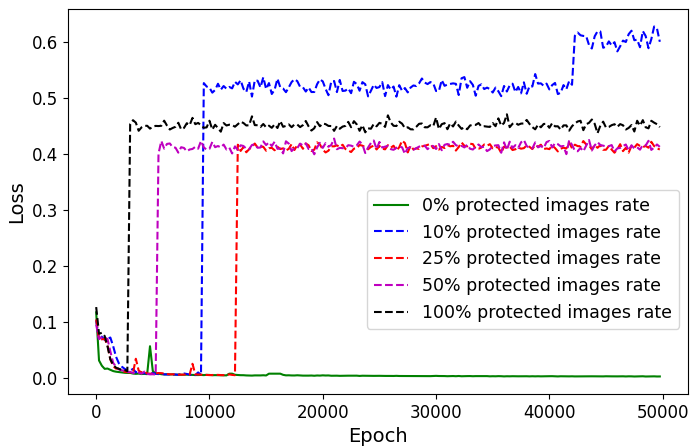}}
    \captionsetup{skip=3pt}
    \caption{Training loss with clean images of $I_A$ and protected images of $I_E$ with different poison ratios.}
    \label{fig:faceswap_loss}
    \Description{}
    \vspace{-3mm}
\end{figure}

\begin{table}[]
  \centering
  \captionsetup{skip=3 pt}
  \setlength\tabcolsep{2.7 pt}
  \caption{Context protection samples against FaceSwap-Deepfake-Pytorch}.
  \label{tab:faceswap_samples}
  
  \begin{tabular}{c  c  c  c  c  c}
    \hline
    & $I_A$ & $I_B$ & $I_C$ & $I_D$ & $I_E$ \\
    \toprule
    \raisebox{-0.5\height}{\rotatebox{90}{$\mathbf{x}$}} & 
    \raisebox{-0.5\height}{\includegraphics[scale=0.62]{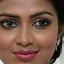}} &
    \raisebox{-0.5\height}{\includegraphics[scale=0.62]{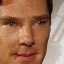}} &
    \raisebox{-0.5\height}{\includegraphics[scale=0.62]{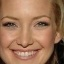}} &
    \raisebox{-0.5\height}{\includegraphics[scale=0.62]{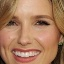}} &
    \raisebox{-0.5\height}{\includegraphics[scale=0.62]{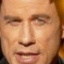}} \\
    \midrule
    \raisebox{-0.5\height}{\rotatebox{90}{w/o Pt.}} &
    \raisebox{-0.5\height}{\includegraphics[scale=0.62]{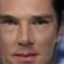}} &
    \raisebox{-0.5\height}{\includegraphics[scale=0.62]{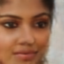}} &
    \raisebox{-0.5\height}{\includegraphics[scale=0.62]{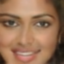}} &
    \raisebox{-0.5\height}{\includegraphics[scale=0.62]{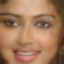}} &
    \raisebox{-0.5\height}{\includegraphics[scale=0.62]{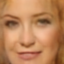}} \\
    \midrule
    \raisebox{-0.5\height}{\rotatebox{90}{$\tilde{\mathbf{x}}$}} &
    \raisebox{-0.5\height}{\includegraphics[scale=0.62]{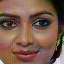}} &
    \raisebox{-0.5\height}{\includegraphics[scale=0.62]{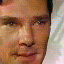}} &
    \raisebox{-0.5\height}{\includegraphics[scale=0.62]{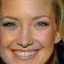}} &
    \raisebox{-0.5\height}{\includegraphics[scale=0.62]{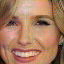}} &
    \raisebox{-0.5\height}{\includegraphics[scale=0.62]{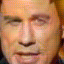}} \\
    \midrule
    \raisebox{-0.5\height}{\rotatebox{90}{w/ Pt.}} &
    \raisebox{-0.5\height}{\includegraphics[scale=0.62]{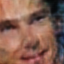}} &
    \raisebox{-0.5\height}{\includegraphics[scale=0.62]{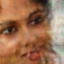}} &
    \raisebox{-0.5\height}{\includegraphics[scale=0.62]{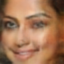}} &
    \raisebox{-0.5\height}{\includegraphics[scale=0.62]{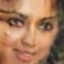}} &
    \raisebox{-0.5\height}{\includegraphics[scale=0.62]{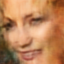}} \\
    \bottomrule
  \end{tabular}
  \vspace{-3mm}
\end{table}

\begin{table}[!ht]
  \centering
  \captionsetup{skip=3 pt}
  \setlength\tabcolsep{2.2 pt}
  \caption{Context protection samples against FaceSwap.}
  \label{tab:faceswap_gui_samples}

  \begin{tabular}{c | c  c  c  c  c  c}
    \toprule
    \multirow{4}{*}[-5.75 em]{$I_A$} &
    \raisebox{-0.5\height}{\rotatebox{90}{$\mathbf{x}$}} & 
    \raisebox{-0.5\height}{\includegraphics[scale=0.171]{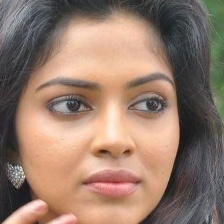}} &
    \raisebox{-0.5\height}{\includegraphics[scale=0.171]{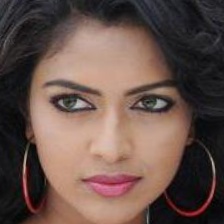}} &
    \raisebox{-0.5\height}{\includegraphics[scale=0.171]{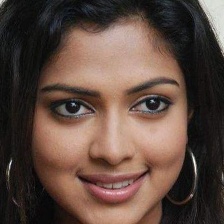}} &
    \raisebox{-0.5\height}{\includegraphics[scale=0.171]{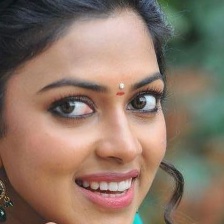}} &
    \raisebox{-0.5\height}{\includegraphics[scale=0.171]{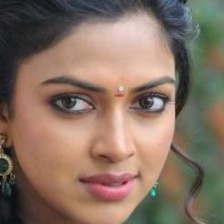}} \\
    \cmidrule(lr){2-7}
    ~ & 
    \raisebox{-0.5\height}{\rotatebox{90}{w/o Pt.}} &
    \raisebox{-0.5\height}{\includegraphics[scale=0.171]{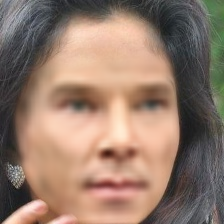}} &
    \raisebox{-0.5\height}{\includegraphics[scale=0.171]{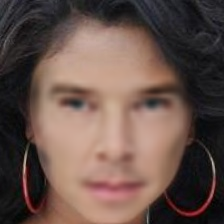}} &
    \raisebox{-0.5\height}{\includegraphics[scale=0.171]{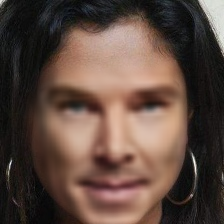}} &
    \raisebox{-0.5\height}{\includegraphics[scale=0.171]{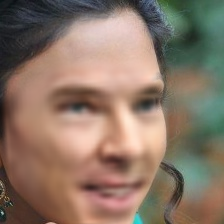}} &
    \raisebox{-0.5\height}{\includegraphics[scale=0.171]{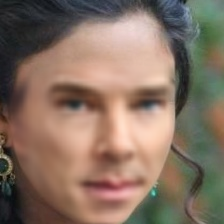}} \\
    \cmidrule(lr){2-7}
    ~ &
    \raisebox{-0.5\height}{\rotatebox{90}{$\tilde{\mathbf{x}}$}} &
    \raisebox{-0.5\height}{\includegraphics[scale=0.15]{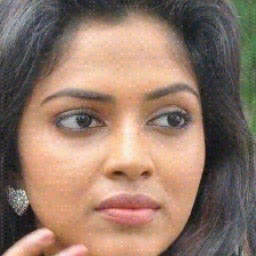}} &
    \raisebox{-0.5\height}{\includegraphics[scale=0.15]{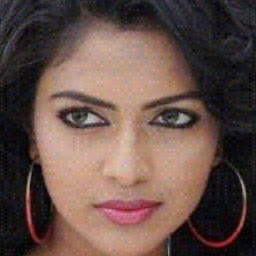}} &
    \raisebox{-0.5\height}{\includegraphics[scale=0.15]{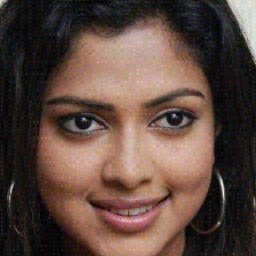}} &
    \raisebox{-0.5\height}{\includegraphics[scale=0.15]{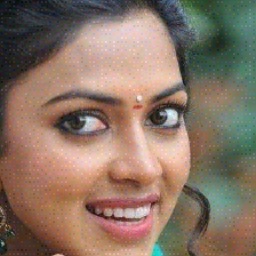}} &
    \raisebox{-0.5\height}{\includegraphics[scale=0.15]{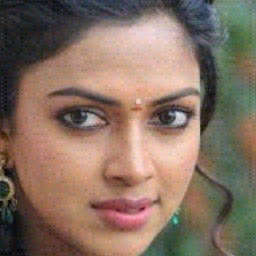}} \\
    \cmidrule(lr){2-7}
    ~ &
    \raisebox{-0.5\height}{\rotatebox{90}{w/ Pt.}} &
    \raisebox{-0.5\height}{\includegraphics[scale=0.15]{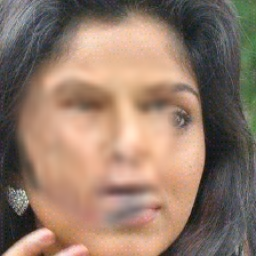}} &
    \raisebox{-0.5\height}{\includegraphics[scale=0.15]{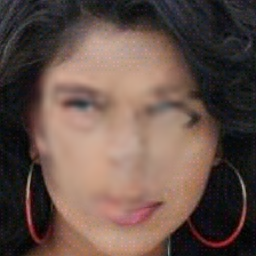}} &
    \raisebox{-0.5\height}{\includegraphics[scale=0.15]{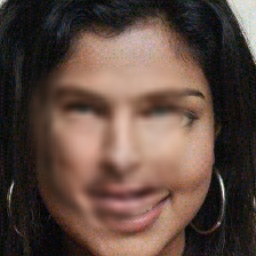}} &
    \raisebox{-0.5\height}{\includegraphics[scale=0.15]{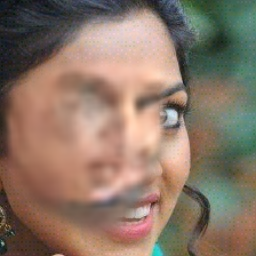}} &
    \raisebox{-0.5\height}{\includegraphics[scale=0.15]{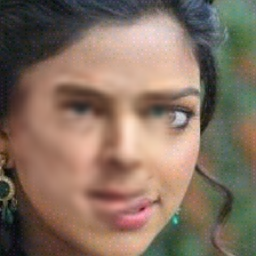}} \\
    
    \midrule

    \multirow{4}{*}[-5.75 em]{$I_B$} &
    \raisebox{-0.5\height}{\rotatebox{90}{$\mathbf{x}$}} & 
    \raisebox{-0.5\height}{\includegraphics[scale=0.171]{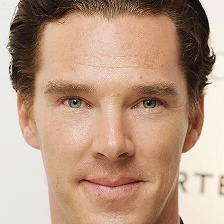}} &
    \raisebox{-0.5\height}{\includegraphics[scale=0.171]{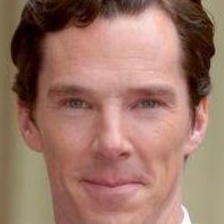}} &
    \raisebox{-0.5\height}{\includegraphics[scale=0.171]{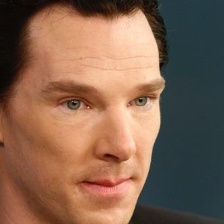}} &
    \raisebox{-0.5\height}{\includegraphics[scale=0.171]{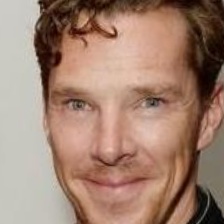}} &
    \raisebox{-0.5\height}{\includegraphics[scale=0.171]{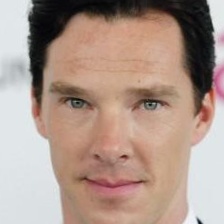}} \\
    \cmidrule(lr){2-7}
    ~ & 
    \raisebox{-0.5\height}{\rotatebox{90}{w/o Pt.}} &
    \raisebox{-0.5\height}{\includegraphics[scale=0.171]{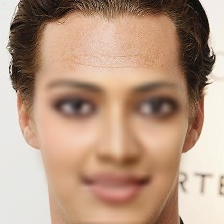}} &
    \raisebox{-0.5\height}{\includegraphics[scale=0.171]{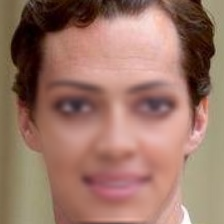}} &
    \raisebox{-0.5\height}{\includegraphics[scale=0.171]{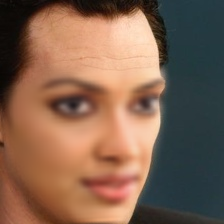}} &
    \raisebox{-0.5\height}{\includegraphics[scale=0.171]{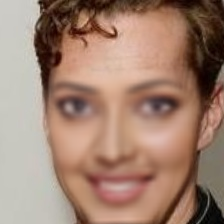}} &
    \raisebox{-0.5\height}{\includegraphics[scale=0.171]{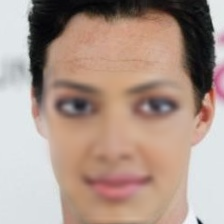}} \\
    \cmidrule(lr){2-7}
    ~ &
    \raisebox{-0.5\height}{\rotatebox{90}{$\tilde{\mathbf{x}}$}} &
    \raisebox{-0.5\height}{\includegraphics[scale=0.15]{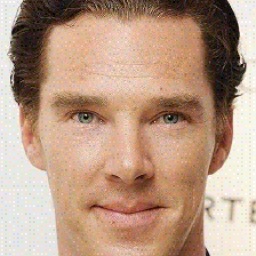}} &
    \raisebox{-0.5\height}{\includegraphics[scale=0.15]{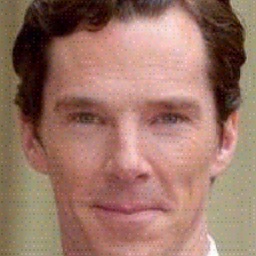}} &
    \raisebox{-0.5\height}{\includegraphics[scale=0.15]{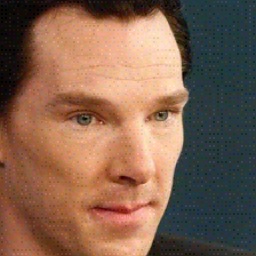}} &
    \raisebox{-0.5\height}{\includegraphics[scale=0.15]{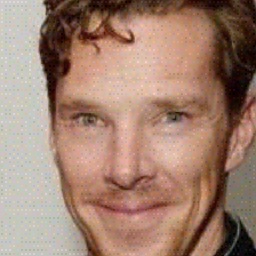}} &
    \raisebox{-0.5\height}{\includegraphics[scale=0.15]{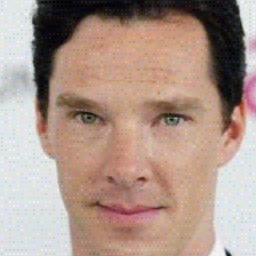}} \\
    \cmidrule(lr){2-7}
    ~ &
    \raisebox{-0.5\height}{\rotatebox{90}{w/ Pt.}} &
    \raisebox{-0.5\height}{\includegraphics[scale=0.15]{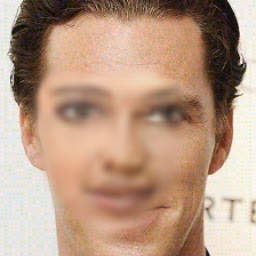}} &
    \raisebox{-0.5\height}{\includegraphics[scale=0.15]{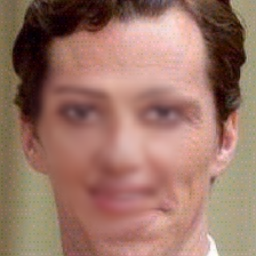}} &
    \raisebox{-0.5\height}{\includegraphics[scale=0.15]{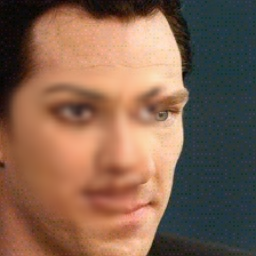}} &
    \raisebox{-0.5\height}{\includegraphics[scale=0.15]{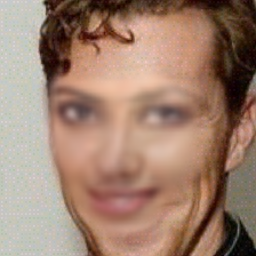}} &
    \raisebox{-0.5\height}{\includegraphics[scale=0.15]{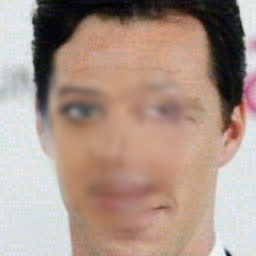}} \\
    \bottomrule
  \end{tabular}
\end{table}

\noindent$\bullet$ Identity Protection. As shown in Fig~\ref{fig:faceswap_loss}, even a small portion of protected images in the autoencoder training data will cause a sharp spike in training loss, and high poison ratios lead to fast failure. The generated deepfake outputs become nearly monochromatic and unrecognizable. We also test an early-stopping strategy by the attacker when he selects the checkpoint with the lowest training loss. As shown in Table~\ref{tab:faceswap_identity}, even with 10\% poisoned data, the best checkpoint is still insufficient: while some identity features remain, contextual and attribute details are severely distorted. The resulting deepfakes are low-quality and highly noticeable.

\noindent$\bullet$ Context Protection. \name\ is also highly effective in context protection. As shown in Table \ref{tab:faceswap_samples}, deepfake outputs generated from protected images suffer severe quality degradation (row 4). Quantitatively, MSE between the successful deepfakes (row 2 of Table \ref{tab:faceswap_samples}) and failed deepfakes (row 4 of Table \ref{tab:faceswap_samples}) increases significantly, e.g., MSE ranges between 806 and 1403 for the samples in Table \ref{tab:faceswap_samples}. 

\subsection{Experiments on FaceSwap and DeepFaceLive}

In this section, we further evaluate the protection performance of \name~on two additional popular autoencoder-based face-swapping pipelines, FaceSwap\cite{faceswap} and DeepFaceLive~\cite{deep_face_live}. We adopt the same protection strategy described in Section~\ref{subsec:faceswap_methodologies}.

FaceSwap provides a graphical user interface (GUI) and supports multiple operating systems and hardware platforms, making it one of the most popular open-source face-swapping tools on GitHub, with more than 55.2 thousand stars. Because FaceSwap only supports face swapping between a single pair of identities, we use identities A and B introduced in Section~\ref{subsec:first_autoencoder} to train the face-swapping model and evaluate the protection performance. As shown in Table~\ref{tab:faceswap_gui_samples}, \name~is also highly effective at protecting image context against FaceSwap. With only minor perturbations (MSE=74, PSNR=29.44, SSIM=0.71, LPIPS=0.31), \name~significantly disrupts both the context extractor and the face-swapping pipeline. In all examples, the face-swapping results generated from protected images exhibit obvious artifacts and severe facial misalignment.

DeepFaceLive~\cite{deep_face_live} also provides a GUI and supports multiple operating systems and hardware platforms. It is a popular open-source face-swapping tool on GitHub, with 30.8 thousand stars. In addition, DeepFaceLive provides several pre-trained models to facilitate usage, allowing users to transfer identities from a predefined set of 31 identities, such as Mr. Bean, to their own images. In our evaluation, we use Mr. Bean as the source identity and report the results in Table~\ref{tab:deepfacelive_samples}. Without protection, DeepFaceLive successfully swaps the identity of Mr. Bean into the target image. In contrast, although the protected images introduce only minor perturbations (MSE=111, PSNR=28.04, SSIM=0.66, LPIPS=0.43), they disrupt 81.97\% of the face-swapping pipeline, causing the process to fail without producing any output.

\begin{table}[!ht]
  \centering
  \captionsetup{skip=3 pt}
  \setlength\tabcolsep{1pt}
  \caption{Context protection samples against DeepFaceLive.}
  \label{tab:deepfacelive_samples}

  \begin{tabular}{c  c  c  c  c  c}
    \toprule
    \raisebox{-0.5\height}{\rotatebox{90}{$\mathbf{x}$}} & 
    \raisebox{-0.5\height}{\includegraphics[scale=0.171]{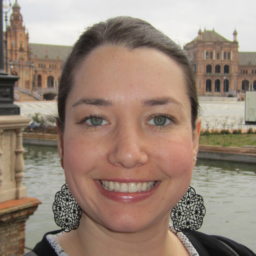}} &
    \raisebox{-0.5\height}{\includegraphics[scale=0.171]{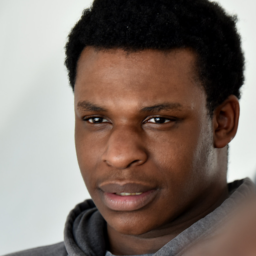}} &
    \raisebox{-0.5\height}{\includegraphics[scale=0.171]{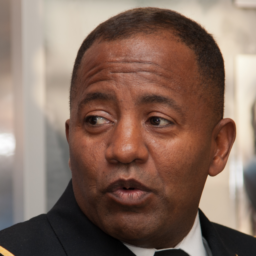}} &
    \raisebox{-0.5\height}{\includegraphics[scale=0.171]{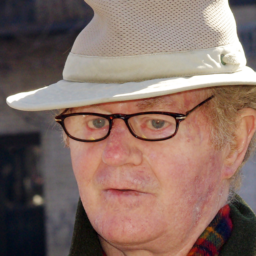}} &
    \raisebox{-0.5\height}{\includegraphics[scale=0.171]{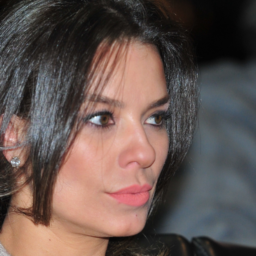}} \\
    \midrule
    \raisebox{-0.5\height}{\rotatebox{90}{w/o Pt.}} &
    \raisebox{-0.5\height}{\includegraphics[scale=0.171]{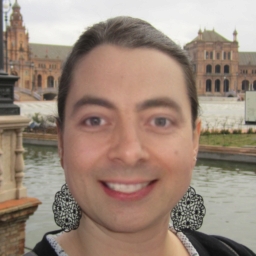}} &
    \raisebox{-0.5\height}{\includegraphics[scale=0.171]{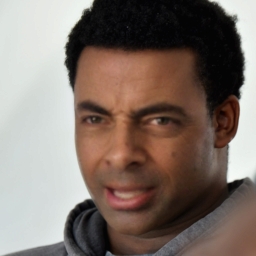}} &
    \raisebox{-0.5\height}{\includegraphics[scale=0.171]{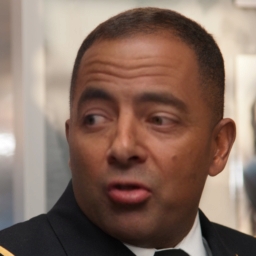}} &
    \raisebox{-0.5\height}{\includegraphics[scale=0.171]{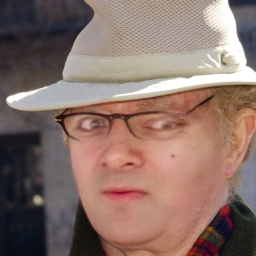}} &
    \raisebox{-0.5\height}{\includegraphics[scale=0.171]{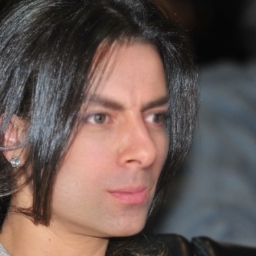}} \\
    \midrule
    \raisebox{-0.5\height}{\rotatebox{90}{$\tilde{\mathbf{x}}$}} &
    \raisebox{-0.5\height}{\includegraphics[scale=0.171]{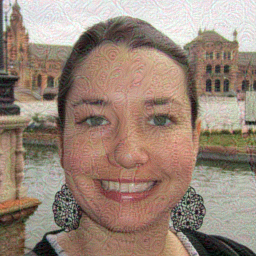}} &
    \raisebox{-0.5\height}{\includegraphics[scale=0.171]{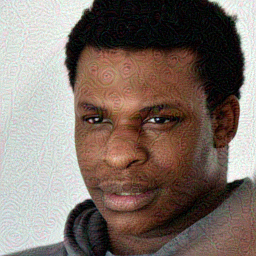}} &
    \raisebox{-0.5\height}{\includegraphics[scale=0.171]{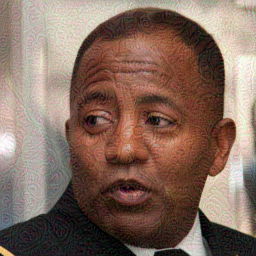}} &
    \raisebox{-0.5\height}{\includegraphics[scale=0.171]{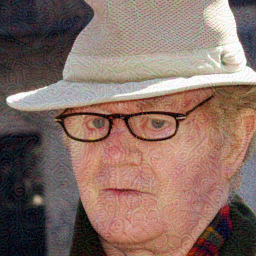}} &
    \raisebox{-0.5\height}{\includegraphics[scale=0.171]{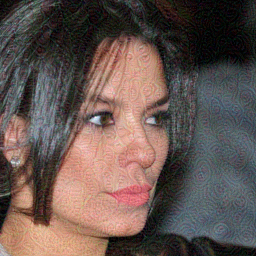}} \\
    \bottomrule
  \end{tabular}
\end{table}

\end{document}